# Mirror vs. inversion symmetry breaking in mesogenic dimers: $N_{TB}$ vs. $N_F$ phase

Małgorzata Bakiera[a], Jakub Karcz[a], Damian Pociecha[b], Katarzyna Kwiatkowska[b], Mateusz Pawlak[b], Przemysław Kula[a], Ewa Gorecka*[b]

[a] M. Bakiera, J. Karcz, prof. P. Kula
Faculty of Advanced Technology and Chemistry,
Military University of Technology
Sylwestra Kaliskiego 2, 00-908 Warsaw, Poland

[b] prof. D. Pociecha, K. Kwiatkowska, dr M. Pawlak, prof. E. Gorecka
Faculty of Chemistry
University of Warsaw
Zwirki i Wigury 101, 02-089 Warsaw, Poland

*E-mail*: gorecka@chem.uw.edu.pl

**Abstract:** The recently discovered twist-bend nematic $N_{TB}$ and ferroelectric nematic $N_F$ phases are distinct examples of spontaneous symmetry breaking in liquid crystals. Here, we report the occurrence of both type nematic phases within the same homologous series of dimers consisting of two strongly dipolar mesogenic units linked by a flexible spacer. The $N_F$ phase, observed for dimers having spacer with even number of atoms, exhibits the strong polar order; in the $N_{TB}$ phase, observed for dimers with odd number of atoms in the spacer, short-pitch heliconical director structure develops.

Mesogenic dimers provide an exceptional platform for investigating the interplay between molecular shape and spontaneous symmetry breaking in liquid crystals [1]. Their molecular geometry can be readily tuned, from bent to nearly linear, simply by varying the parity of the flexible spacer linking mesogenic cores. Whereas nearly rod-like dimers typically exhibit only the conventional nematic phase, bent dimers stabilize the twist-bend nematic ($N_{TB}$) phase, characterized by a heliconical director modulation with a nanoscale pitch [2–6]. Introducing an additional design parameter, a large longitudinal dipole moment of the mesogenic units, raises the possibility of a competition between chiral and polar symmetry breaking. Here, we demonstrate that the relative orientation of the mesogenic core dipoles, together with the parity of the spacer, determines whether the system adopts a ferroelectric nematic ($N_F$) or a heliconical twist-bend nematic ($N_{TB}$) ground state. The discovery of the ferroelectric state driven by intermolecular dipole–dipole interactions in liquid state [7] represented a major breakthrough, as such interactions had long been considered too weak to overcome thermal fluctuations and establish long-range polar order. To date, the $N_F$ phase has been reported primarily for low-molar-mass mesogens with exceptionally large longitudinal dipole moments (typically around 10 D) [8–10] and only in a limited number of oligomeric systems [11]. In contrast, the relationship between polar ordering and spontaneous bend deformation in liquid-crystalline dimers remains largely unexplored. To study the competition between polar and helical order, we synthesized two homologous series of dimers composed of highly polar mesogenic units derived from the prototype ferroelectric nematogens, DIO [9] and UUQU [12]. In the **n-I** series, mesogenic cores were linked with their longitudinal dipoles aligned in the same direction, whereas in the **II-m** series two mesogenic units were connected in an antiparallel arrangement (Figure 1). In the **n-I** series, the parity of the flexible spacer was varied to control the molecular shape: even-membered spacer (**n** being odd number) produced nearly linear dimers, whereas odd-membered spacer (**n** being even number) generated bent molecular conformations. In the **II-m** series, the spacer parity was kept constant (9 atoms in the spacer) to preserve the bent molecular shape, while the length of the terminal alkyl chains (**m**) was varied.

The series **n-I** exhibit typical odd-even effect of clearing point temperature, the rod-like dimers have significantly larger thermal stability of the mesophase than the bent ones. Moreover, elongation of linking group has an opposite effect on the stability of nematic phase for bent- and rod-like dimers, in case of bent dimers clearing temperature decreases with elongation of the linkage, while it increases for linear ones. These changes can be related to different influence of the spacer length on the overall molecular shape anisotropy [1]. All compounds of series **n-I** form a conventional nematic phase; compounds with **n** = 8, 10 exhibit also the twist-bend nematic phase, while for compounds with **n** = 7, 9 the ferroelectric $N_F$ phase is found. In all nematic phases X-ray diffraction studies (XRD) show only diffused signals, pointing to lack of long range positional order of molecules (Fig. S1 and S2). Position of the low angle diffraction signal corresponds to half molecular length, evidencing local intercalated structure.

Homologues of series **II-m** with short terminal chains (**m**=3,4) exhibit the N and $N_{TB}$ phases, whereas compound with longer terminal chain (**m**=5) forms the smectic phase below N phase. Smectic layer spacing was determined to be half the molecular length (intercalated structure) and a diffused signal at high diffraction angle was registered, confirming liquid like order inside the layers (Fig. S3). The optical textures (Fig. S4) and the decrease of the layer spacing with lowering temperature (Fig. S5) suggested the tilted-type smectic phase. The enhanced thermal stability of the intercalated lamellar structure for the longest homologue can be attributed to better match between the length of lateral tails and the inner flexible spacer [13].

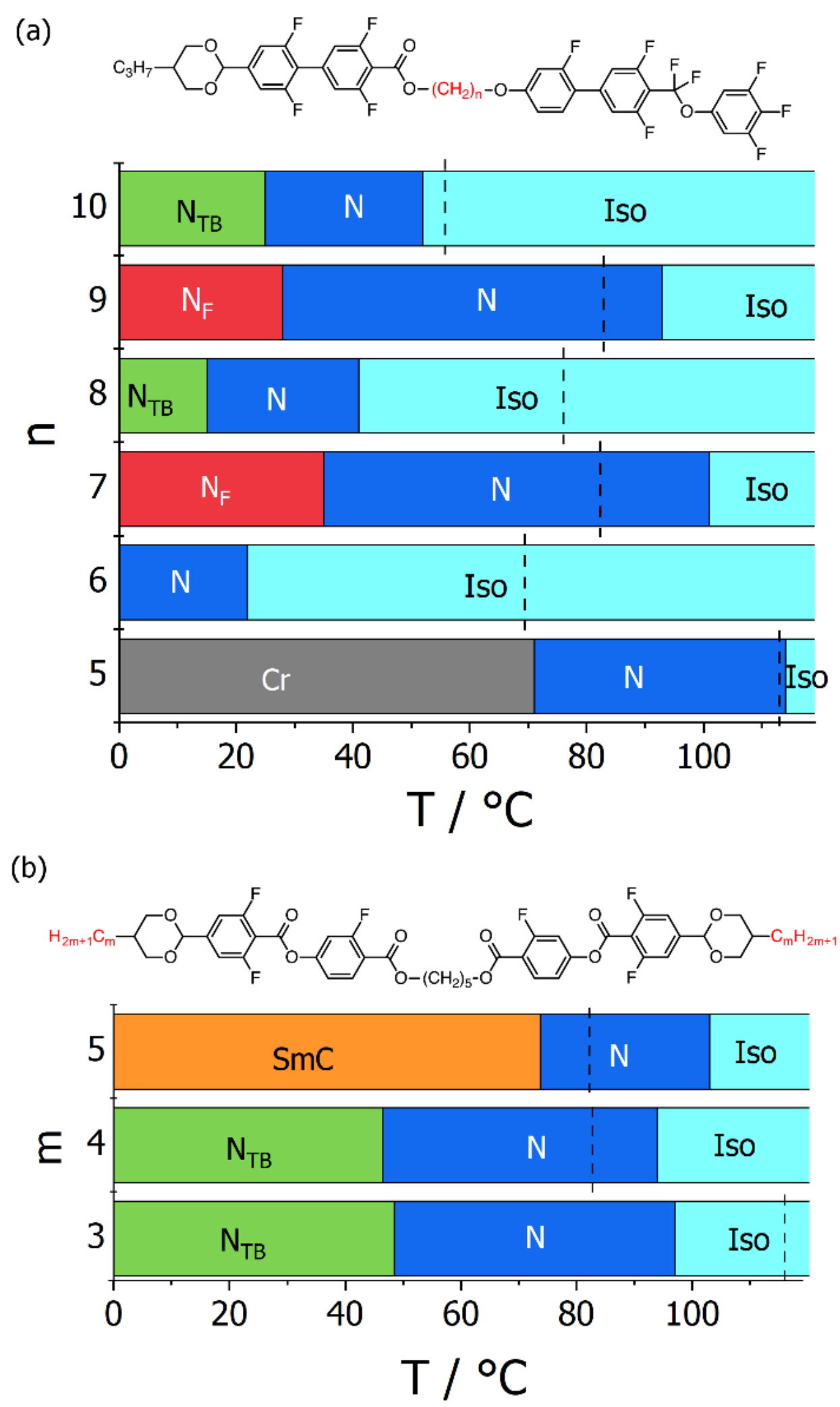

**Figure 1.** Molecular structure and observed phase sequences for homologues of series **n-I** (a) and **II-m** (b). The dashed lines show the melting temperature.

The phase sequence N-$N_{TB}$ was identified based on observation of characteristic textures and changes in optical birefringence (Fig. 2). The $N_{TB}$ phase, when observed in thin cells with planar anchoring condition, showes stripe patterns, stripes with the periodicity determined by the cell thickens are due to the formation of in-plane chevron structure [14]. The optical birefringence increases monotonically on cooling in the N phase, following the increase of orientational order of molecules. In the $N_{TB}$ phase it decreases, as formation of heliconical structure lowers the refractive index along the helix axis and increases refractive index perpendicular to it (Fig. 2c) [15]. For compound **II-3** AFM measurements, performed at room temperature, revealed regular array of parallel lines, corresponding to the helical structure; the determined helical pitch ~100 nm (Fig. S7) was somehow longer than measured previously for other $N_{TB}$ materials [5,16,17].

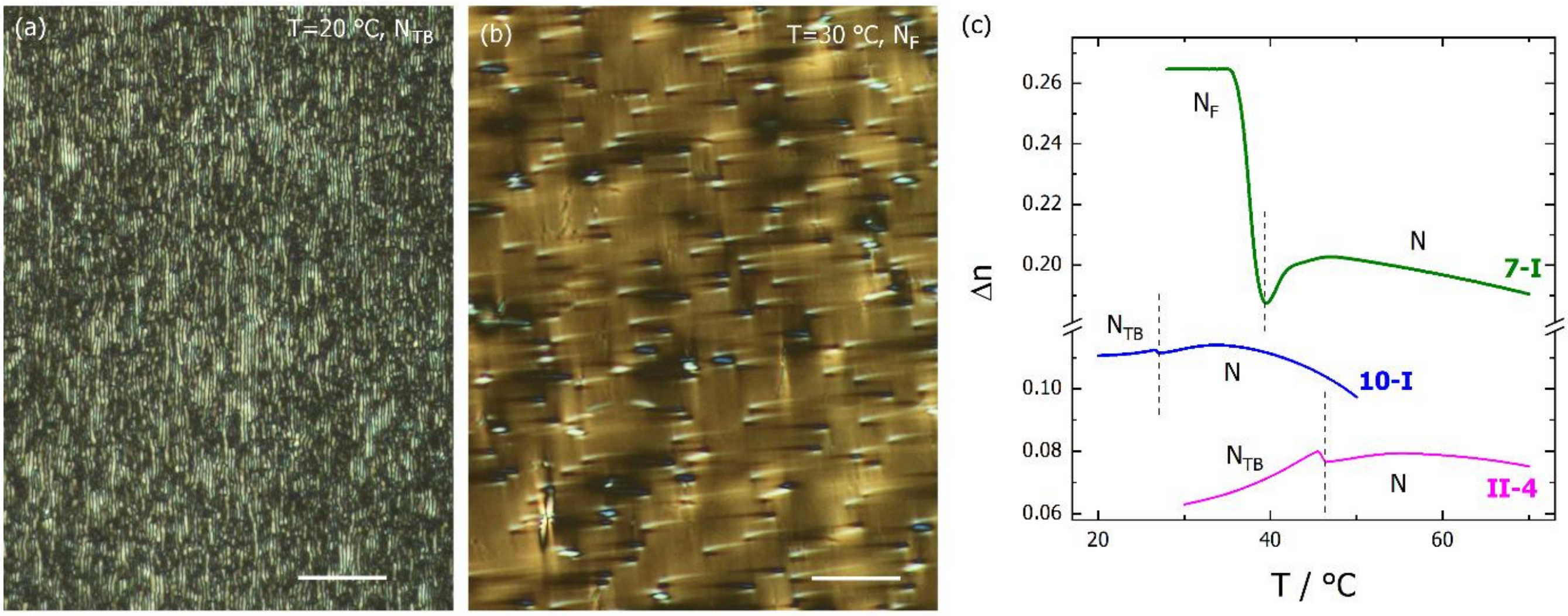

**Figure 2.** Optical texture of $N_{TB}$ phase of compound **10-I** (a) and $N_F$ phase of compound **7-I** (b), observed between crossed polarizers in 1.6-μm-thick cells with planar anchoring condition; scale bars correspond to 50 μm. (c) Optical birefringence vs. temperature for linear dimer **7-I** (green line) and bent dimers: **10-I** (blue line) and **II-4** (magenta line).

As expected, the birefringence of linear dimers in N phase was significantly higher than those of bent ones (Fig. 2c). For linear dimers at N-$N_F$ phase transition a blocky-type optical texture is formed (Fig. 2b), and the birefringence increases in $N_F$ phase (Fig. 2c), showing that the polar order is positively coupled to orientational order of molecules.

We have also studied elastic properties of these materials, by performing Fréedericksz transition experiments. For homogeneous, planar-aligned nematic cells, the capacitance as a function of the applied voltage, defined by the effective dielectric permittivity, $\varepsilon_{LC}(U)$, can be described within the continuum theory of director reorientation [18,19]. Within this framework, the splay elastic constant, $K_{11}$, is determined from the threshold voltage of the Fréedericksz transition, whereas the bend elastic constant, $K_{33}$, primarily affects the shape of the capacitance response at higher voltages, where strong bend deformations of the director field develop. By fitting the entire experimental $\varepsilon_{LC}(U)$ dependence with the theoretical model [18], both elastic constants were determined simultaneously, as well as dielectric anisotropy (Fig. S8). In N phase of linear dimer **9-I**, (having N-$N_F$ phase sequence) a splay Fréedericksz transition with a well-defined threshold at low voltage (in the range 0.4-0.5 V) occurs, that for the given dielectric anisotropy ($\varepsilon_{\parallel}/\varepsilon_{\perp}\sim$ up to15) provides $K_{11}$ in the range 10-20 pN (Fig.3), significantly higher than reported for RM-734, the prototype $N_F$ material [20-22] and for weakly polar 5CB mesogen [23]. This enhancement may be attributed to the larger molecular anisotropy of the dimeric molecules, since the elastic constants are expected to scale approximately with the square of the molecular length ($K \propto L^2$) [24, 25]. Comparable values of $K_{11}$ in N phase were found also for bent dimer **10-I** with parallel arrangement of mesogenic core dipoles ($K_{11}$ in the range 5-15 pN), and slightly smaller for bent dimer **II-4** having antiparallel arrangement of mesogenic core dipoles ($K_{11}$ in the range 1-12 pN), both bent dimers undergo the N-$N_{TB}$ phase transition upon cooling. For the linear dimer, the bend elastic constant, $K_{33}$, exceeds the splay elastic constant, with $K_{33}/K_{11} \approx 2$. The $\varepsilon_{LC}(U)$ curve differed markedly at high voltage for the non-linear and linear dimers (insets in Fig. 3). For bent dimers, above the threshold voltage the $\varepsilon_{LC}(U)$ increases steeply, indicating a low value bend elastic constant $K_{33}$. In contrast to linear dimers, no perfect agreement between the experimental data and the continuum model could be achieved even when very small positive values of $K_{33}$<0.1pN is assumed. This suggests that, for the investigated bent dimers, the $K_{33}$ is close to zero or may even become negative. Behavior corresponding to $K_{33} \ll K_{11}$ is consistent with our expectations for bent dimeric liquid crystals as the intrinsic molecular curvature reduces the bend elastic constant of the phase. The resulting preference for bend deformations is a characteristic feature of non-linear dimers and underlies the stabilization of twist-bend nematic phases at lower temperature [26, 27].

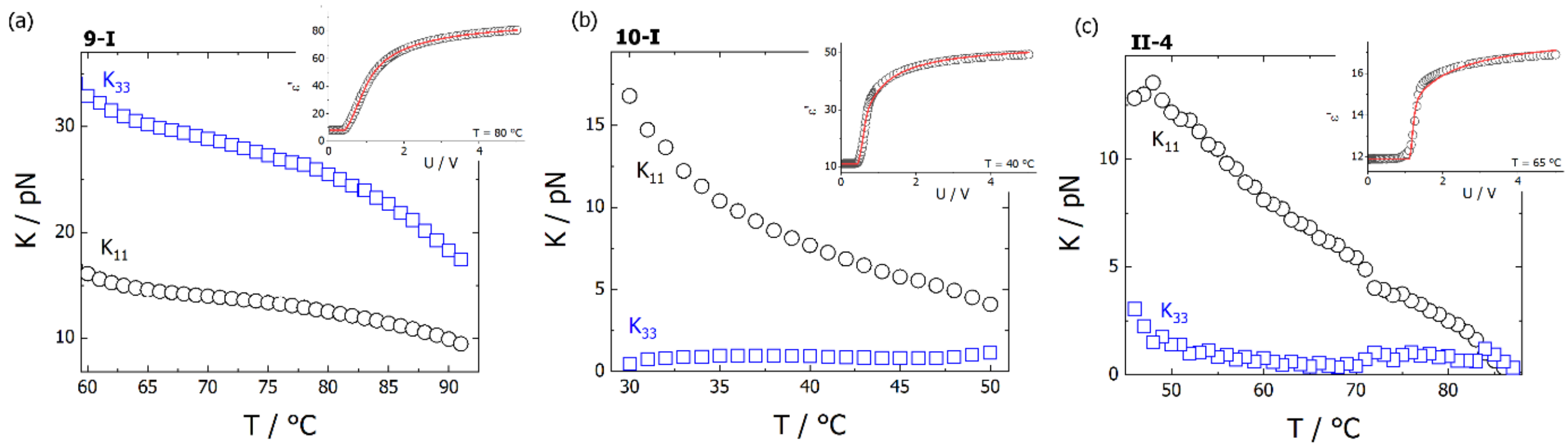

**Figure 3.** Splay $K_{11}$ and bent $K_{33}$ elastic constants for linear dimer **9-I** (a) and bent dimers **10-I** (b) and **II-4** (c). In the insets, the $\varepsilon_{LC}$ vs. applied voltage U, measured in chosen temperatures in 5-μm-thick cells with planar anchoring conditions. Red lines present the best fit to model of continuum director reorientation, equations given in SI.

Since for all pure materials $N_{TB}$ or $N_F$ phases were monotropic, to stabilize them over broader temperature range (suppress recrystallization) the mixtures of homologues were prepared: **mixture 1** with 50% wt. of **9-I** with **7-I** showing $N_F$ phase, and **mixture 2** of 20% wt. **8-I** with **10-I** showing $N_{TB}$ phase, that allowed for more detailed studies of phase properties. In $N_F$ phase of **mixture 1** clear repolarization current peak was observed upon application of ac voltage, that together with strong dielectric response at low frequency and SHG activity, evidenced unambiguously the ferroelectric nature of the phase (Fig. 4).

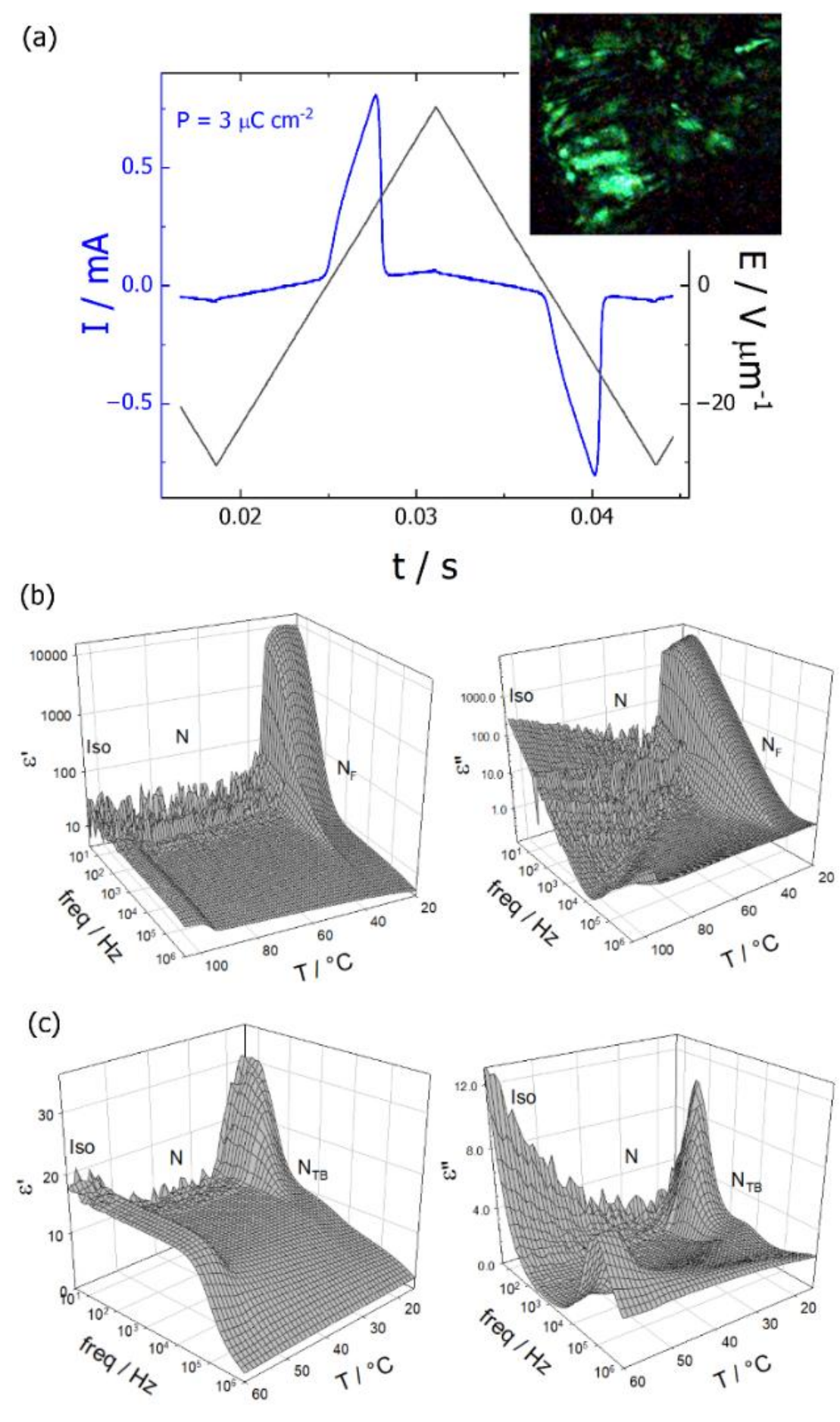

**Figure 4.** Polar properties of mixtures: (a) Repolarization current profile recorded in $N_F$ phase in **mixture 1** (50-50 wt.% of **9-I** and **7-I** , blue line) under application of triangular-wave voltage, with single peak per half a cycle of applied voltage; in the inset, SHG-microscopy image taken under illumination of the sample with IR light (λ=1064 nm), green light emission proves non-cetnrosymmetric, polar structure of the phase; (b,c) real and imaginary parts of dielectric permittivity measured vs. temperature and frequency for (b) **mixture 1** and (c) **mixture 2** of 20 wt.% **8-I** with **10-I**.

The spontaneous electric polarization determined by integration of switching current peak, $P_s$~3 μC $cm^{-2}$, was somehow lower than typically reported for $N_F$ materials [8-10]. In $N_F$ phase strong relaxation mode was observed at ~ 10 Hz frequency (Fig. 4b), in preceding nematic phase the weak mode is observed that softens (lowers relaxation frequency) on approaching the $N_F$ phase, a typical behavior upon building the polar order [28, 29]. In $N_{TB}$ phase of the **mixture 2** the dielectric response was much weaker (Fig. 4c), however stronger than in N phase, showing only short range correlations of dipoles. Two relaxation modes were found that reflect non-collective rotational motions along short and long molecular axis [30], lower frequency mode that is due to the rotation around short axis being stronger, as the molecule has strong longitudinal dipole moment and nearly compensated transverse component of dipole moment.

Summarizing: New dimeric molecules were synthesized, in which two highly polar mesogenic units were linked by a flexible spacer in a parallel arrangement. This molecular design enabled, for the first time, occurrence of either the ferroelectric nematic ($N_F$) phase and the non-polar twist-bend nematic ($N_{TB}$) phase within a single homologous series. The type of the nematic phase was found to depend solely on the parity of the spacer connecting the mesogenic cores. These findings demonstrate that the parity of the spacer provides an efficient molecular design parameter for controlling phase structure, the symmetry of the nematic ground state can be switched between the $N_F$ and $N_{TB}$ phases without changing the chemical nature of the mesogenic units. The results provides useful guidelines for the rational design of advanced liquid-crystalline materials with tunable electro-optic and electromechanical functionalities.

## Acknowledgements:

Authors acknowledge the financial support of the National Science Centre (Poland) 2024/53/B/ST5/03275.

*Supporting Information for*

# Mirror vs. inversion symmetry breaking in mesogenic dimers: $N_{TB}$ vs. $N_F$ phase

Małgorzata Bakiera[a], Jakub Karcz[a], Damian Pociecha[b], Katarzyna Kwiatkowska[b], Mateusz Pawlak[b], Przemysław Kula[a], Ewa Gorecka*[b]

[a] M. Bakiera, J. Karcz, prof. P. Kula
Faculty of Advanced Technology and Chemistry,
Military University of Technology
Sylwestra Kaliskiego 2, 00-908 Warsaw, Poland

[b] prof. D. Pociecha, K. Kwiatkowska, dr M. Pawlak, prof. E. Gorecka
Faculty of Chemistry
University of Warsaw
Zwirki i Wigury 101, 02-089 Warsaw, Poland

*e-mail*: gorecka@chem.uw.edu.pl

## Contents

## Additional results

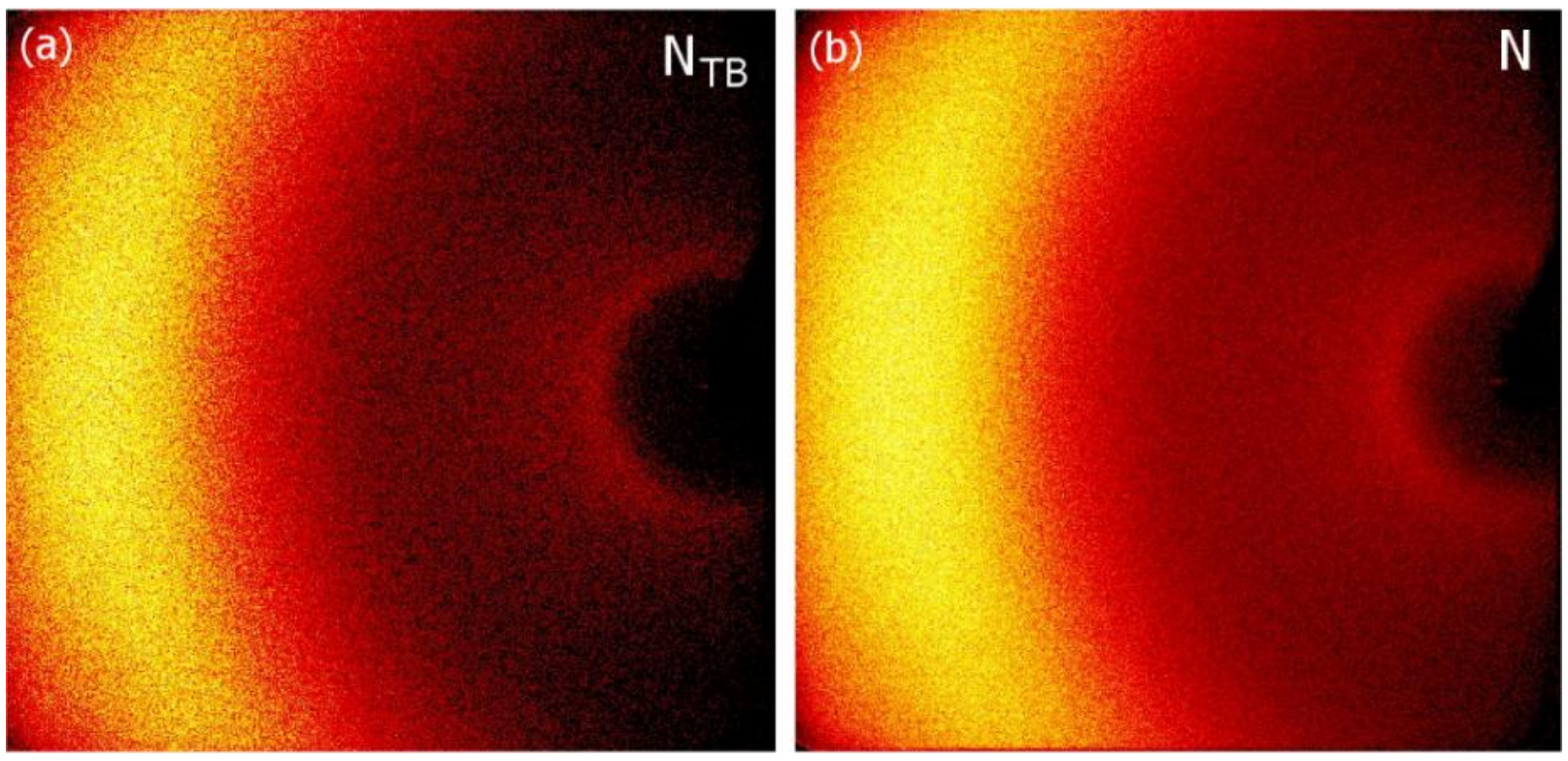

**Figure S1**. 2D X-ray diffraction patterns recorded in (a) $N_{TB}$ and (b) N phases of compound **10-I**. Lack of sharp, Bragg-type signals points to short-range positional order in both phases.

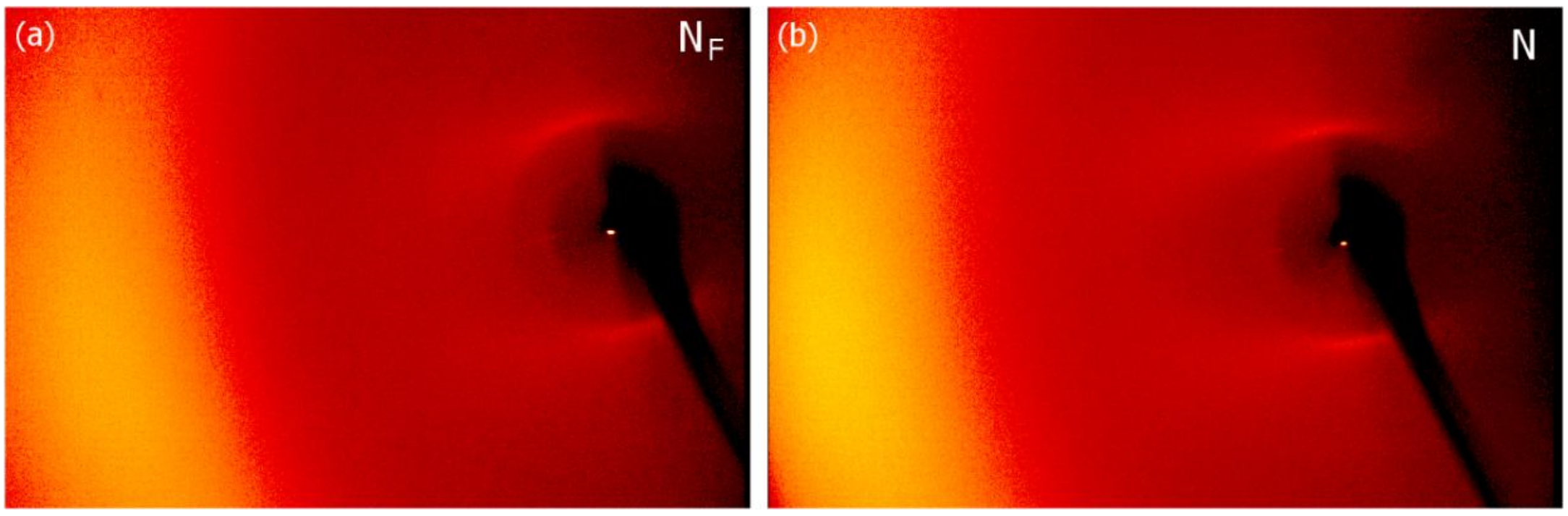

**Figure S2**. 2D X-ray diffraction patterns recorded in (a) $N_F$ and (b) N phases of **mixture 1** (50% wt. of **9-I** with **7-I**). Lack of sharp, Bragg-type signals points to short-range positional order in both phases.

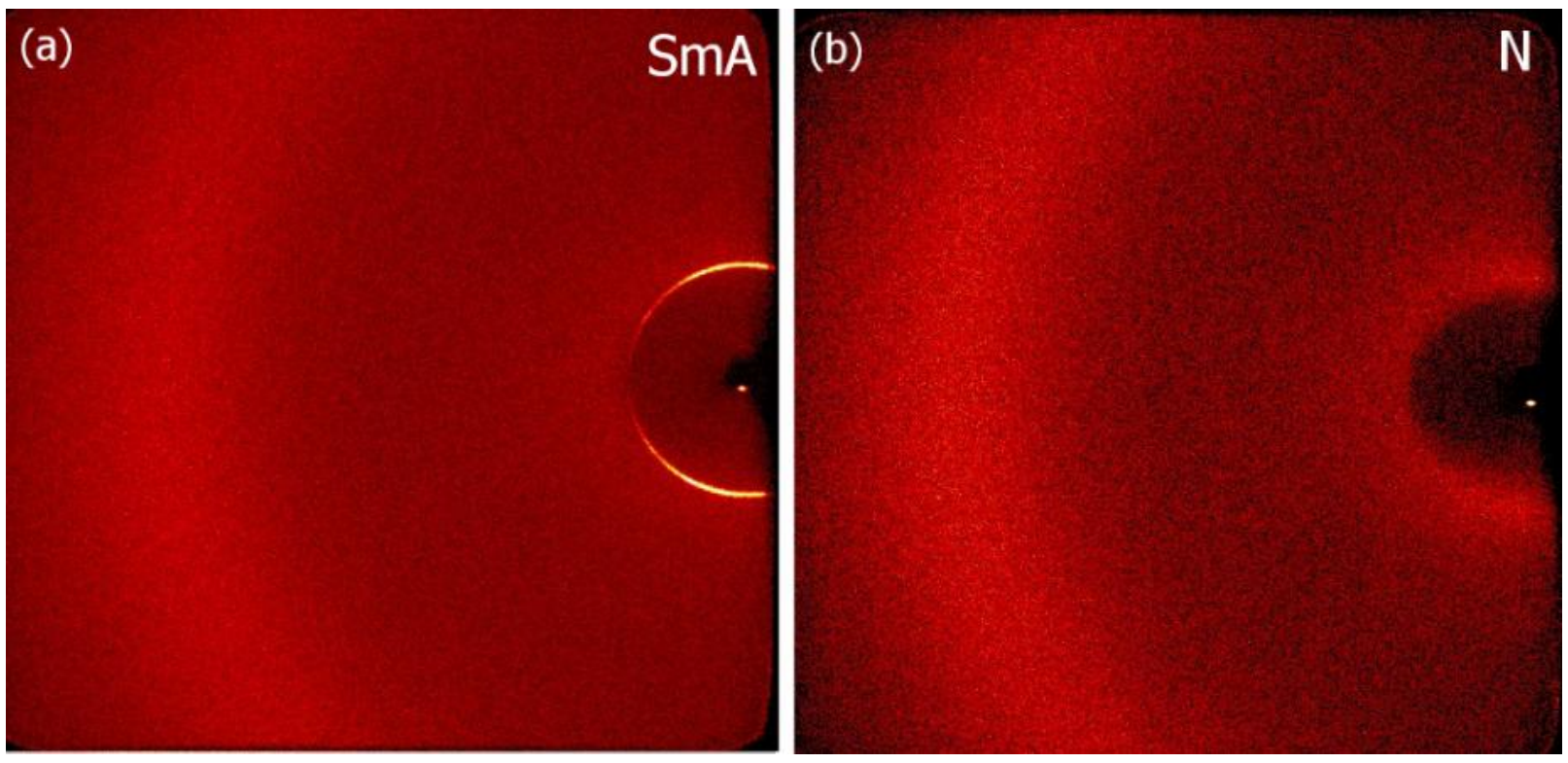

**Figure S3**. 2D X-ray diffraction patterns recorded in (a) Smectic and (b) N phases of compound **II-5**. Sharp, Bragg-type low-angle signal recorded in smectic phase evidences lamellar structure of the phase, with one-dimensional long-range positional order of molecules.

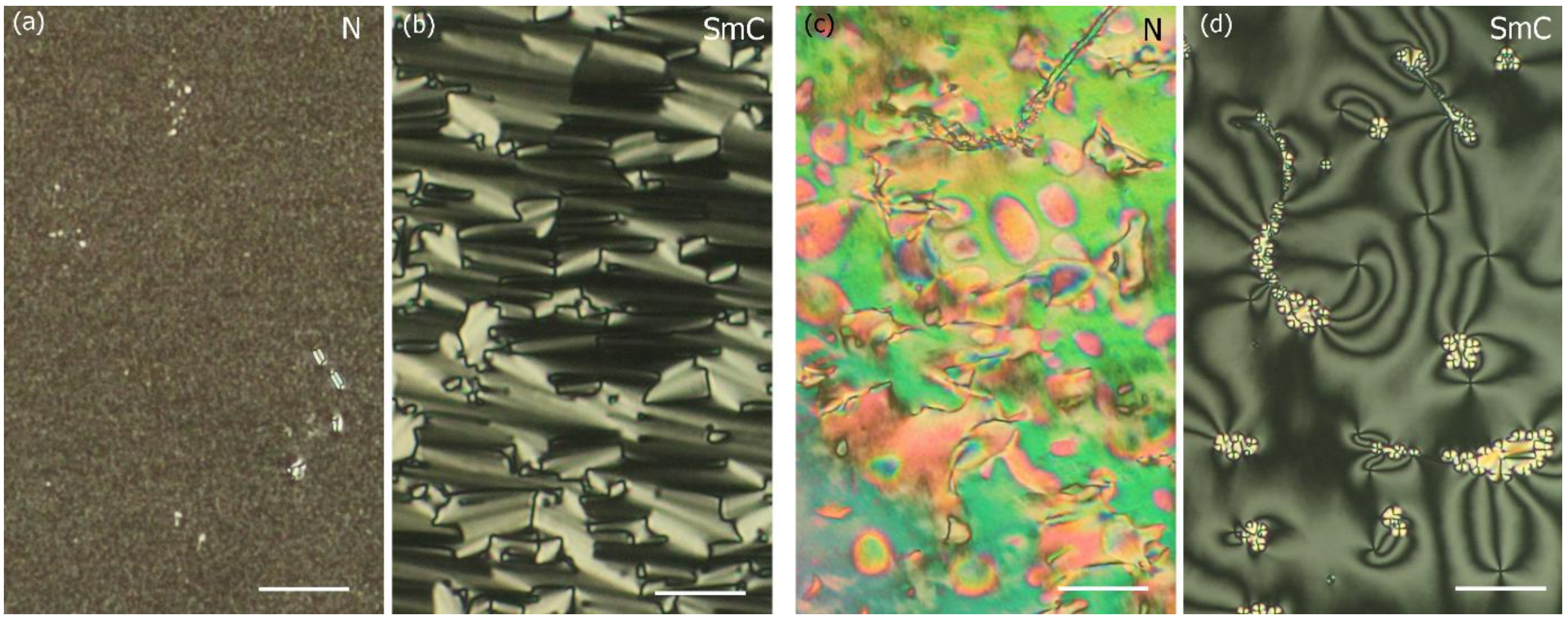

**Figure S4**. Optical textures of N (a,c) and smectic (b,d) phases of compound **II-5**, taken in 1.6-μm-thick cell with planar anchoring condition (a,b) or for one-surface-free sample (b,d). Scale bar corresponds to 50 μm. Schlieren texture (d) suggest the tilted type smectic.

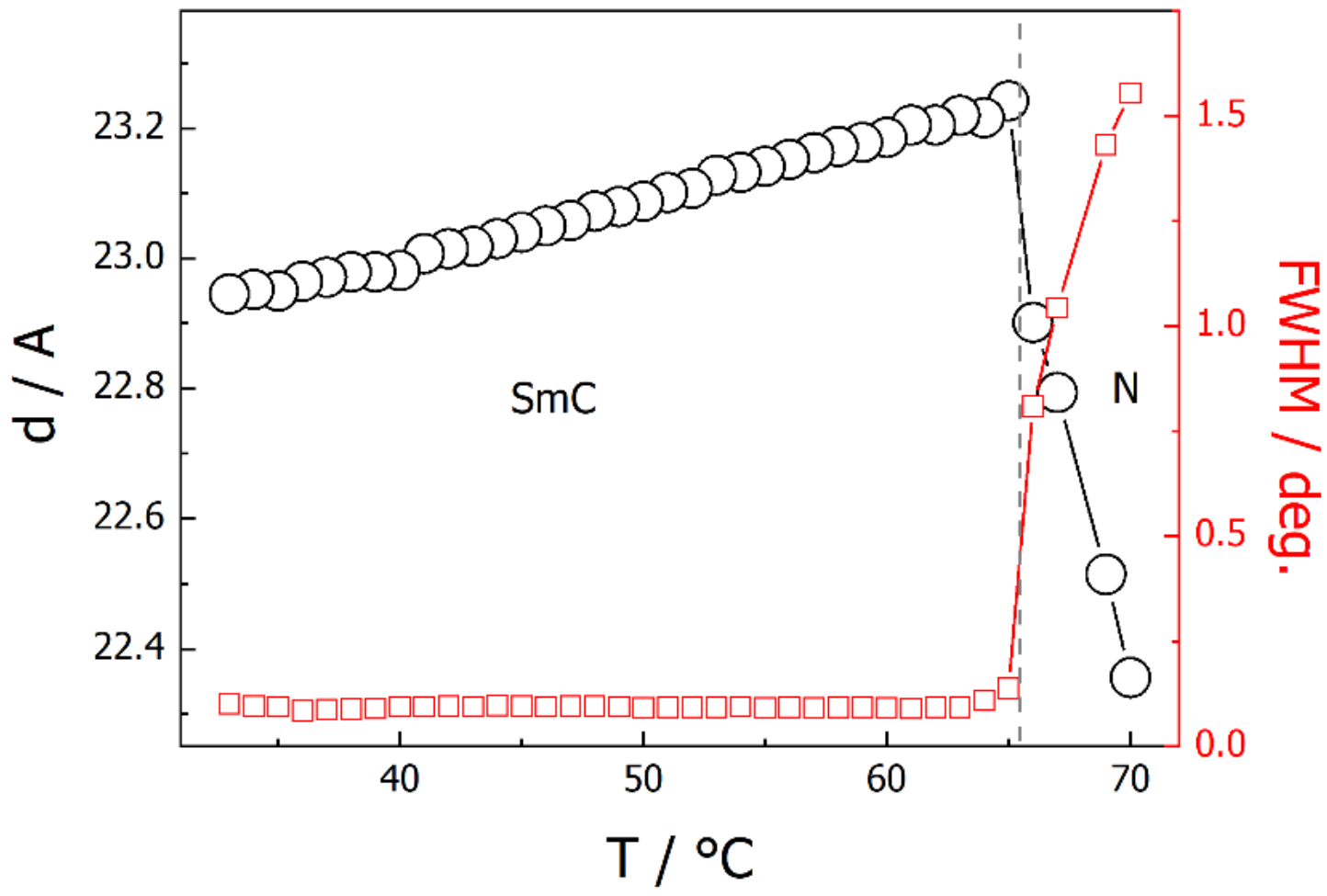

**Figure S5**. Layer thickness in SmC phase and local periodicity in N phase (black circles) for compound **II-5**. Red squares present width of the related diffraction signal; its gradual narrowing in N phase on approaching smectic phase evidences growing correlation length of positional order of molecules.

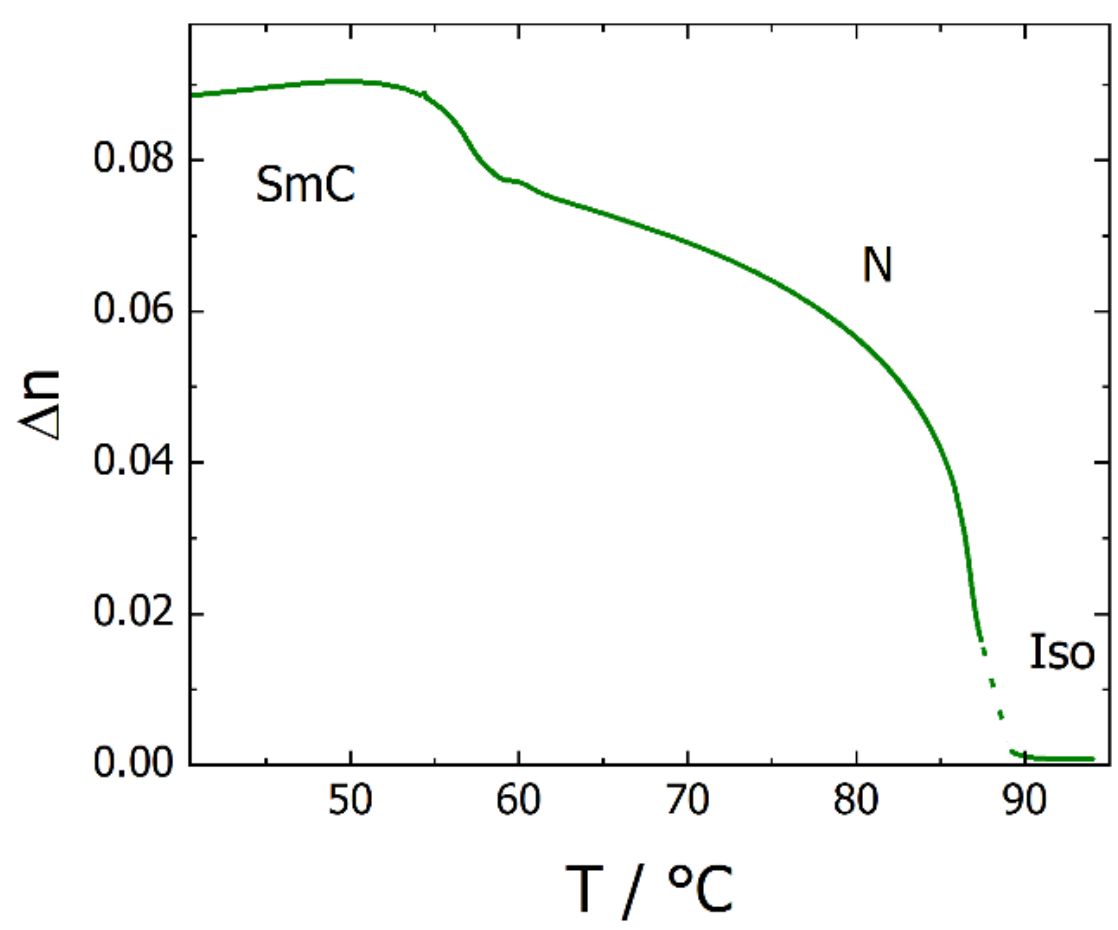

**Figure S6**. Optical birefringence for compound **II-5**, measured in 1.6-μm-thick cell with planar anchoring condition. Dashed line present a temperature range, in which due to sample misalignment at Iso-N phase transition no reliable values can be determined.

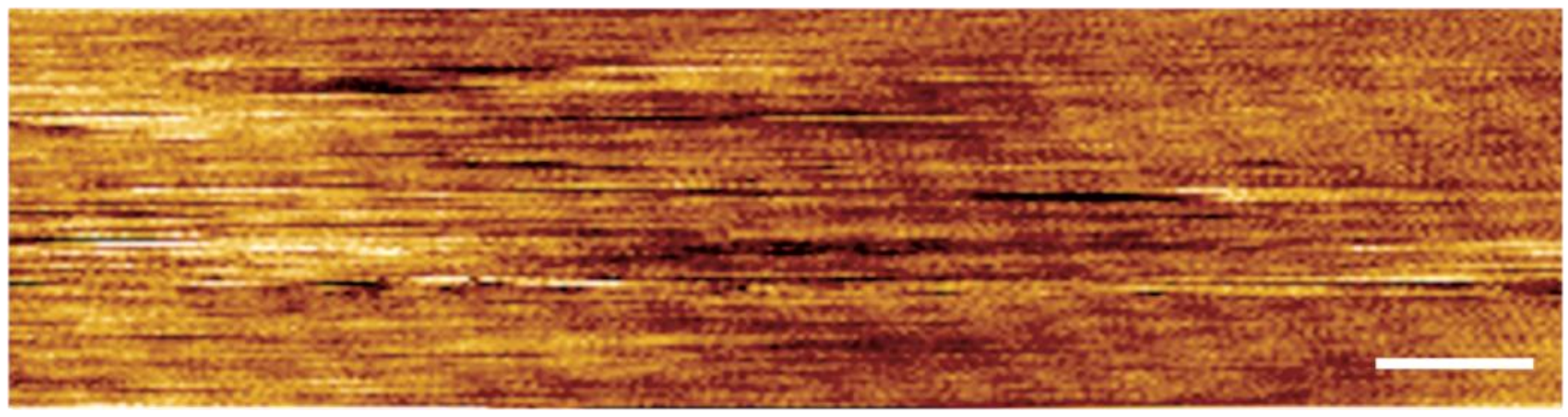

**Figure S7**. AFM image of $N_{TB}$ phase of compound **II-3**, taken at room temperature. Scale bar corresponds to 200 nm. Visible periodic structure is due to heliconical arrangement of molecules, with pitch of order of 100 nm.

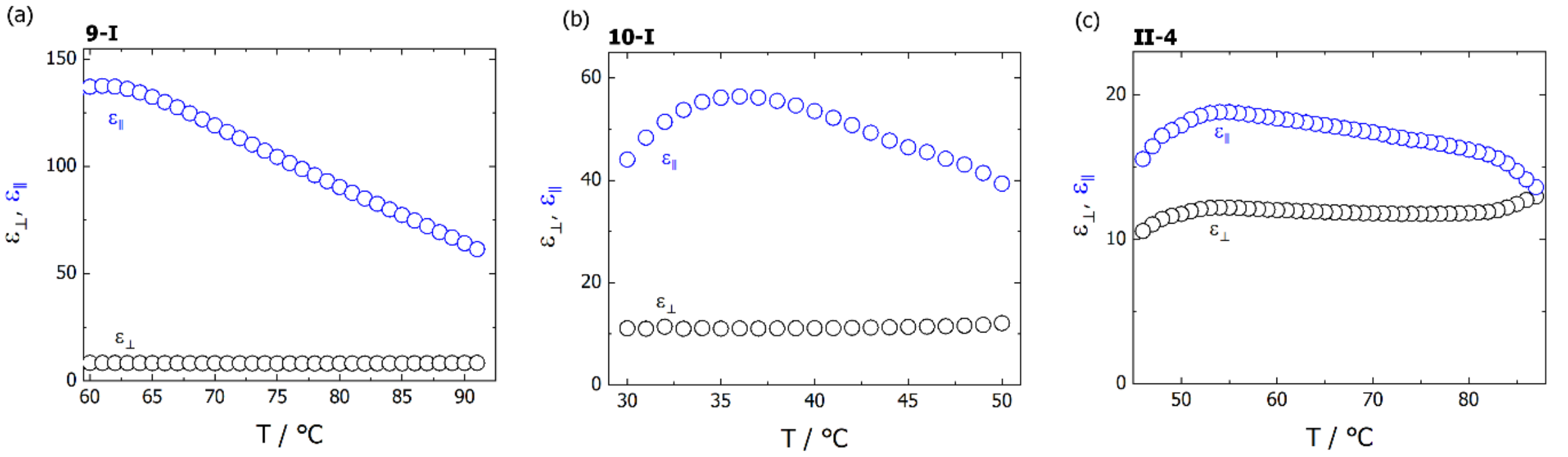

**Figure S8.** Parallel ($\epsilon_{\|}$) and perpendicular ($\epsilon_{\perp}$) components of apparent static dielectric permittivity in nematic phase vs. temperature, determined from Fréedericksz transition experiments for linear dimer 9-I (a) and bent dimers 10-I (b) and II-4 (c).

## Experimental methods

*Differential scanning calorimetry* has been utilized for determination of phase transition temperatures and the associated enthalpy changes. Measurements were performed differential scanning calorimeter DSC 204 F1 Phoenix instrument (Netzsch, Selb, Germany), under a nitrogen atmosphere with a heating/cooling rate of 2 K min$^{-1}$, unless otherwise specified.

*Polarised-light optical microscopy* observations were carried out with a Zeiss AxioImager.A2m microscope equipped with a Linkam heating stage.

*Optical birefringence* as a function of temperature has been measured with a setup based on a photoelastic modulator (PEM-90, Hinds) working at a modulation frequency $f$ = 50 kHz; with LED illuminator Photonic F5100, equipped with narrow bandpass filters, as a light source. Intensity of transmitted light was measured with a photodiode (FLC Electronics PIN-20) and de-convoluted with a lock-in amplifier (EG&G 7265) into 1$f$ and 2$f$ components to yield a retardation induced by the sample. Knowing the sample thickness, the retardation was recalculated into optical birefringence. Samples were prepared in 1.6-μm-thick cells with planar anchoring. The alignment quality was checked prior to measurement by inspection under the polarised-light optical microscope.

*X-ray diffraction measurements* of samples in liquid crystalline phases were carried out using a Bruker D8 GADDS system, equipped with micro-focus-type X-ray source with Cu anode and dedicated optics and VANTEC2000 area detector.

*Spontaneous electric polarisation* was determined by integration of the current peaks recorded during polarization switching upon applying a triangular-wave voltage. Cells with gold electrodes and no polymer aligning layers were used, and the switching current was determined by recording the voltage drop on a resistor connected in series with the sample, using Siglent SDS2104X oscilloscope.

*The SHG activity* has been checked using a microscopic setup based on a solid-state laser EKSPLA NL202. Laser pulses (9 ns) at a 10 Hz repetition rate and max. 2 mJ pulse energy at λ=1064 nm were applied. An IR pass filter was placed at the entrance to the sample and a green pass filter at the exit of the sample.

*The complex dielectric permittivity*, ε*, was measured using a Solartron 1260 impedance analyser, in the 1 Hz −10 MHz frequency range, and a probe voltage of 50 mV. The material was placed in a 5- or 10-μm-thick glass cell with gold electrodes and no polymer aligning layers. Lack of a surfactant layer resulted in a random configuration of the director in the LC phases.

*For elastic properties of nematic phase* the capacitance changes of 5-μm-thick cell with ITO electrodes and planar anchoring condition, filled with LC material were measured as a function of the applied AC voltage, using Precision Component Analyzer, Wayne-Kerr 6425, and recalculated to dielectric permittivity, $\varepsilon_{LC}(U) = C_{LC}/C_o$, where $C_o$ is capacitance of the empty cell. The frequency of the applied voltage was chosen to avoid effects related ion conductivity, but below the relaxation frequency of the dielectric modes. The $\varepsilon_{LC}(U)$ curves were fitted to the equations describing the director realignment under the applied voltage [18]:

$$U = \frac{2U_{th}}{\pi}\sqrt{1+\gamma\sin^2\theta_m}\int_0^{\frac{\pi}{2}}\sqrt{\frac{1+\kappa sin^2\theta_m sin^2\psi}{(1+\gamma sin^2\theta_m sin^2\psi)(1-sin^2\theta_m sin^2\psi)}}\,d\psi$$

$$\epsilon_{LC} = \epsilon_\perp \frac{\int_0^{\frac{\pi}{2}}\sqrt{\frac{(1+\gamma sin^2\theta_m sin^2\psi)(1+\kappa sin^2\theta_m sin^2\psi)}{(1-sin^2\theta_m sin^2\psi)}}\,d\psi}{\int_0^{\frac{\pi}{2}}\sqrt{\frac{1+\kappa sin^2\theta_m sin^2\psi}{(1+\gamma sin^2\theta_m sin^2\psi)(1-sin^2\theta_m sin^2\psi)}}\,d\psi}$$

where $U_{th}$, $\kappa$ and $\gamma$ are fitting parameters related to material properties as:

$$U_{th} = \pi\sqrt{\frac{K_{11}}{\epsilon_0(\epsilon_\parallel - \epsilon_\perp)}}$$

$$\kappa = \frac{K_{33}}{K_{11}} - 1$$
$$\gamma = \frac{\epsilon_{\parallel}}{\epsilon_{\perp}} - 1$$

The fitting procedure was performed using the custom-written Python script. Briefly, for given $U_{th}$, $\kappa$ and $\gamma$ first formula was numerically inverted using the bisection method to obtain $\theta_m(U)$ function, which was then substituted into the second equation to calculate $\epsilon_{LC}(U)$. The resulting theoretical curve was compared to experimental data. The fitting parameters were iteratively adjusted using the least-squares method, until the difference between theoretical and experimental dependences was minimized.

To determine static dielectric anisotropy, the perpendicular component of dielectric permittivity $\varepsilon_{\perp}$ was taken as value of $\varepsilon_{LC}$(U) below the threshold voltage, and parallel component $\varepsilon_{\parallel}$ was recalculated from fitted parameter γ.

Although, the influence of the aligning polymer layers is discussed in the literature with respect to their effect on the measured permittivity in the $N_F$ phase [29], we assumed that in N phase, the $\varepsilon_{LC}(U)$ is close to real values. Unfortunately no data could be obtained close to N-$N_F$ or N-$N_{TB}$ transition due to the tendency for the samples to re-crystallize.

## Synthetic Procedures and Structural Characterisation

### Series n-I

The molecular structure of the dimers of the homologous series n-I constitutes a structural link to the nematogen exhibiting the properties of the unique $N_{TBF}$ phase – MUT_JK103, as well as its modifications, characterized by the presence of structural elements typical for dimers. The varied length of the linking chain (n = 5–10) and the presence of numerous functional groups in the structure of asymmetric dimers necessitated the use of protecting groups in the synthetic pathway, which dictated the development of different synthetic approaches. Figure S9 depicts the synthesis of dimers whose structures incorporate 5, 6, and 7 methylene units in the spacer. The synthetic pathway was divided into two stages, with the intermediate step consisting of the incorporation of the linking chain into the molecular framework, a characteristic element of dimeric structures. In the first part, as a result of the Miyaura borylation reaction applied to the bromo derivative (**1**), a boronate ester was obtained (**2**), which was subsequently subjected to a Suzuki–Miyaura cross-coupling reaction with 4-bromo-3-fluorophenol. Phenol with an elongated core (**3**) was subjected to a substitution reaction with an aliphatic iodo derivative containing a hydroxyl group protected by a benzyl group (**5**) which was subsequently deprotected under high pressure (**7**). In the second part, 1,3-dioxane ring was introduced into the molecule (**9**) which was subsequently lithiated to afford the iodo derivative (**10**), employed in Suzuki–Miyaura cross-coupling reaction (**11**), with the following step consisting of lithiation and subsequent carboxylation (**12**). Finally, dimers was synthesized in a Steglich esterification reaction between (**7**) and (**12**) with the presence of DCC and DMAP. The final products were purified by flash column chromatography and recrystallization from appropriate solvent.
For dimers having the spacer with a methylene chain of 8 to 10 carbon atoms (Fig. S10), the phenol (**3**) underwent a substitution reaction with a bromo derivative with a hydroxyl group protected as a tetrahydropyranyl (THP) ether (**14**). During this process, the resulting product underwent partial deprotection. The obtained alcohols (**15**) were subjected to esterification and purified to afford homologs with different linking chain lengths, obtained via an alternative synthetic pathway.

**Figure S9.** Synthesis of Series **n-I** (n=5-7) Dimers

**Figure S10.** Synthesis of Series **n-I** (n=8-10) Dimers

### *2-(4-(difluoro(3,4,5-trifluorophenoxy)methyl)-3,5-difluorophenyl)-4,4,5,5-tetramethyl-1,3,2-dioxaborolane (2)*

5-((4-bromo-2-fluorophenyl)difluoromethoxy)-1, 2,3-trifluorobenzene (1) (119.81 g; 0.308 mol), bis(pinacolato)diboron (82.14 g; 0.323 mol), potassium acetate (90.55 g; 0.924 mol), toluene (500ml) and dioxane (500ml) were placed in flask. The reaction was refluxed under a nitrogen atmosphere. $PdCl_2$(dppf) (6.75 g; 0.00924 mol) was then added. The reaction was refluxed for 48 h under a nitrogen atmosphere. The reaction was poured into water and filtered under reduced pressure through an activated carbon plate. The phases were separated by washing the organic phase with water and the aqueous phase with toluene. The organic phase was dried over anhydrous $MgSO_4$ and concentrated. The residue was recrystallized from ethanol.

Yield 127.9 g (95%)

Purity 97.7% (GC-MS);

MS(EI) m/z: 437 (M+); 421; 289; 207; 189

$^1$H NMR (500 MHz, $CDCl_3$) δ: 7.40 (d, *J*=9.46 Hz, 2 H); 6.98 (dd, *J*=7.78, 5.95 Hz, 2 H); 1.37 (s, 12 H)

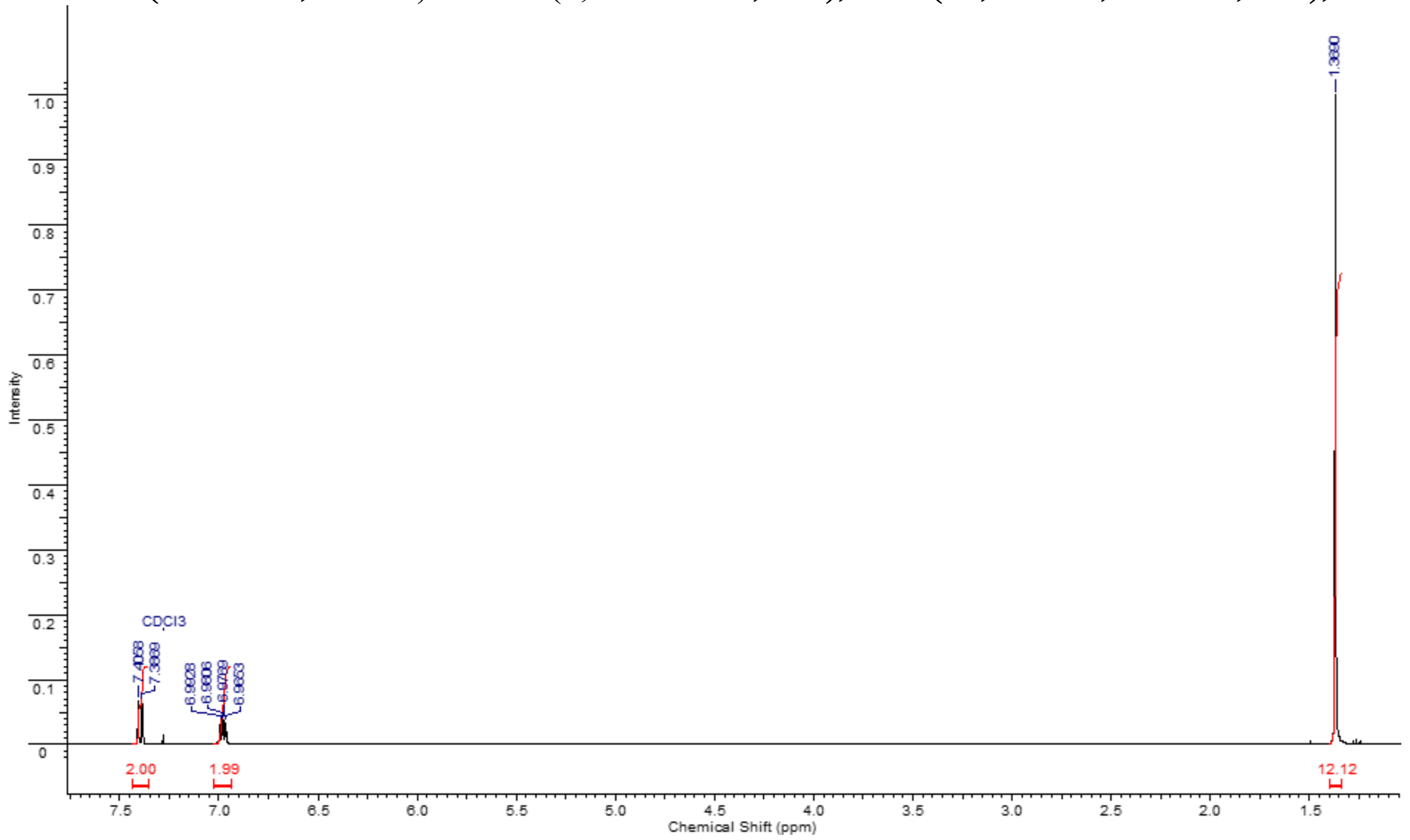

$^{13}C$ NMR (126 MHz, $CDCl_3$) δ: 159.55 (dd, *J*=259.32, 4.09 Hz); 150.98 (ddd, *J*=250.69, 5.45 Hz); 144.69 (m); 138.43 (dt, *J*=249.78, 15.44 Hz); 120.24 (t, *J*=266.12 Hz); 118.07 (dd, *J*=20.89, 3.63 Hz); 111.69 (m); 107.42 (m); 84.90; 24.80

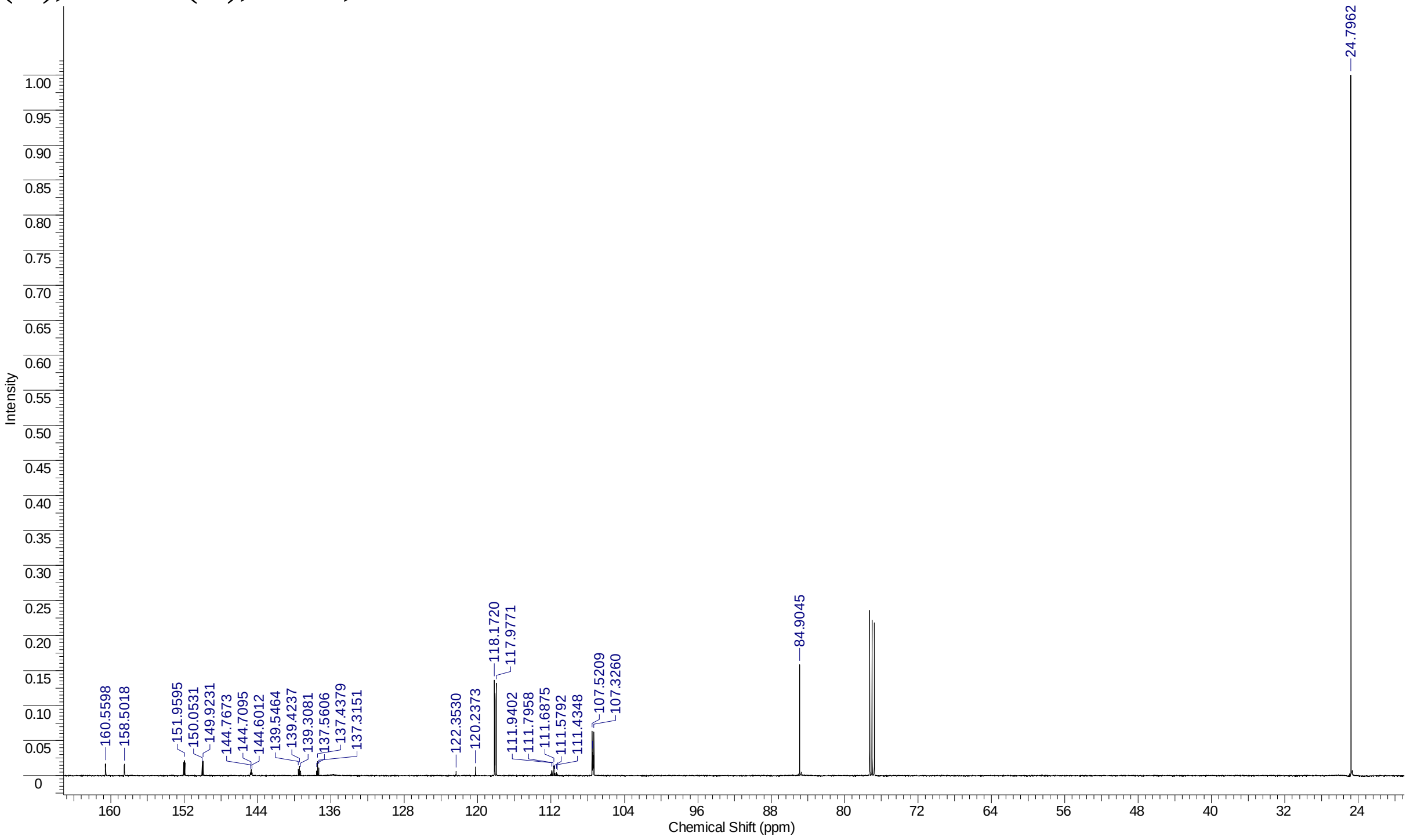

***4'-(difluoro(3,4,5-trifluorophenoxy)methyl)-2,3',5'-trifluoro-[1,1'-biphenyl]-4-ol (3)***

The mixture of 2-(4-(difluoro(3,4,5-trifluorophenoxy)methyl)-3,5-difluorophenyl)-4,4,5,5- tetramethyl-1,3,2-dioxaborolane (20 g; 0.0473 mol), 4-bromo-3-fluorophenol (8.15 g; 0.0429 mol), potassium phosphate trihydrate (39.92 g; 0.15 mol) in anhydrous THF (250 ml) was refluxed for 1h under nitrogen atmosphere. Then $Pd(OAc)_2$ and SPhos were added and the mixture was stirred at reflux for 2h. Later, it was cooled down, acidified using 10% HCl solution and extracted with DCM. The organic layer was then dried over $MgSO_4$, and concentrated under vacuum. The residue was recrystallized from hexane to give an off-white solid.

Yield 15.39 (85.4%)

Purity 94.1 % (GC-MS)

MS(EI) m/z: 420 (M+); 401; 273; 244; 224; 204; 175; 156; 136; 119;

[1]H NMR (500 MHz, DMSO-d6) δ: 10.41 (s, 1 H); 7.49 (t, *J*= 9.15 Hz,1 H); 7.42 (d, *J*=11.60 Hz, 2 H); 7.36 (dd, *J*=7.93, 6.10 Hz, 2 H); 6.74 (dd, *J*=8.55, 2.14 Hz, 1 H); 6.69 (dd, *J*=13.12, 2.44 Hz, 1 H)

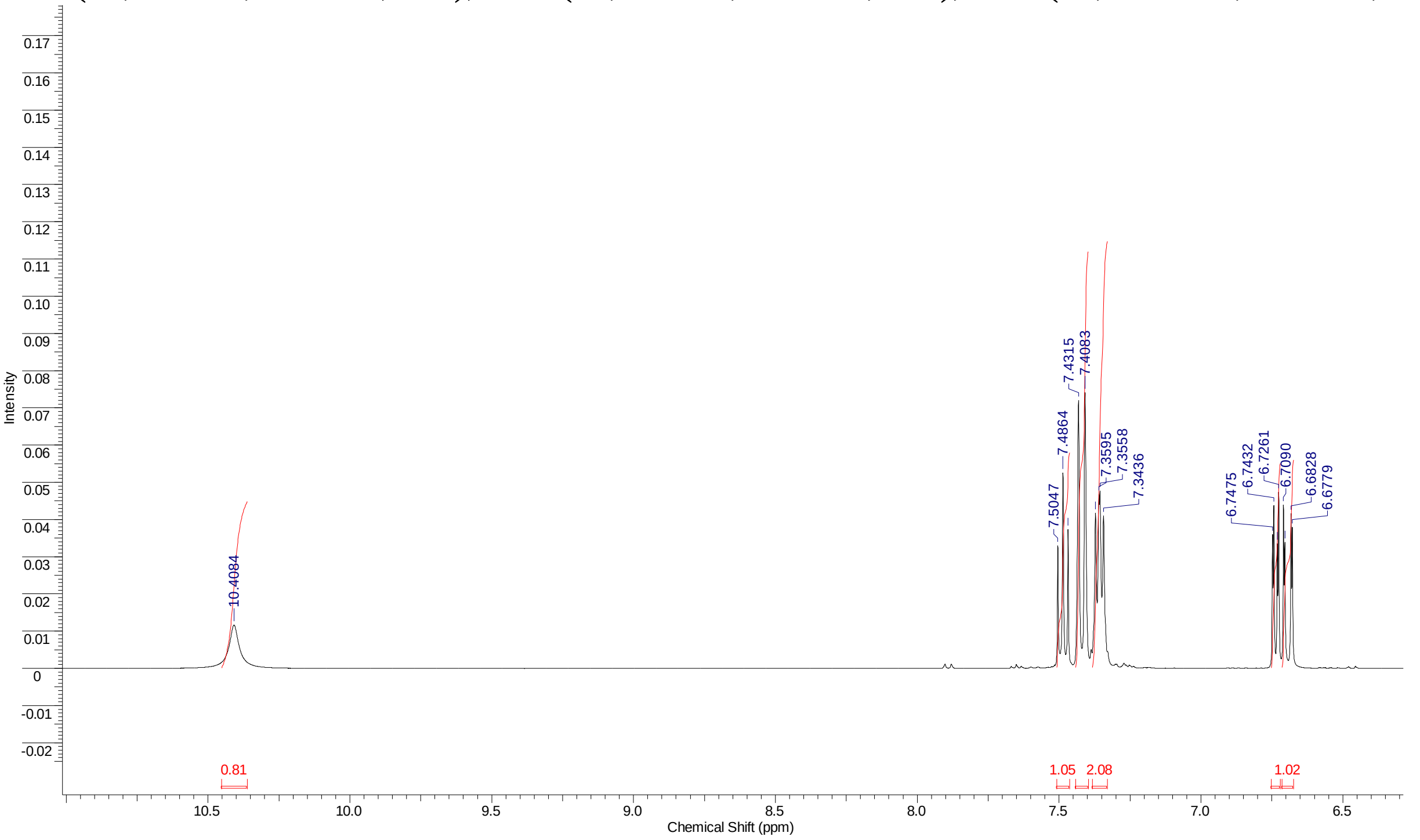

[13]C NMR (125 MHz, DMSO-*d6*) δ: 160.84 (d, *J*=11.81 Hz); 160.43 (d, *J*=247.96 Hz); 159.49 (dd, *J*=255.23, 6.36 Hz); 150.71 (ddd, *J*=248.87, 10.90, 5.45 Hz); 144.62 (m); 142.52 (t, *J*=11.35 Hz); 138.26 (dt, *J*=247.96, 15.44 Hz); 131.69 (d, *J*=4.54 Hz); 120.45 (t, *J*=264.31 Hz); 115.49 (d, *J*=11.81 Hz); 112.96 (m); 108.25 (m); 106.90 (m); 103.70 (d, *J*=25.43 Hz)

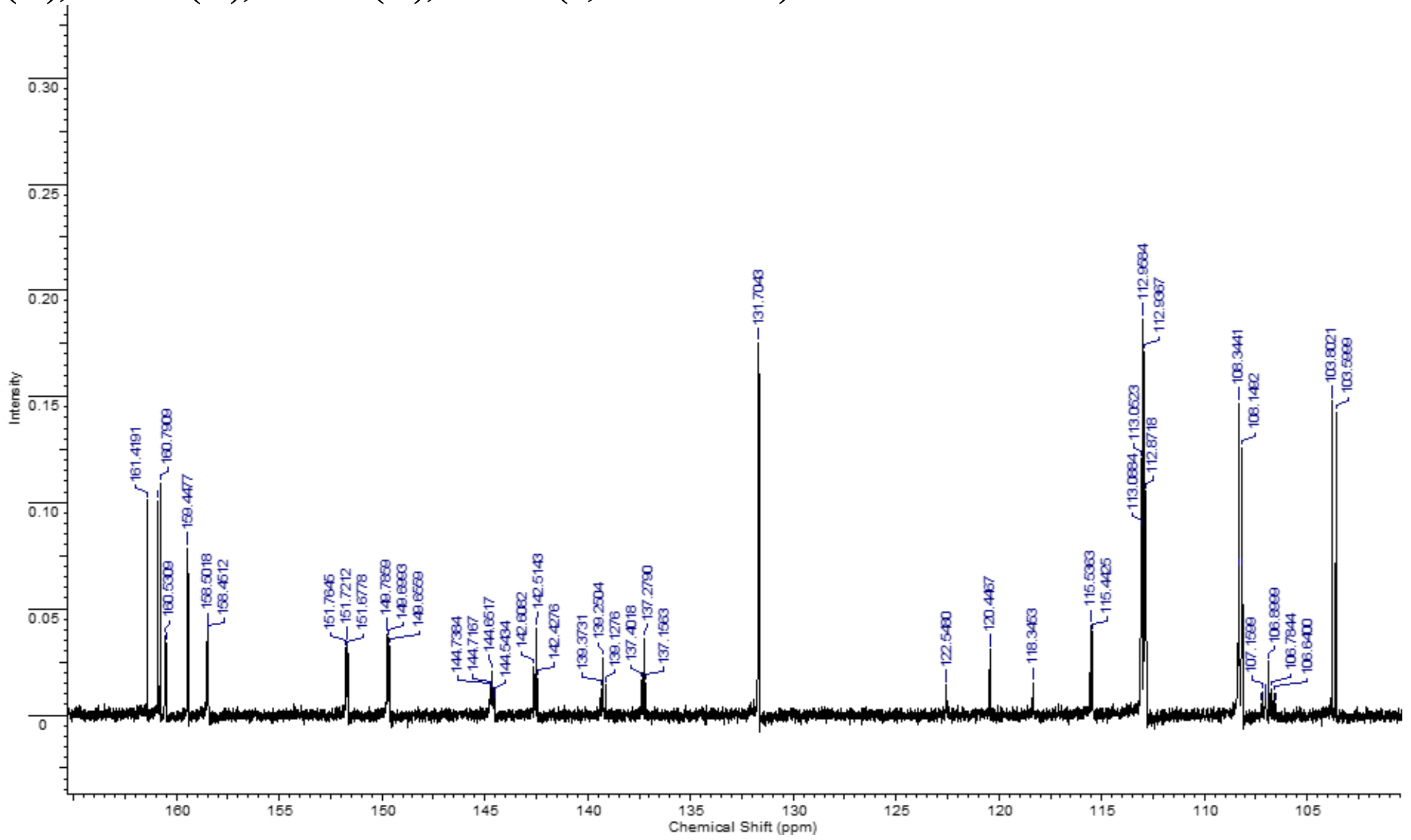

***(((7-iodoheptyl)oxy)methyl)benzene (5c)***

The mixture of 7-(benzyloxy)heptan-1-ol (40 g; 0.18 mol), triethylamine (27.32 g; 0.27 mol) in dry THF (200 ml) was cooled to 0°C and the mixture of mesyl chloride (21.65 g; 0.189 mol) in dry THF (50 ml) was added dropwise. After addition, the mixture was stirred at 0°C for 1h. The mixture was acidified using hydrochloric acid and the mixture was extracted with toluene. Organic phase was washed twice with water, dried over $MgSO_4$ and solvent was evaporated.

Crude product was then dissolved in butan-2-one (500 ml) and KI (50.3 g; 0.303 mol) was added. The solution was refluxed until all of the mesylate was consumed. Water was then added to the mixture and it was extracted with hexane. Organic phase was dried over $MgSO_4$ and solvent was evaporated. Crude product was purified on column chromatography using hexane as the eluent.

Yield 56.14 g (93.9%)

Purity 99.1% (GC-MS);

MS(EI) m/z: 332; 241; 223; 205; 187; 169; 155; 131; 107, 91;

$^1$H NMR (500 MHz, $CDCl_3$) δ: 7.22 (m, 5 H); 4.41 (s, 2 H); 3.37 (t, *J*=6.56 Hz, 2 H); 3.09 (m, 2 H); 1.71 (dd, *J*=14.50, 7.17 Hz, 2 H); 1.52 (m, 2 H); 1.28 (m, 6 H)

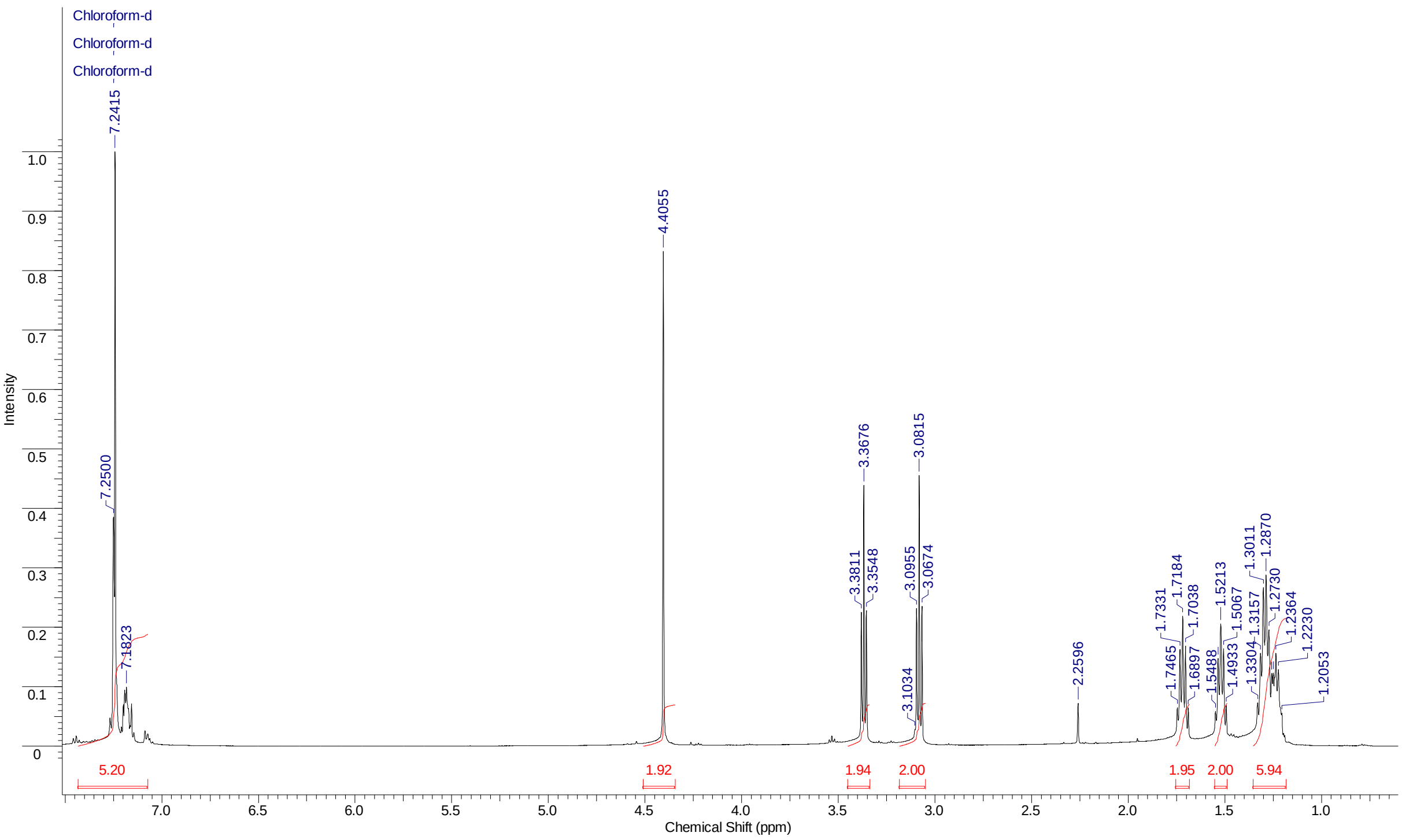

$^{13}C$ NMR (126 MHz, $CDCl_3$) δ: 138.62; 128.31; 127.57; 127.44; 72.85; 70.29; 33.43; 30.39; 29.62; 28.32; 25.97; 7.14

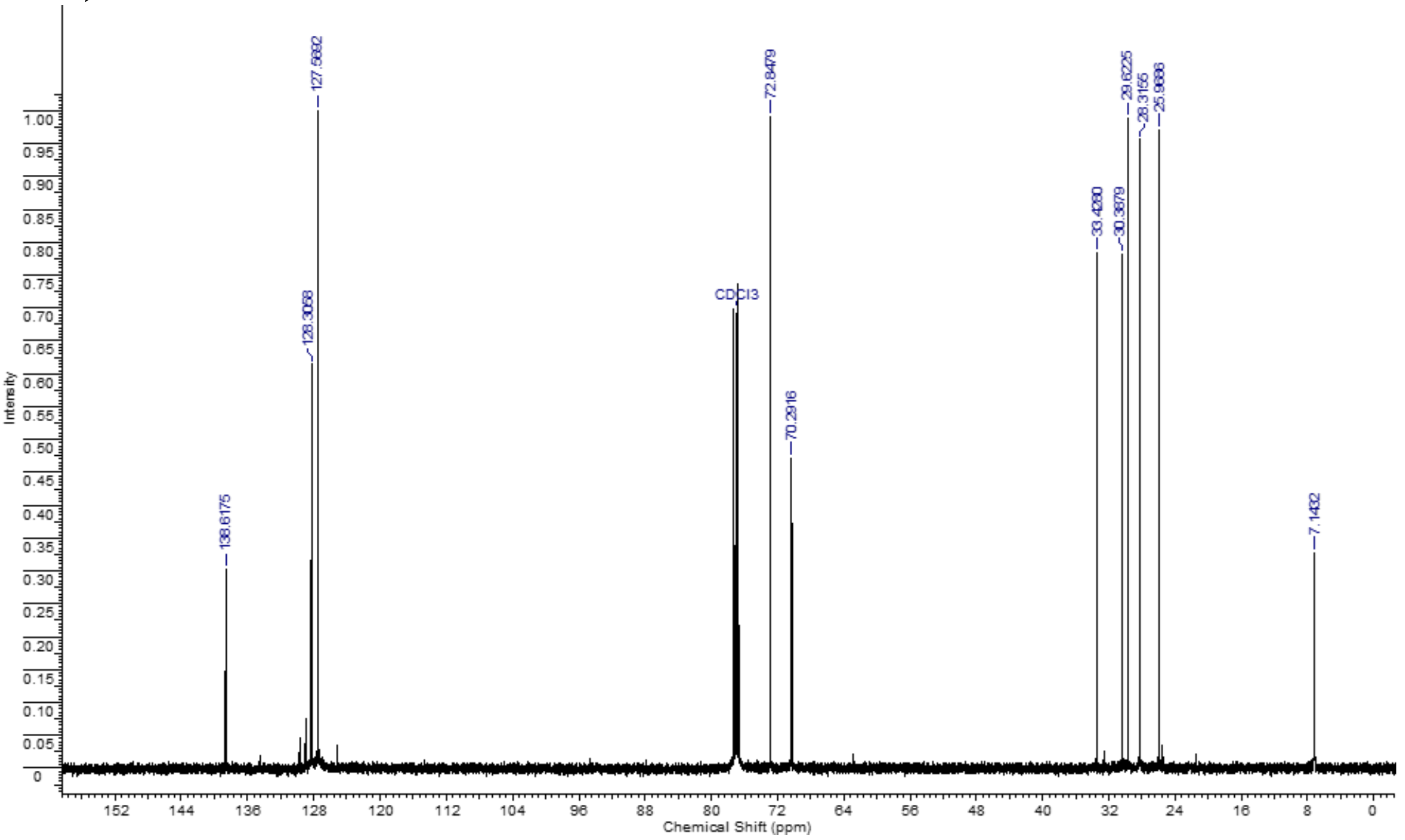

Compounds 5a and 5b were synthesized in the same manner as 5c.

***4-((7-(benzyloxy)heptyl)oxy)-4'-(difluoro(3,4,5-trifluorophenoxy)methyl)-2,3',5'-trifluoro-1,1'-biphenyl (6c)***

The mixture of 4'-(difluoro(3,4,5-trifluorophenoxy)methyl)-2,3',5'-trifluoro-[1,1'-biphenyl]-4-ol (6.65 g; 0.0158 mol), (((7-iodoheptyl)oxy)methyl)benzene (4.78 g; 0.0144 mol), potassium carbonate (3.97 g; 0.0288 mol) in butan-2-one (150 ml) was refluxed for 5h, solution was warmed up to room temperature. The mixture was poured into 500 ml of water and 30 ml of 10% HCl. Crude product was extracted with Dichloromethane. Organic layer was washed three times with water, dried over $MgSO_4$ and solvent was evaporated. Crude product was purified by liquid column chromatography (Hexane as a mobile phase)
Yield 6 g (67%)
Purity 91 % (GC-MS);
MS(EI) m/z: 624 (M+); 586; 513;477; 385; 355; 327; 299; 273; 244; 91;
$^{1}H$ NMR (500 MHz, $CDCl_3$) δ: 7.40 (m, 5 H); 7.34 (m, 1 H); 7.24 (d, *J*=10.99 Hz, 2 H); 7.05 (m, 2 H); 6.85 (dd, *J*=8.55, 2.14 Hz, 1 H); 6.78 (dd, *J*=12.82, 2.44 Hz, 1 H); 4.58 (s, 2 H); 4.05 (t, *J*=6.41 Hz, 2 H); 3.55 (t, *J*=6.56 Hz, 2 H); 1.89 (m, 2 H); 1.72 (m, 2 H); 1.52 (m, 6 H)

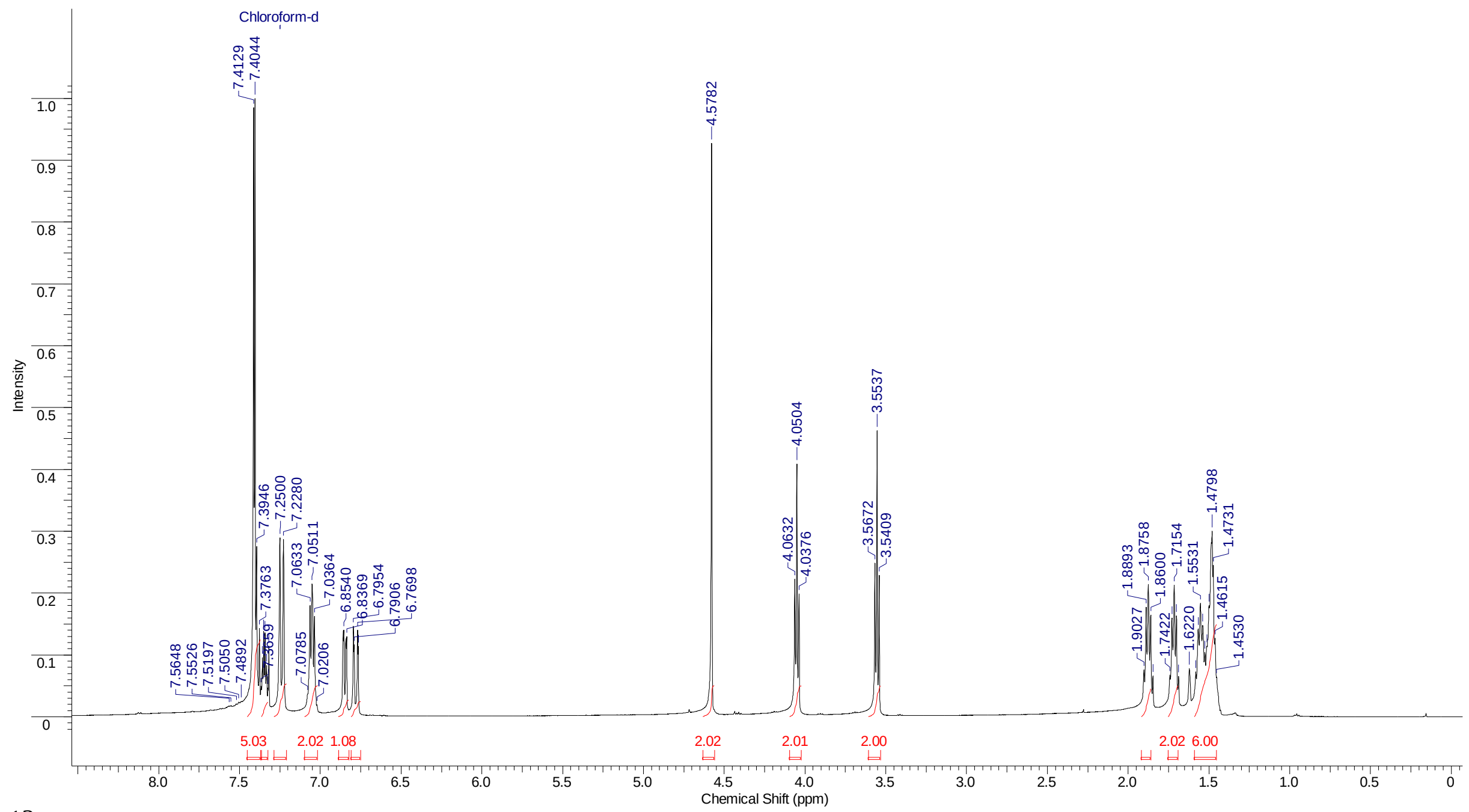

$^{13}$C NMR (126 MHz, $CDCl_3$) δ: 161.64 (d, *J*=2.72 Hz); 160.60 (d, *J*=237.06 Hz); 160.13 (dd, *J*=257.50, 5.90 Hz); 151.24 (ddd, *J*=250.91, 10.22, 4.99 Hz); 144.96 (m); 141.90 (t, *J*=11.35 Hz); 138.91; 138.66 (dt, *J*=250.69, 15.44 Hz); 130.67 (d, *J*=4.54 Hz, 1 C); 128.58; 127.84; 127.73; 120.52 (t, *J*=265.22 Hz); 117.63 (d, *J*=12.72 Hz); 112.79 (dd, *J*=27.25, 3.63 Hz); 111.64 (d, *J*=2.72 Hz); 107.65 (m); 103.02 (d, *J*=26.34 Hz); 73.13; 70.62; 68.80; 29.94; 29.40; 29.20; 26.38; 26.16

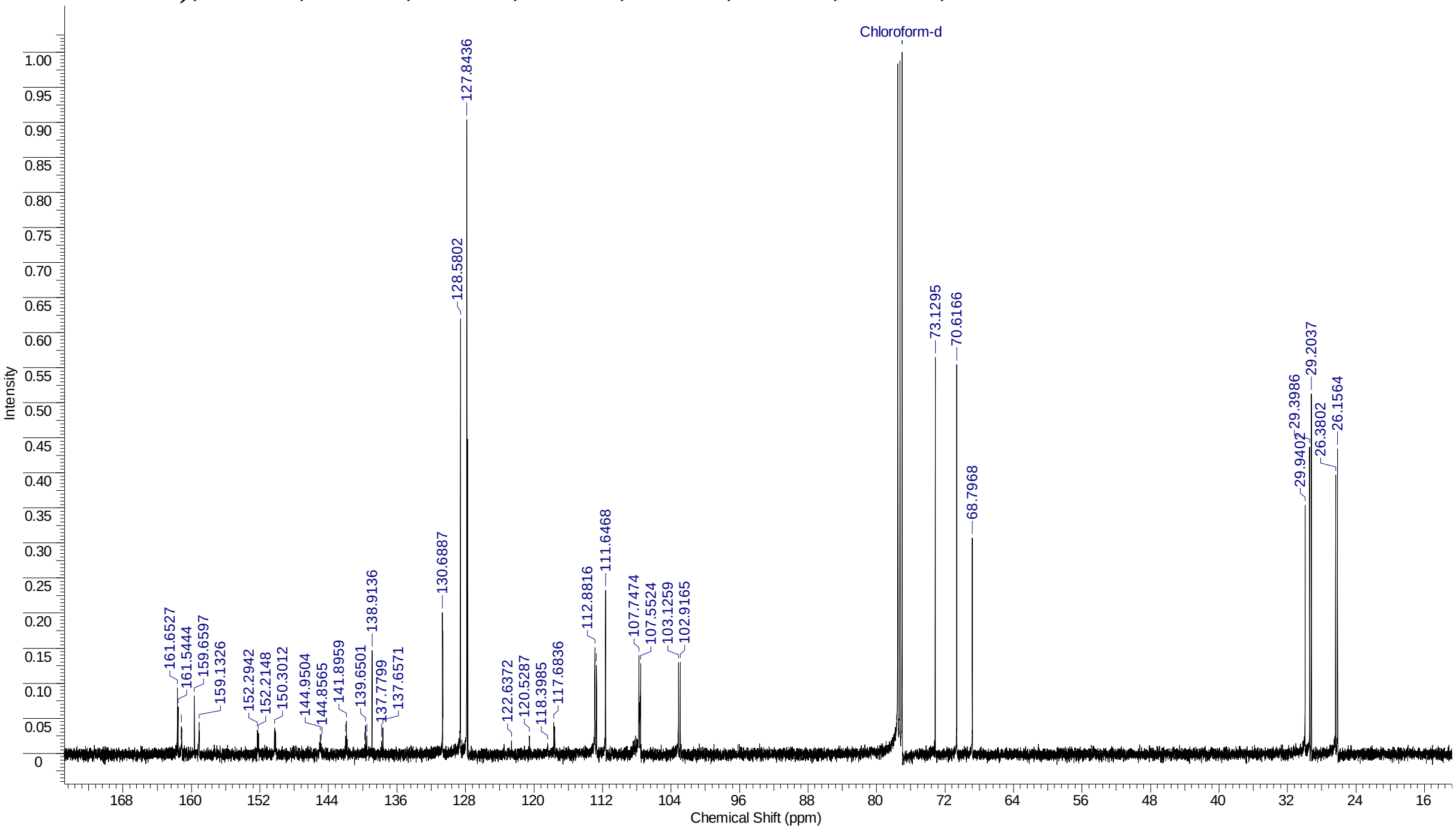

Compounds 6a and 6b were synthesized in the same manner as 6c.

***7-((4'-(difluoro(3,4,5-trifluorophenoxy)methyl)-2,3',5'-trifluoro-[1,1'-biphenyl]-4-yl)oxy)heptan-1-ol (7c)***

The 4-((7-(benzyloxy)heptyl)oxy)-4'-(difluoro(3,4,5-trifluorophenoxy)methyl)-2,3',5'-trifluoro-1,1'-biphenyl (6 g; 0.0096 mol) Pd/C catalyst (10%mol; 0,1 g; 0.000096 mol) and Tetrahydrofuran ( 50 ml) were placed in a pressure reactor. The reactor was purged with hydrogen (the procedure was repeated twice). The reactor was filled with hydrogen (up to a pressure of 50 bar), and the reaction was carried out at 50 °C for 6 hours. After completion of the reaction, the pressure in the reactor was slowly reduced. The reactor was then purged twice with an inert gas (nitrogen) and subsequently opened. To remove the catalyst from the post-reaction mixture, the mixture was filtered through a sintered glass funnel, solvent was evaporated. Crude product was purified by liquid column chromatography (Dichloromethane as a mobile phase).

Yield 4.72 g (92%)

Purity >99.9% (GC-MS);

MS(EI):534 (M+); 504; 463;387; 357; 273; 244; 204; 175;

$^{1}$H NMR (500 MHz, DMSO-d6) δ: 7.58 (t, *J*=9.00 Hz, 1 H); 7.47 (d, *J*=11.60 Hz, 2 H); 7.38 (m, 2 H); 6.95 (d, *J*=13.43 Hz, 1 H); 6.89 (dd, *J*=8.55, 2.14 Hz, 1 H); 4.33 (s, 1H); 4.02 (t, *J*=6.56 Hz, 2 H); 3.39 (m, 3 H); 1.71 (m, 2 H); 1.37 (m, 8 H)

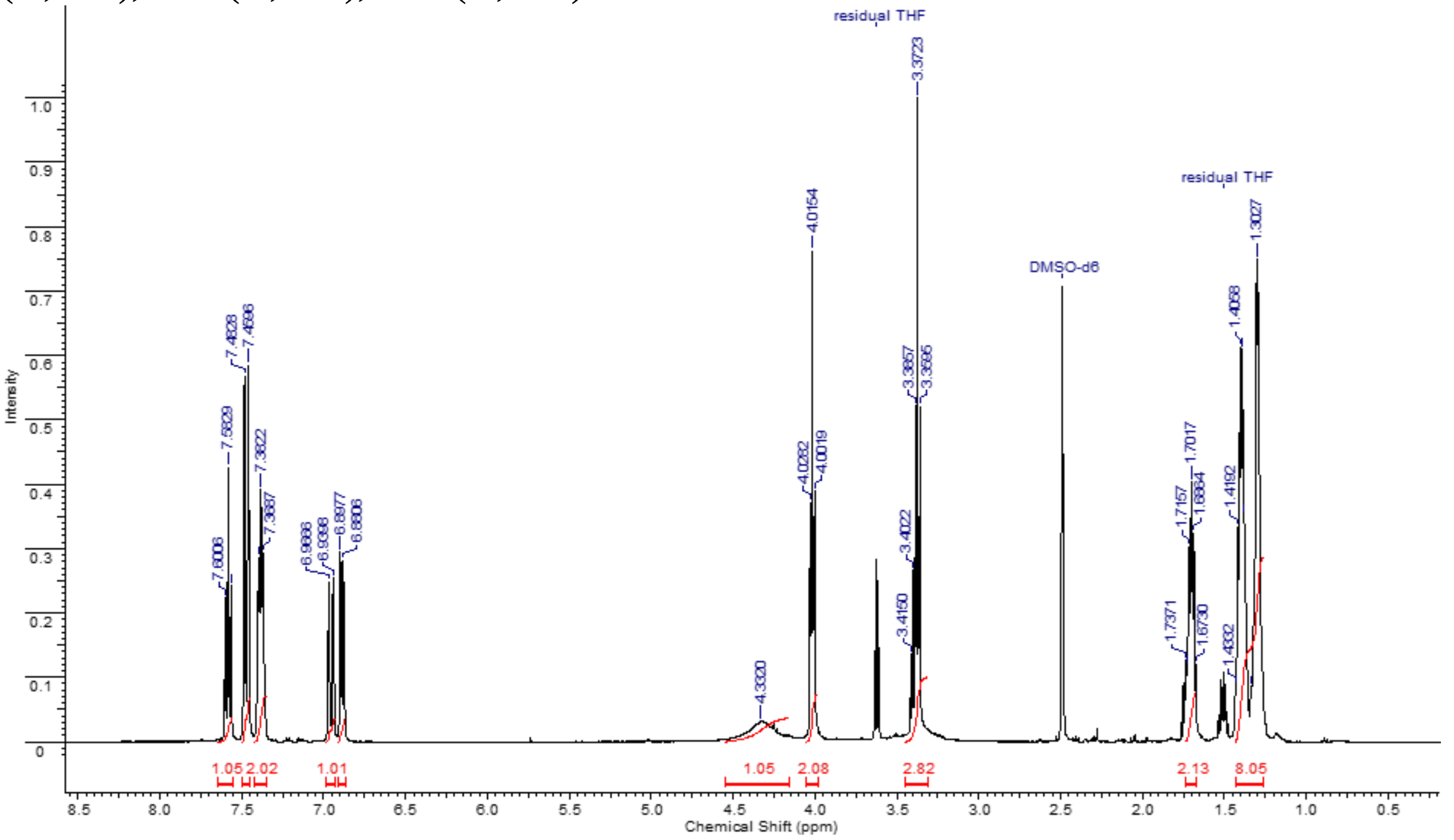

$^{13}C$ NMR (125 MHz, DMSO-*d6*) δ: 161.59 (d); 160.40 (d, *J*=247.96 Hz); 159.50 (dd, *J*=254.32, 6.36 Hz); 150.73 (ddd, *J*=250.91, 10.22, 4.99 Hz); 144.61 (m); 142.18 (t, *J*=10.90 Hz); 138.31 (dt, *J*=247.96, 15.44 Hz); 131.70 (d, *J*=3.63 Hz); 120.44 (t, *J*=264.77 Hz); 116.91 (m); 113.28 (m); 112.17 (d, *J*=2.72 Hz); 108.34 (m); 107.23 (m); 103.08 (d, *J*=25.43 Hz); 68.77; 61.17; 32.96; 30.23; 29.11; 28.91; 25.94

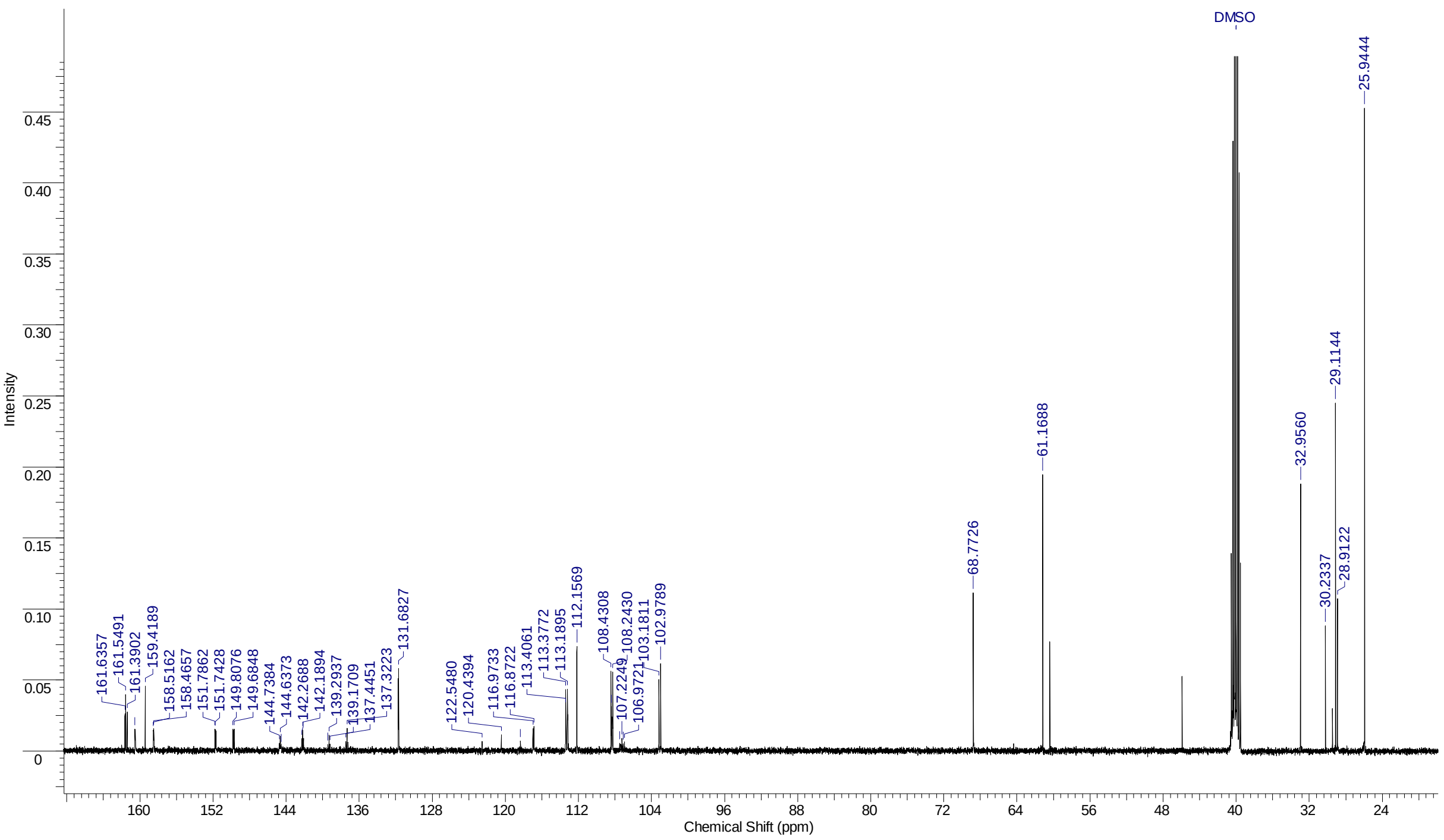

Compounds 7a and 7b were synthesized in the same manner as 7c.

***2-(3,5-difluorophenyl)-5-propyl-1,3-dioxane (9)***

A solution 3,5-difluorobenzaldehyde (43 g; 0.275 mol), 2-propylpropane-1,3-diol (40 g; 0.303 mol), p-toluenesulfonic acid (0.45 g; 0.0026 mol) in toluene (250 ml) was refluxed for 4h, while collecting water using Dean-Stark apparatus. After the reaction was completed, mixture was washed three times with water, organic layer was dried over $MgSO_4$ and concentrated.

Yield 60.98 g (90%)

Purity 99.29% (GC-MS) (65% of trans isomer)

MS(EI) m/z: 241 (M+); 211; 141; 113; 83; 55;

$^{1}H$ NMR (500 MHz, $CDCl_3$) δ: 7.05 (d, *J*=5.80 Hz, 2 H); 6.79 (tt, *J*=8.85, 2.44 Hz, 1 H); 5.38 (s, 1 H); 4.25 (dd, *J*=11.60, 4.58 Hz, 2 H); 3.53 (t, *J*=11.44 Hz, 2 H); 2.15 (m, 1 H); 1.35 (m, 2 H); 1.10 (m, 2 H); 0.95 (t, *J*=7.32 Hz, 3 H)

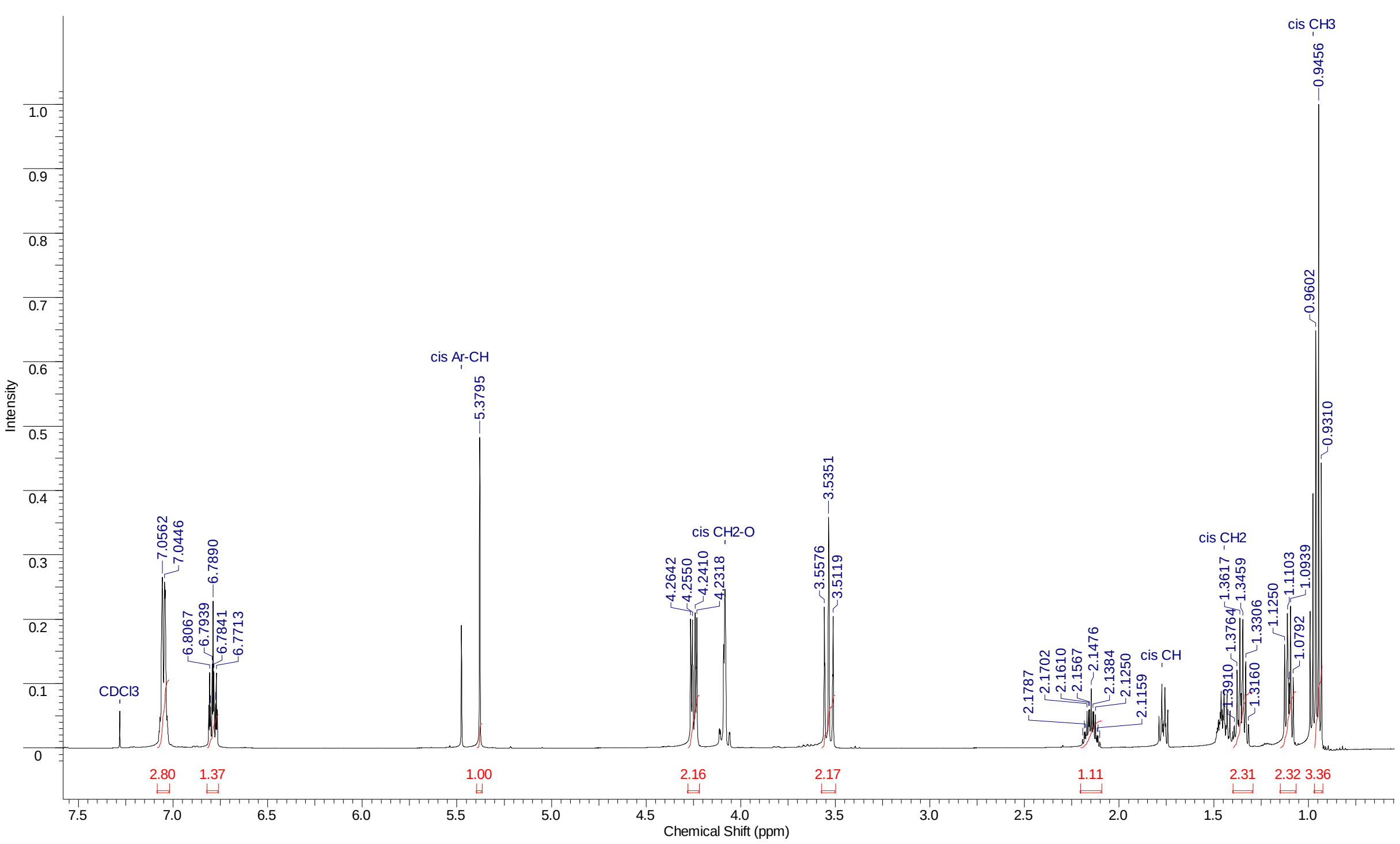

$^{13}$C NMR (126 MHz, $CDCl_3$) δ: 162.89 (dd, $J$=247.97, 11.81 Hz); 142.25 (t, $J$=9.08 Hz); 109.27 (m); 103.97 (t, $J$=25.43 Hz); 99.65 (t, $J$=2.73 Hz); 72.56; 33.93; 30.28; 19.55; 14.20

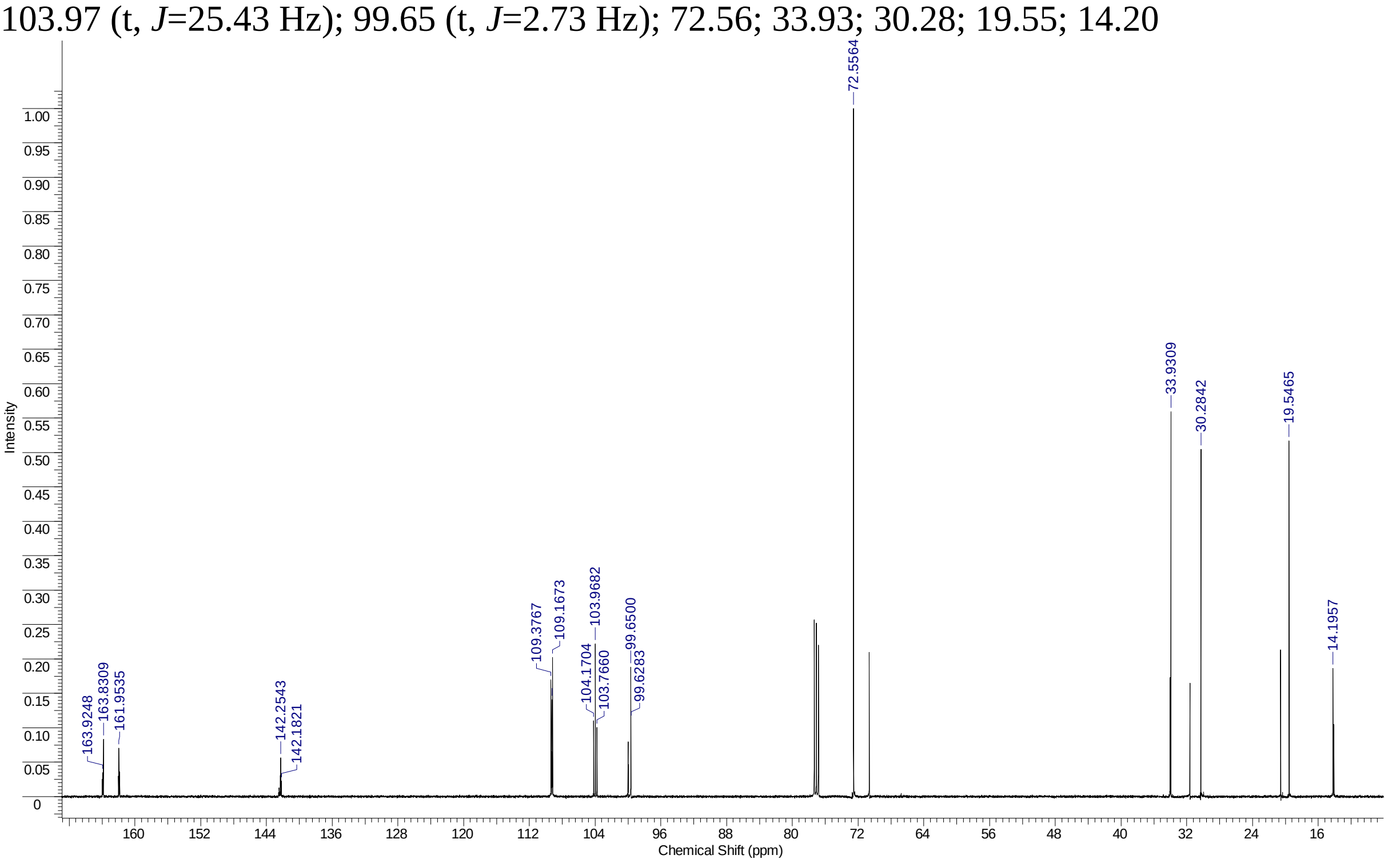

***2-(3,5-difluoro-4-iodophenyl)-5-propyl-1,3-dioxane (10)***

2-(3,5-difluorophenyl)-5-propyl-1,3-dioxane (3) (45 g; 0.186 mol) was mixed with anhydrous THF (300 ml ) under nitrogen and cooled to −78°C in an acetone/dry ice bath. Solution of n-butyllithium dissolved in cyclohexane–hexane mixture (0.2 mol, 2.5M) was added dropwise, and temperature was kept below −70°C. The reaction mixture was stirred for 2.5h in −78°C. Then a solution of iodine (50.8 g; 0.02 mol) in anhydrous tetrahydrofuran (150 ml) were added. After the reaction mixture was warmed to room temperature, sodium sulfite was added for discoloration. The solvent was evaporated. The crude product was extracted with dichloromethane; mixture was washed three times with water, organic layer was dried over $MgSO_4$ and concentrated. The crude product was used in subsequent reactions.

Yield: 40.36 g (89.7%)

Purity (GCMS): 97% (84.2% of trans isomer)

MS (EI) m/z: 368 (M+); 337; 267; 240; 211; 180; 141

$^1$H NMR (500 MHz, $CDCl_3$) δ: 7.06 (d, *J*=6.71 Hz, 2 H, overlapping aromatic proton signals of cis isomer); 5.36 (s, 1 H); 4.24 (dd, *J*=11.90, 4.58 Hz, 2 H); 3.53 (t, *J*=11.44 Hz, 2 H); 2.14 (m, 1 H); 1.35 (m, 2 H); 1.10 (m, 2 H); 0.94 (t, *J*=7.32 Hz, 3 H)

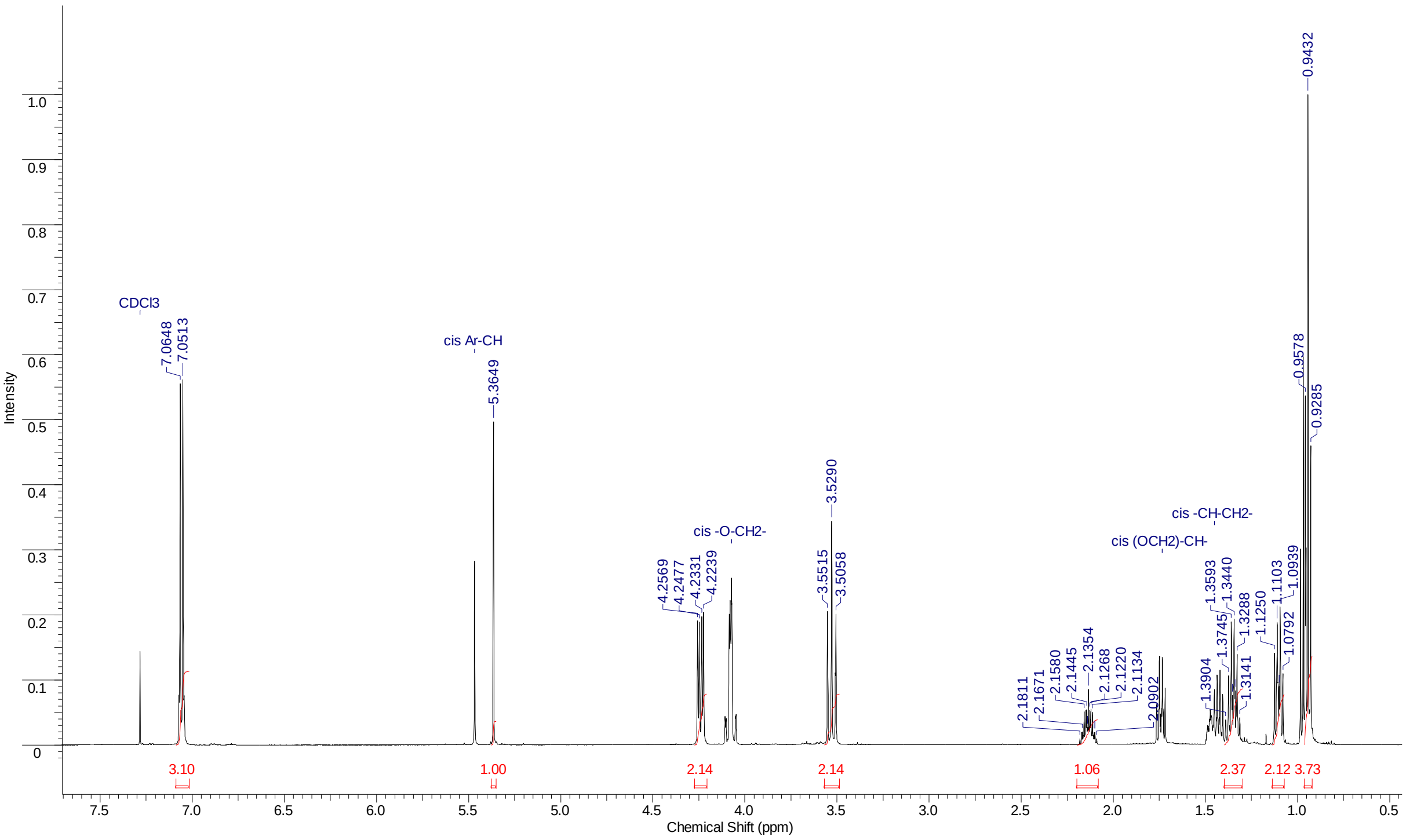

$^{13}C$ NMR (125 MHz, $CDCl_3$) δ: 162.50 (dd, *J*=245.53, 5.45 Hz);142.41 (t, *J*=9.08 Hz); 109.28 (m); 99.18 (t, *J*=2.27 Hz); 72.55; 70.93 (t, *J*=29.52 Hz); 33.90; 30.27; 19.55; 14.22

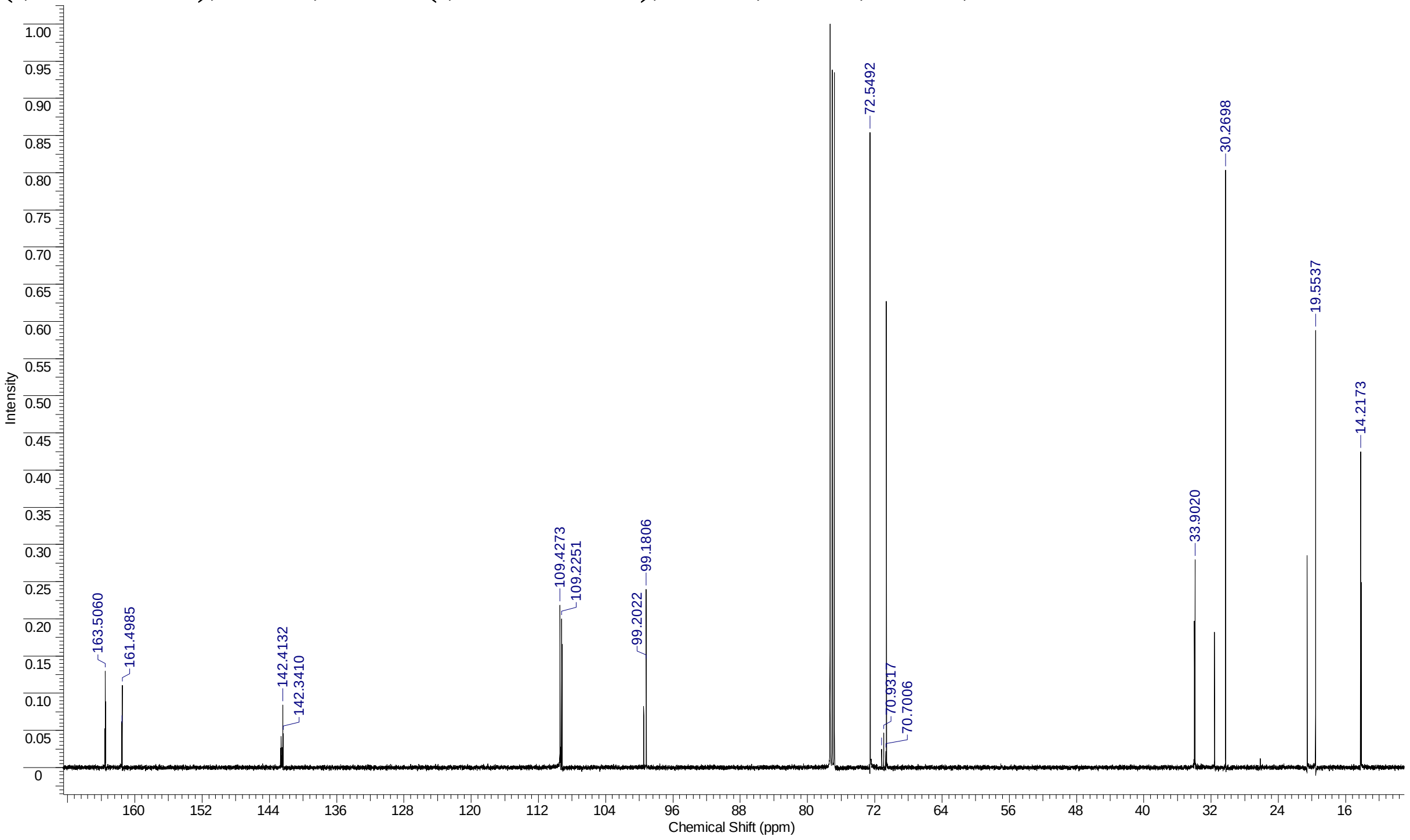

***5-propyl-2-(2,3',5',6-tetrafluoro-[1,1'-biphenyl]-4-yl)-1,3-dioxane (11)***

A mixture of 2-(3,5-difluoro-4-iodophenyl)-5-propyl-1,3-dioxane (40.36 g; 0.109 mol), the 2-(3,5-difluorophenyl)-1,3,2-dioxaborinane (23.78 g; 0.121 mol), anhydrous, $Cs_2CO_3$ (124,79 g; 0.383 mol), acetone (300 ml), and water (100 ml) was heated to reflux (~55 °C) under nitrogen and maintained at that temperature for 5 minutes. The reaction mixture was then cooled (~40 °C). Then $Pd(OAc)_2$ (3 %mol) was added and the mixture was stirred at reflux for 2h. Later, it was cooled down, acidified using 10% HCl solution and extracted with DCM. The organic layer was then dried over $MgSO_4$ and concentrated under vacuum. The residue was recrystallized from ethanol.

Yield 23.4 (60%)

Purity 99.7% (GC-MS) (97.8% of trans isomer)

MS(EI) m/z: 354 (M+); 335; 253; 237; 224 ;206; 175;

$^{1}H$ NMR (500 MHz, $CDCl_3$) δ: 7.18 (d, *J*=8.24 Hz, 2 H); 7.02 (dd, *J*=6.87, 1.07 Hz, 2 H); 6.88 (tt, *J*=8.96, 2.33 Hz, 1 H); 5.42 (s, 1 H); 4.27 (dd, *J*=11.90, 4.58 Hz, 2 H); 3.56 (t, *J*=11.44 Hz, 2 H); 2.17 (m, 1 H); 1.37 (m, 2 H); 1.12 (m, 2 H); 0.96 (t, *J*=7.32 Hz, 3 H)

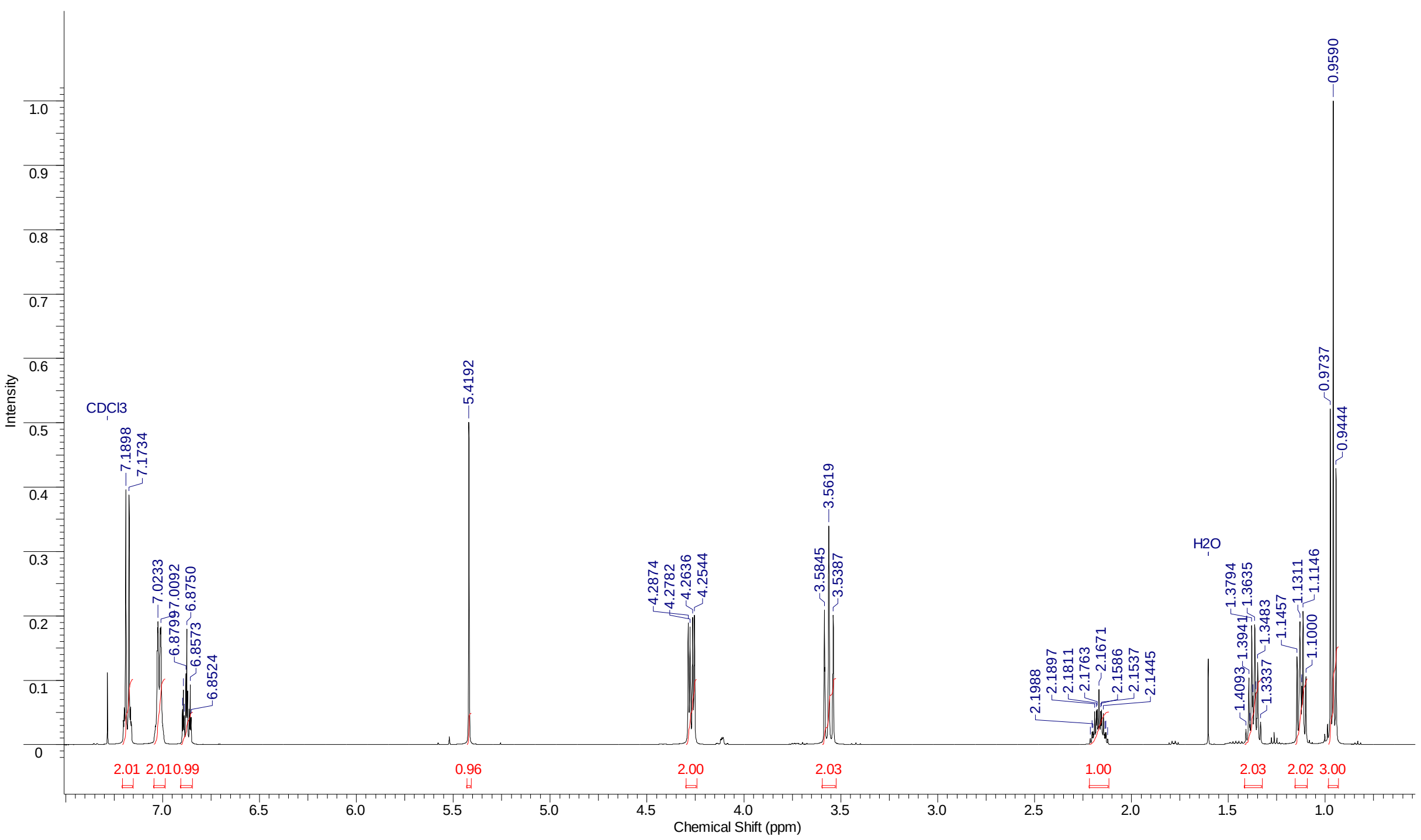

$^{13}$C NMR (126 MHz, $CDCl_3$) δ: 162.69 (dd, *J*=247.96, 12.72 Hz); 159.52 (dd, *J*=248.68, 6.81 Hz); 141.45 (t, *J*=9.54 Hz); 131.93 (t, *J*=10.44 Hz); 116.33 (m); 113.41 (m); 109.79 (m); 103.76 (t, *J*=25.43 Hz); 99.31; 72.54; 33.90; 30.25; 19.51; 14.16

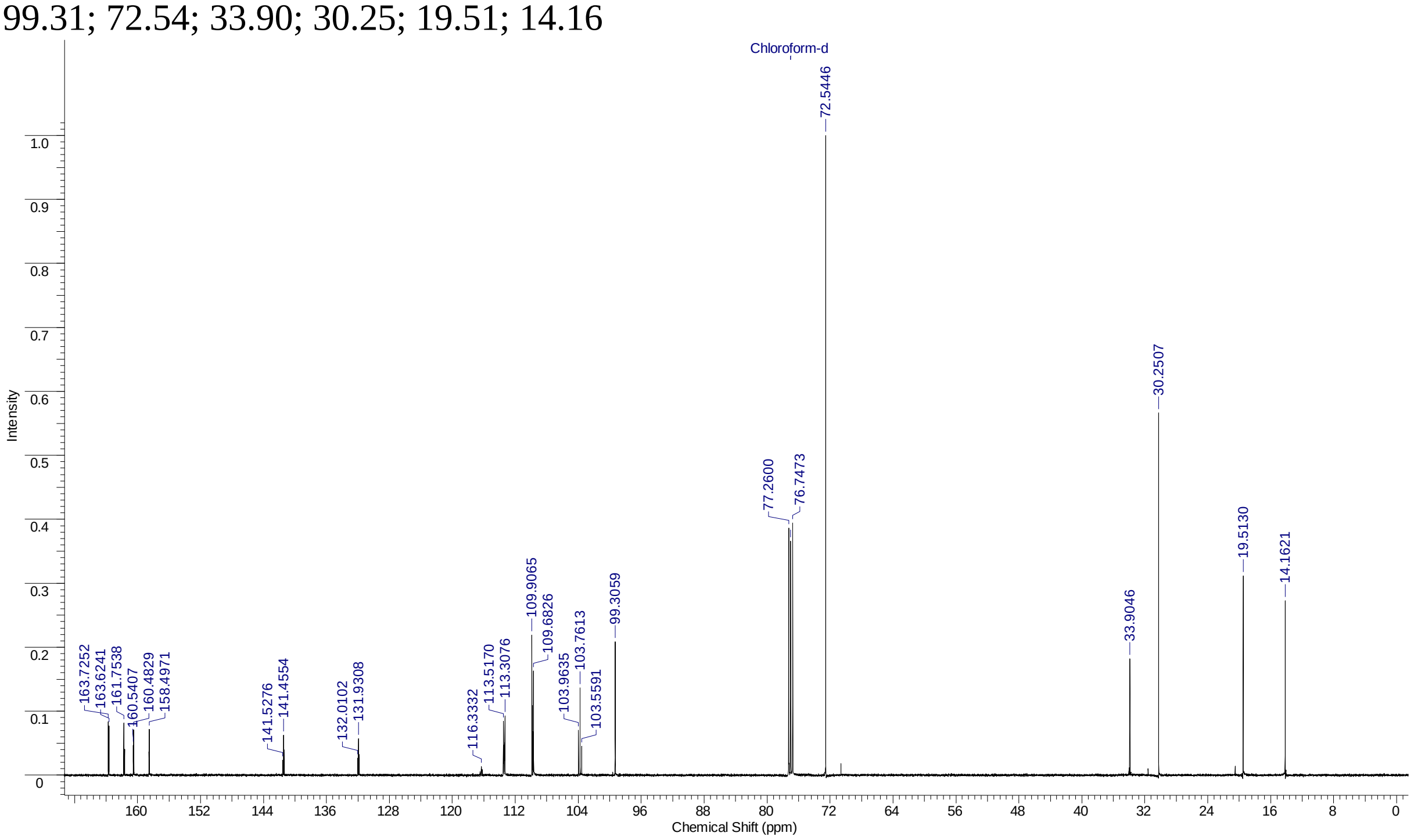

***2',3,5,6'-tetrafluoro-4'-(5-propyl-1,3-dioxan-2-yl)-[1,1'-biphenyl]-4-carboxylic acid (12)***

5-propyl-2-(2,3',5',6-tetrafluoro-[1,1'-biphenyl]-4-yl)-1,3-dioxane (3) (23.36 g; 0.066 mol) was mixed with anhydrous THF (250 ml) under nitrogen and cooled to -78°C in an acetone/dry ice bath. Solution of n-butyllithium dissolved in hexane (0.073 mol, 2.5M) was added dropwise, and temperature was kept below -70°C. The reaction mixture was stirred for 2.5h in −78°C. Then the reaction was flushed with gaseous carbon dioxide (not exceeding the -65°C). After the reaction was completed, it was allowed to reach room temperature. 10% solution of $H_2SO_4$ was added to the mixture. Then the crude product was filtered off, washed with water and hexane. Product was recrystallized three times from hexane-toluene mixture (1:1 v/v).

Yield 13.7 g (52%)

Purity 99.2% (GC-MS) (97.8% of trans isomer)

MS(EI) (MS spectra of corresponding methyl ester) m/z: 412(M+); 381; 353; 311; 281; 253; 224;

$^1$H NMR (500 MHz, THF-*d8*) δ: 7.20 (m, *J*=8.09, 8.09 Hz, 4 H); 5.43 (s, 1 H); 4.20 (dd, *J*=11.75, 4.73 Hz, 4 H); 3.53 (t, *J*=11.44 Hz, 3 H); 2.05 (m, 1 H); 1.35 (m, 2 H); 1.09 (m, 2 H); 0.93 (t, *J*=7.32 Hz, 3 H)

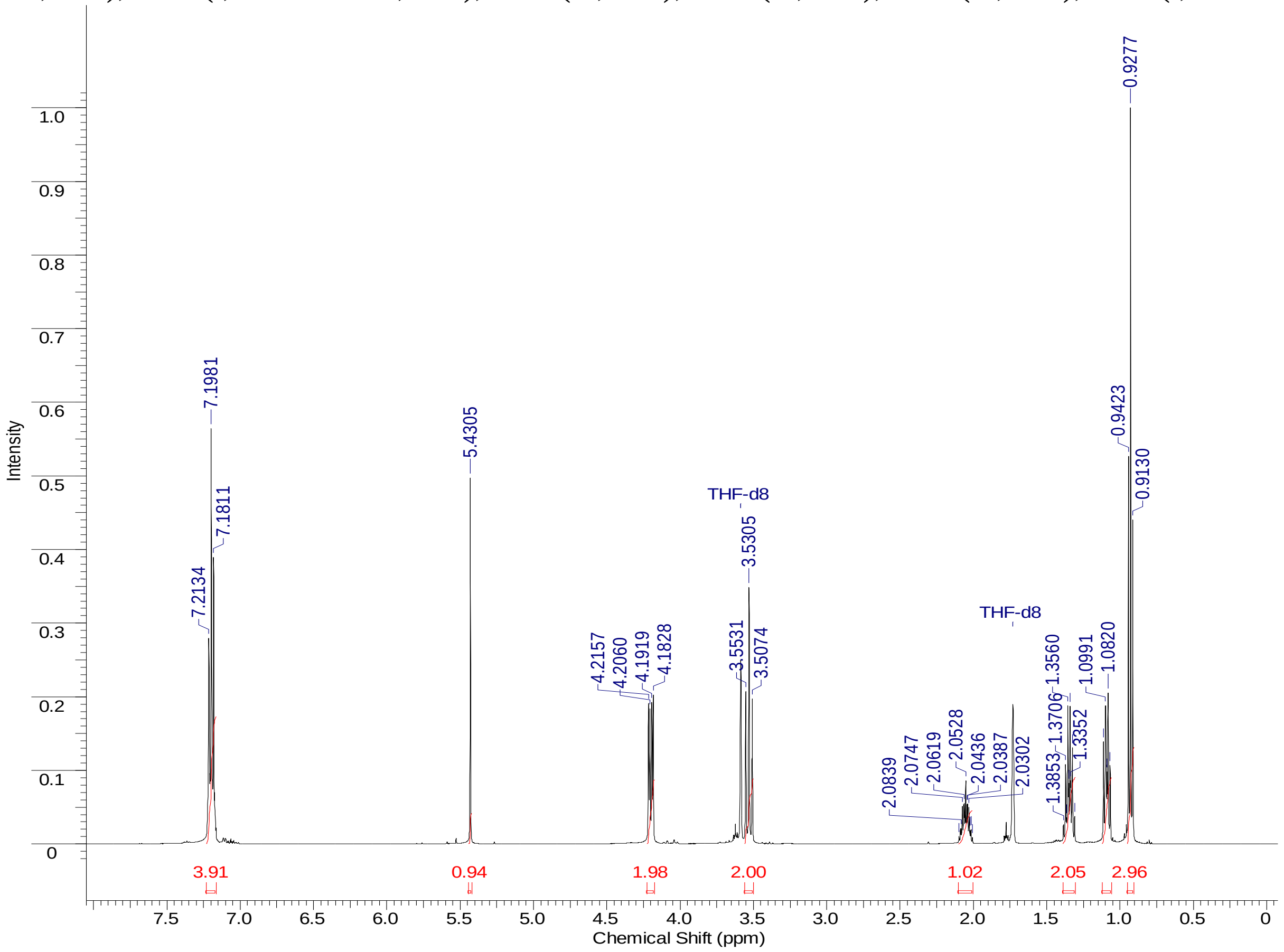

$^{13}C$ NMR (126 MHz, THF-*d8*) δ: 160.06 (dd, *J*=251.83, 7.26 Hz, overlapping with carboxylic carbon); 159.32 (dd, *J*=249.33, 6.81 Hz); 143.14 (t, *J*=9.99 Hz); 133.63 (t, *J*=10.90 Hz); 115.25 (m); 113.85 (d, *J*=27.25 Hz); 111.82 (t, *J*=19.53 Hz); 109.66 (m); 98.95 (t, *J*=2.27 Hz); 72.15; 34.03; 30.21; 19.45; 13.63

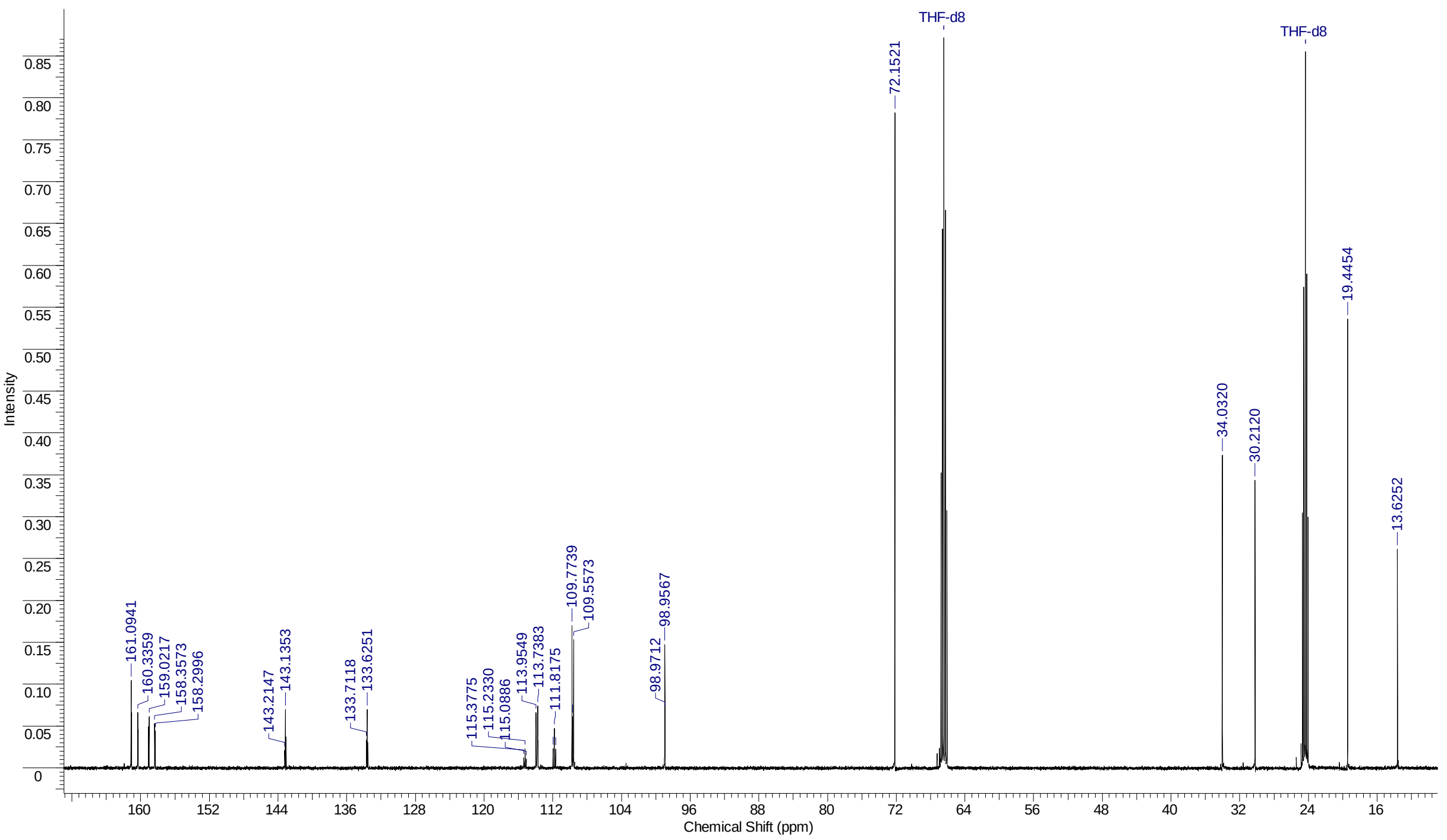

***2-((10-bromodecyl)oxy)tetrahydro-2H-pyran (14c)***

10-bromodecan-1-ol (74,9 g; 0.313 mol) and p-toluenesulfonic acid (73.9 g; 0.313 mmol) was mixed with anhydrous THF (200 ml). Under nitrogen, 3,4-dihydropyran (31.46 ml; 0.345 mol) was added dropwise to the mixture at room temperature. The reaction was stirred at room temperature for 8 hours. The solvent was evaporated. The crude product was dissolved in dichloromethane; mixture was washed three times with water, organic layer was dried over $MgSO_4$ and concentrated. The crude product was purified by flash column chromatography (DCM, hexane as a mobile phase.)

Yield:84.7 g (84.87 %)

Purity 98.21% (GC-MS)

MS(EI) m/z: 319 (M+); 247; 150; 135;101; 85;

$^{1}H$ NMR (500 MHz, $CDCl_3$) δ: 4.55 (dd, *J*=4.43, 2.90 Hz, 1 H); 3.85 (m, 1 H); 3.70 (m, 1 H); 3.48 (m, 1 H); 3.36 (m, 3 H); 1.82 (m, 3 H); 1.69 (m, 1 H); 1.53 (m, 6 H); 1.39 (dd, *J*=14.50, 7.17 Hz, 2 H); 1.31 (m, 10 H)

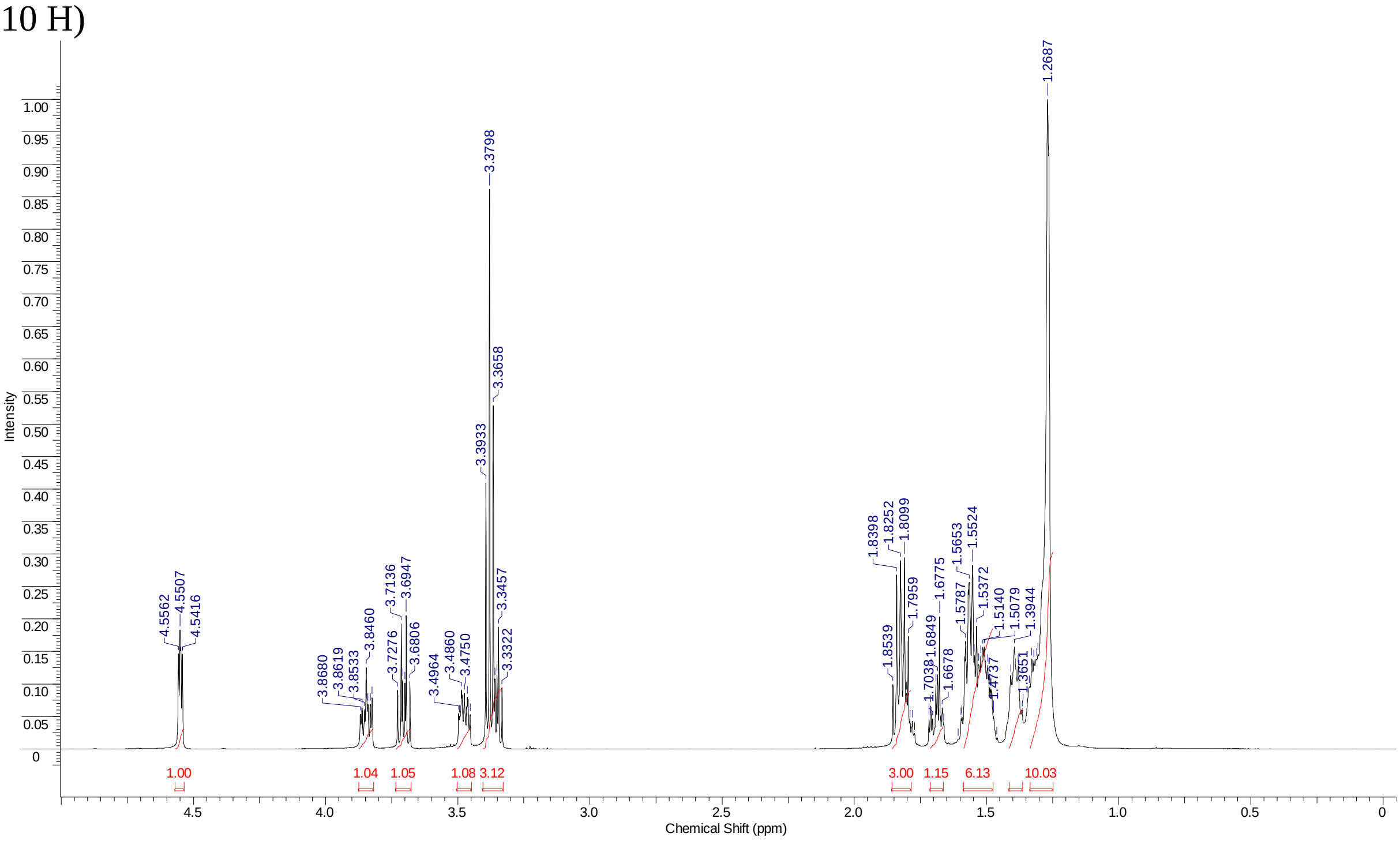

$^{13}C$ NMR (126 MHz, $CDCl_3$) δ: 98.80; 67.61; 62.30; 33.95; 32.79; 30.76; 29.70; 29.41; 29.38; 29.33; 28.70; 28.12; 26.18; 25.48; 19.67

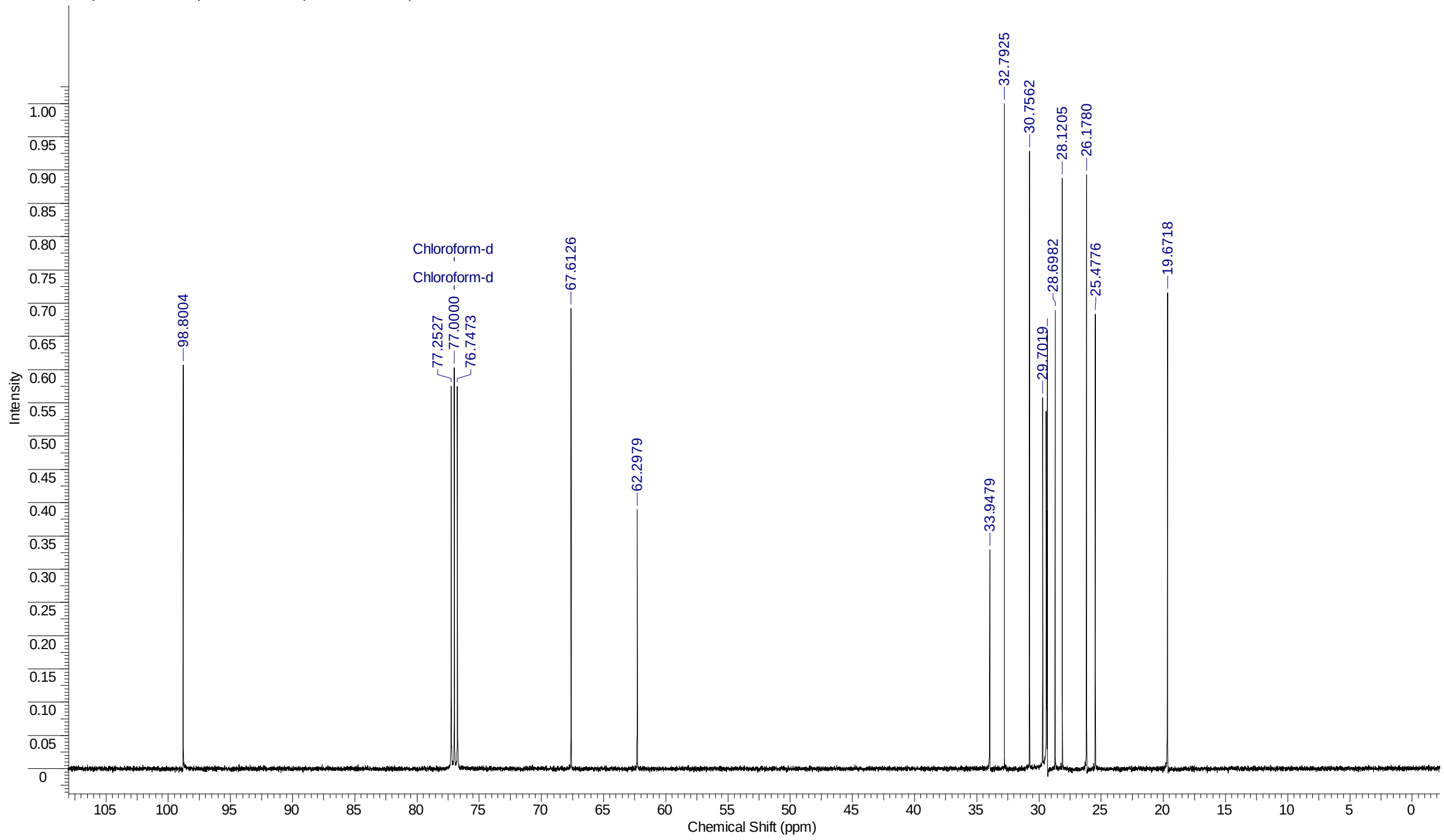

Compounds 14a and 14b were synthesized in the same manner as 14c.

***10-((4'-(difluoro(3,4,5-trifluorophenoxy)methyl)-2,3',5'-trifluoro-[1,1'-biphenyl]-4-yl)oxy)decan-1-ol (15c)***

The mixture of 4'-(difluoro(3,4,5-trifluorophenoxy)methyl)-2,3',5'-trifluoro-[1,1'-biphenyl]-4-ol (4.1 g; 0.00963 mol), 2-((10-bromodecyl)oxy)tetrahydro-2H-pyran (3.4 g; 0.01059 mol), potassium carbonate (2.65 g; 0.0193 mol) in butan-2-one (130 ml) was refluxed for 5h, solution was warmed up to room temperature. The mixture was poured into 80 ml of water and 10 ml of 10% HCl. Crude product was extracted with Dichloromethane. Organic layer was washed three times with water, dried over $MgSO_4$ and solvent was evaporated. The crude product was used in subsequent reactions

Product of the previous synthesis (6.36 g; 0.00963 mol), p-toluenesulfonic acid (pTsOH) (1.63 g; 0.00963 mol) and methanol (50 ml) were stirred at room temperature for 24h. The solvent was evaporated. The crude product was extracted with dichloromethane; mixture was washed three times with water, organic layer was dried over $MgSO_4$ and concentrated.

Yield 3.34 (60.3%)

Purity 98.21% (GC-MS) ( Aliphatic impurities visible in $^{1}$HNMR and $^{13}$CNMR spectra do not affect the subsequent synthetic steps)

MS(EI) m/z: 576 (M+); 493; 429; 383; 341; 303; 273; 244

$^{1}$H NMR (500 MHz, DMSO-*d6*) δ: 7.48 (d, J=11.29 Hz, 1 H); 7.40 (m, 1 H); 6.92 (m, 1 H); 4.32 (s, 1 H); 4.01 (t, J=6.56 Hz, 1 H); 3.35 (t, J=6.41 Hz, 3 H); 3.24 (t, J=7.02 Hz, 1 H); 1.71 (m, 2 H); 1.31 (m, 17 H)

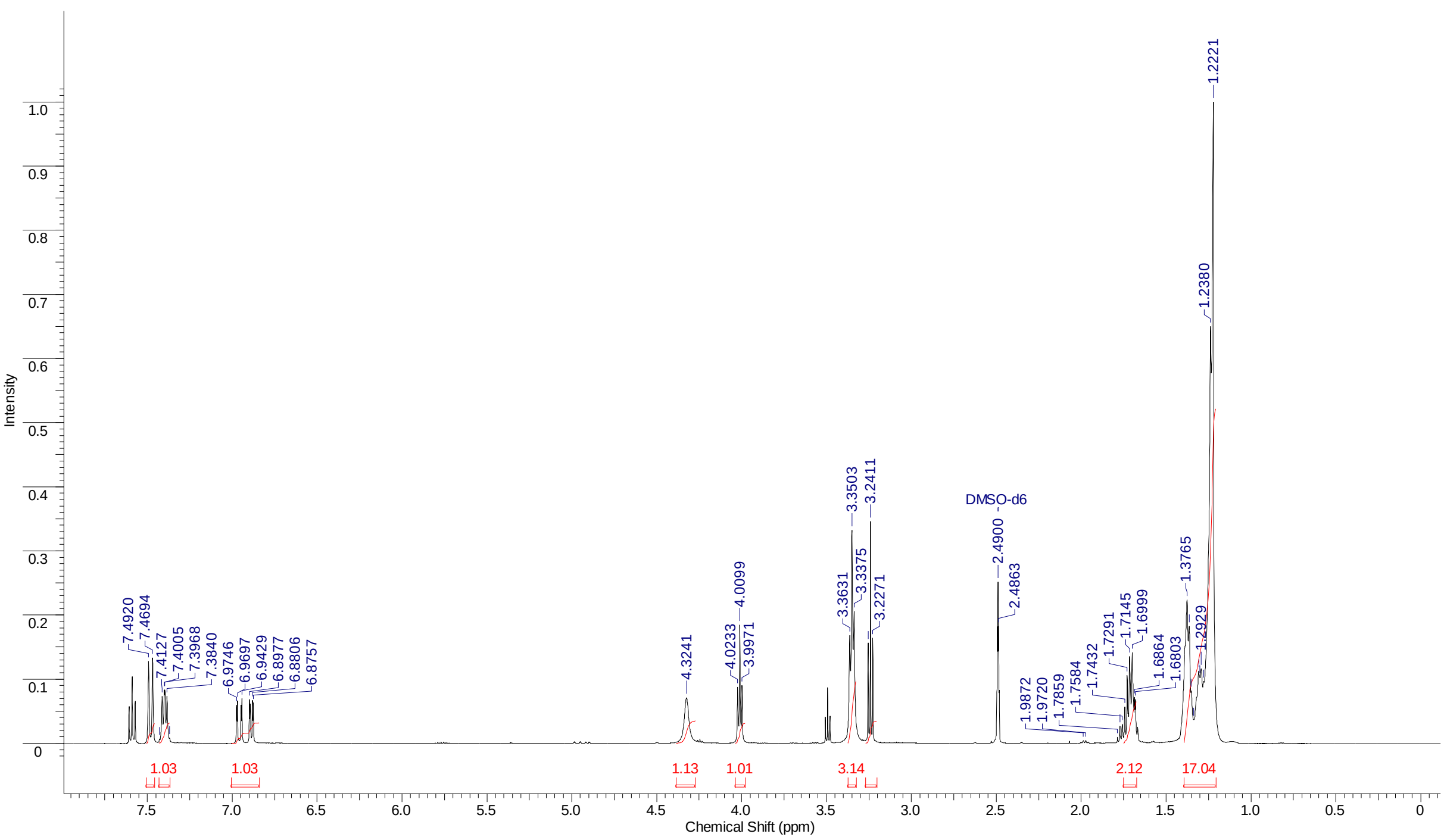

$^{13}$C NMR (126 MHz, DMSO-*d6*) δ: 161.29; 161.19; 161.04; 159.06; 158.14 (m, *J*=6.36 Hz); 151.41 (m, *J*=5.45 Hz); 149.41 (m) 141.83; 136.97; 131.35 (m) 116.55 (m); 113.04 (m); 111.83; 108.00 (m); 102.72 (m); 68.42; 60.85; 33.04; 32.68; 29.98; 29.09 (m); 28.58; 28.25; 28.02; 27.65 ; 25.63; 25.54; 9.10

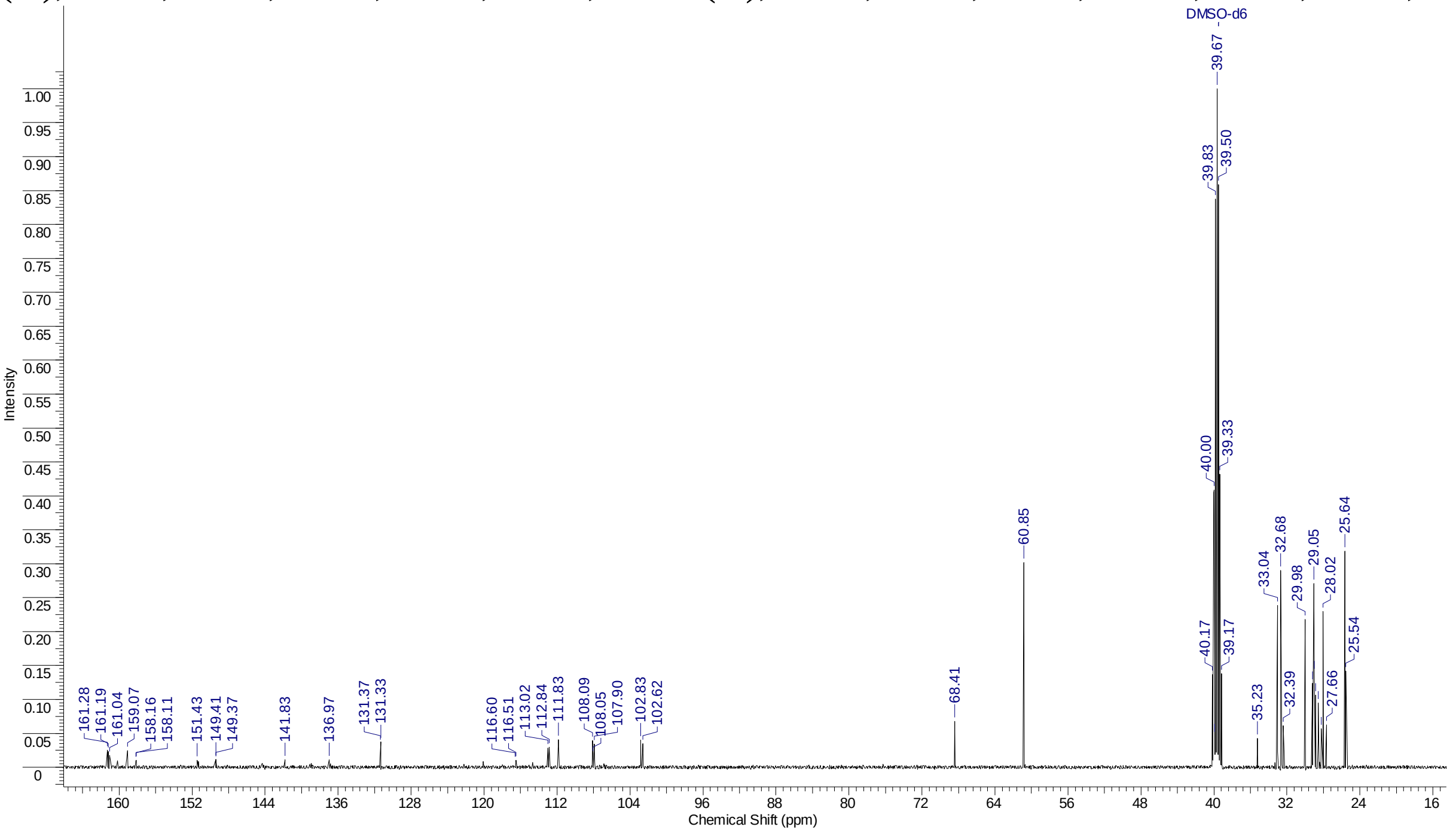

Compounds 15a and 15b were synthesized in the same manner as 15c.

**7-((4'-(difluoro(3,4,5-trifluorophenoxy)methyl)-2,3',5'-trifluoro-[1,1'-biphenyl]-4-yl)oxy)heptyl 2',3,5,6'-tetrafluoro-4'-(5-propyl-1,3-dioxan-2-yl)-[1,1'-biphenyl]-4-carboxylate *(I-7)***

To a stirred solution of 2',3,5,6'-tetrafluoro-4'-(5-propyl-1,3-dioxan-2-yl)-[1,1'-biphenyl]-4-carboxylic acid (1.93 g; 0.00486 mol), ((4'-(difluoro(3,4,5-trifluorophenoxy)methyl)-2,3',5'-trifluoro-[1,1'-biphenyl]-4-yl)oxy)methanol (2 g; 0.00476 mol) and DCC (1 g; 0.00486 mol) in DCM (50 ml), DMAP (0.1 g) was added and the solution was stirred at room temperature. The reaction mixture was filtered through silica pad and the filtrate was concentrated under vacuum. Crude product was purified by flash column chromatography (DCM, hexane as a mobile phase). The residue was then recrystallized from ethanol to give white solid.

Yield 2.1g (48.2%)

Purity 99.4% (HPLC-MS)

MS(EI) m/z: 914; 896; 873; 835; 812;794;767; 666; 645; 571; 381; 281; 273; 253; 224; 83;

$^{1}H$ NMR (500 MHz, $CDCl_3$) δ: 7.32 (t, *J*=8.85 Hz, 1 H); 7.16 (m, 4 H); 7.06 (d, *J*=8.55 Hz, 2 H); 6.98 (m, 2 H); 6.74 (m, 2 H); 5.38 (s, 1 H); 4.39 (t, *J*=6.56 Hz, 2 H); 4.23 (dd, *J*=11.90, 4.58 Hz, 2 H); 3.98 (t, *J*=6.41 Hz, 2 H); 3.53 (t, *J*=11.44 Hz, 2 H); 2.13 (m, 1 H); 1.80 (m, 4 H); 1.46 (m, 5 H, overlapping $H_2O$ signal); 1.33 (m, 2 H); 1.09 (m, 2 H); 0.93 (t, *J*=7.32 Hz, 3 H)

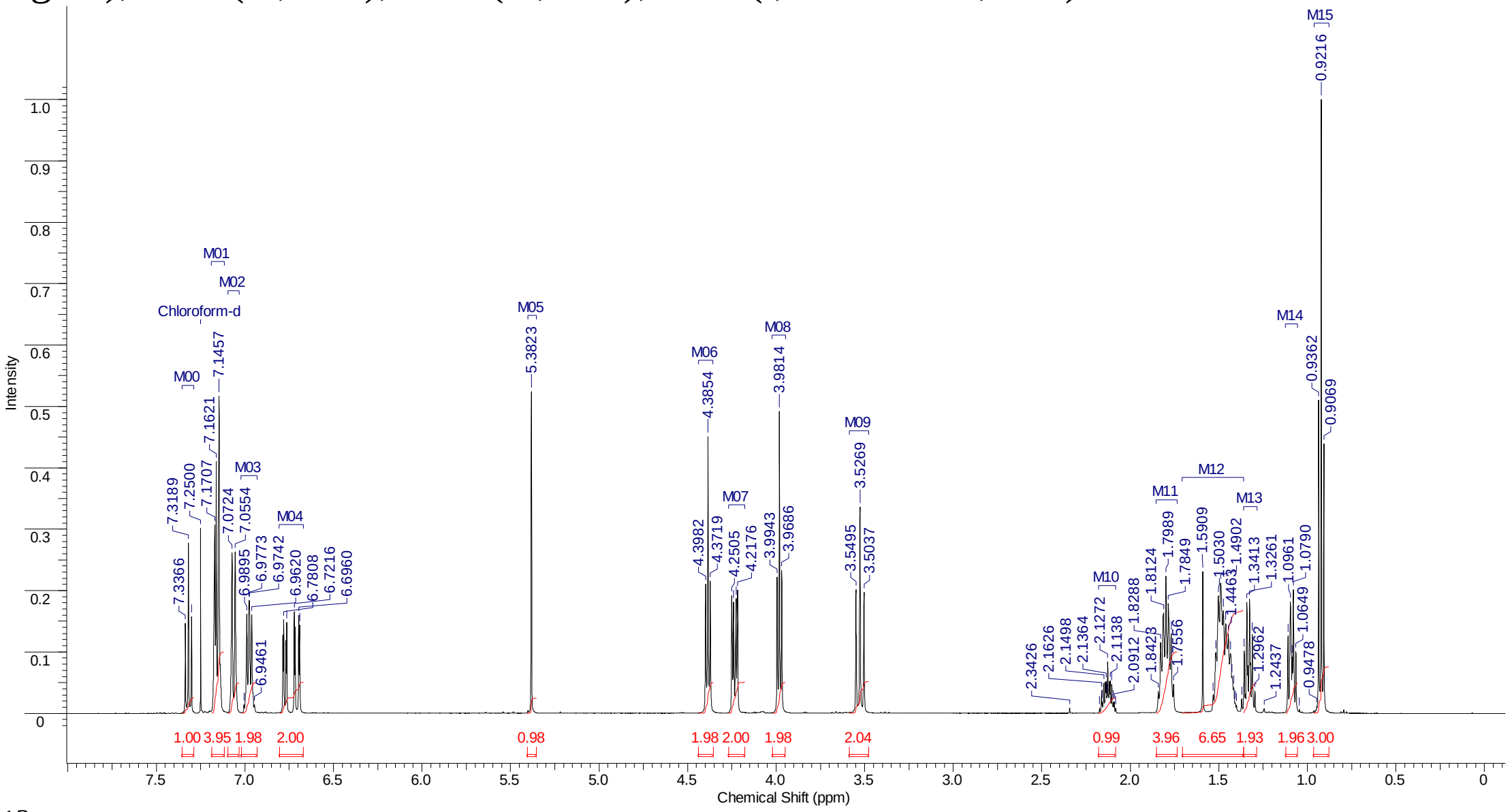

$^{13}C$ NMR (126 MHz, $CDCl_3$) δ: 161.39; 160.40 (dd, *J*=249.78, 6.36 Hz); 159.88 (dd, *J*=256.14, 6.36Hz); 159.92 (dd, *J*=255.33, 7,27Hz); 159.39 (dd, *J*=250.69, 6.36 Hz); 151.02 (ddd, *J*=251.6, 10.97, 5.47 Hz); 144.73 (m); 142.14 (t, *J*=9.54 Hz); 141.65 (t, *J*=11.81 Hz);138.44 (dt, *J*=249.79, 15.44 Hz); 134.16 (t, *J*=10.90 Hz); 130.41 (d, *J*=4.54 Hz); 120.25 (t, *J*=117.5 Hz); 117.37 (d, *J*=11.81 Hz); 115.43 (t, *J*=18.62 Hz); 114.15 (dd, *J*=24.98, 2.27 Hz); 112.43 (m); 111.38 (m); 110.84 (t, *J*=18.17 Hz); 109.94 (m); 107.44 (m); 102.77 (d, *J*=26.34 Hz); 99.15 (t, *J*=2.27 Hz); 72.54; 68.47; 66.01; 33.89; 30.23; 28.89; 28.81; 28.42; 25.82; 25.72; 19.51; 14.15

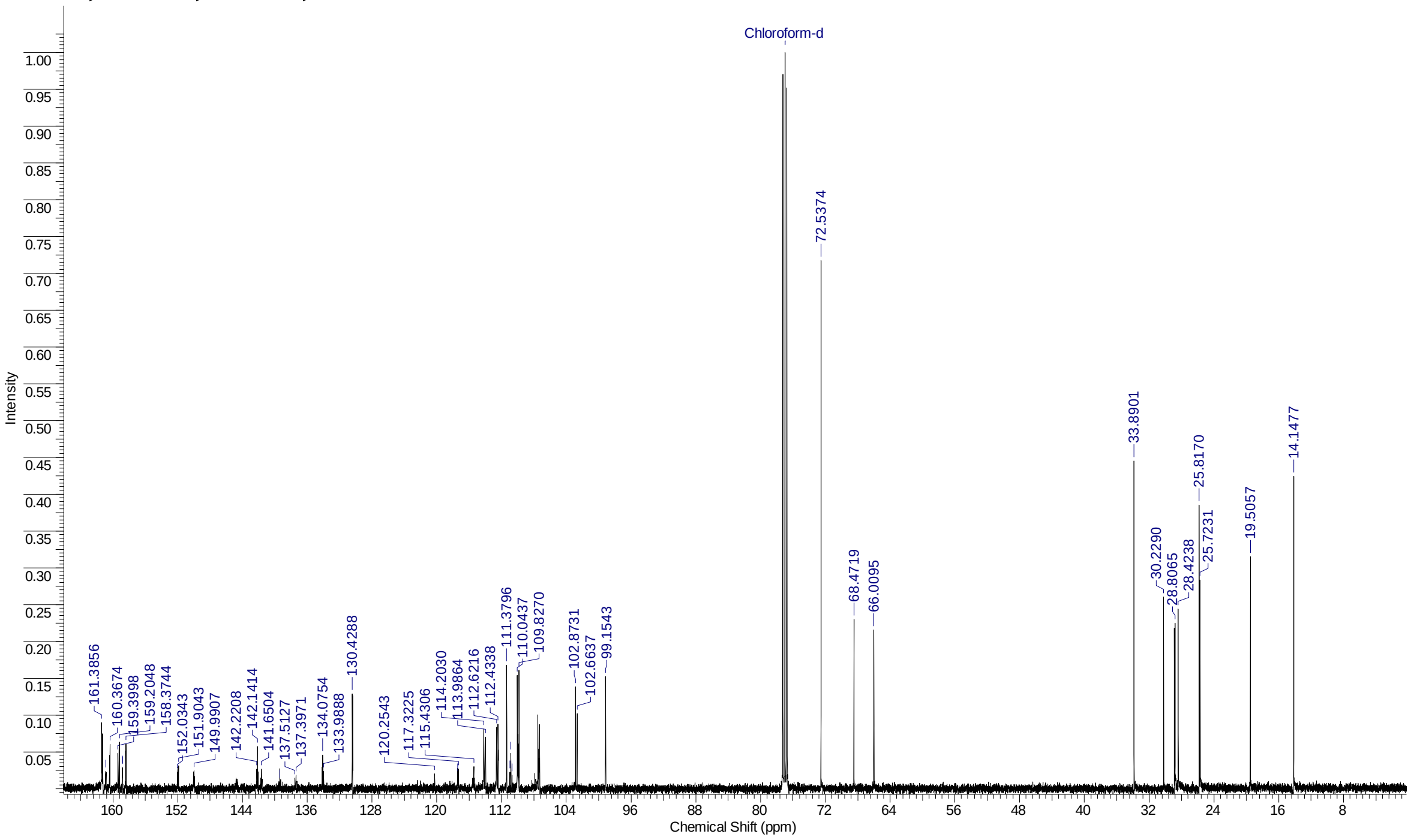

***5-((4'-(difluoro(3,4,5-trifluorophenoxy)methyl)-2,3',5'-trifluoro-[1,1'-biphenyl]-4-yl)oxy)pentyl 2',3,5,6'-tetrafluoro-4'-(5-propyl-1,3-dioxan-2-yl)-[1,1'-biphenyl]-4-carboxylate (I-5)***

The synthesis was carried out in the same manner as in I-7.

Yield 0.4g (48.6 %)

Purity 99.44% (HPLC-MS)

MS(EI) m/z: 886; 867; 822; 785; 739; 689; 639; 617; 566; 511; 381; 281; 273

$^{1}$H NMR (500 MHz, $CDCl_3$) δ: 7.32 (t, *J*=8.70 Hz, 1 H); 7.16 (d, 4 H); 7.07 (d, *J*=8.85 Hz, 2 H); 6.97 (t, *J*=13.58, 7.48 Hz, 2 H); 6.78 (dd, *J*=8.55, 2.44 Hz, 1 H); 6.71 (dd, *J*=12.82, 2.44 Hz, 1 H); 5.38 (s, 1 H); 4.42 (t, *J*=6.41 Hz, 2 H); 4.24 (dd, *J*=11.75, 4.43 Hz, 2 H); 4.01 (t, *J*=6.26 Hz, 2 H); 3.53 (t, *J*=11.44 Hz, 2 H); 2.13 (m, 1 H); 1.87 (m, 4 H); 1.64 (m, 2 H, overlapping $H_2O$ singal); 1.34 (m, 2 H); 1.08 (m, 2 H); 0.92 (t, *J*=7.32 Hz, 3 H)

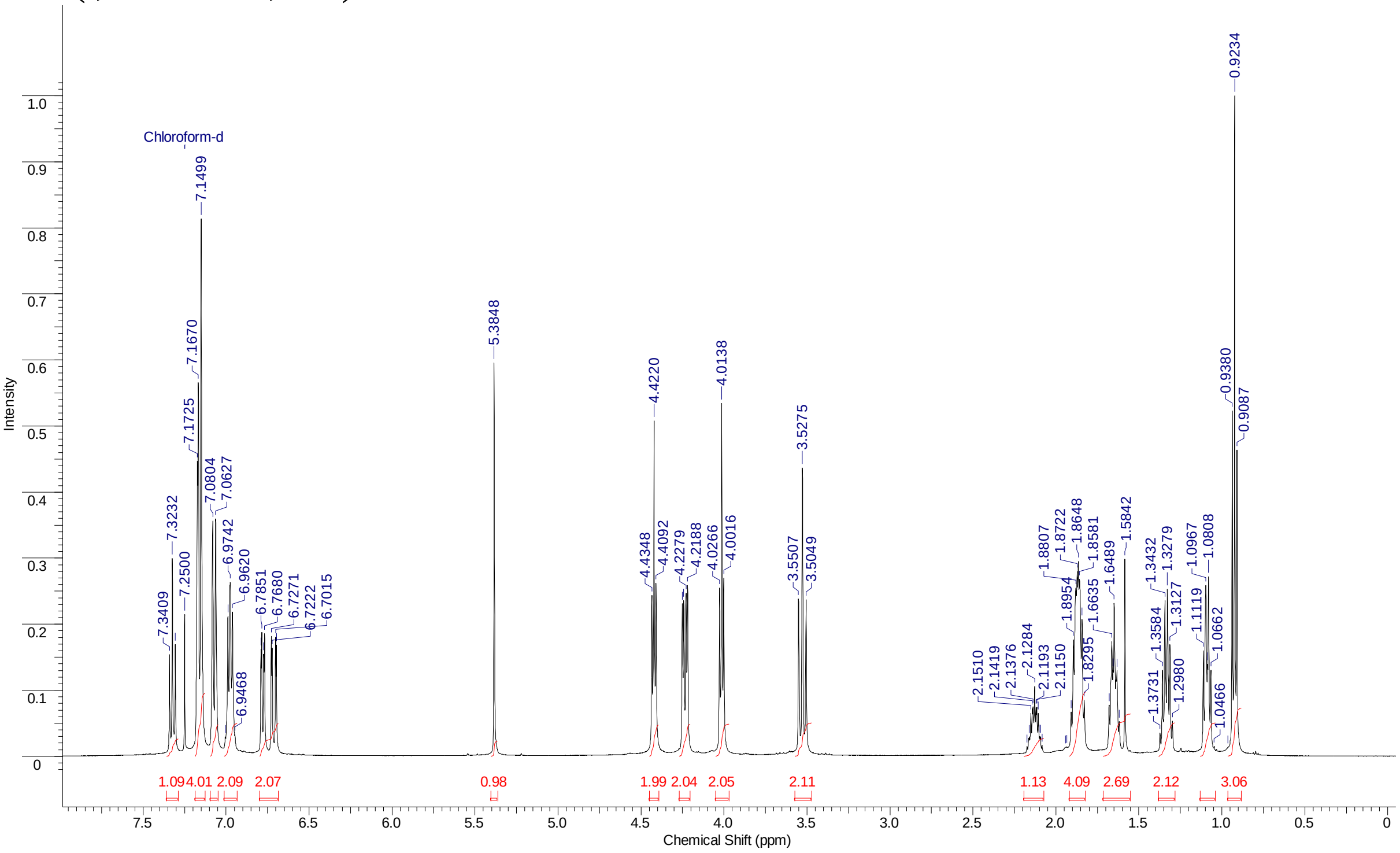

$^{13}C$ NMR (126 MHz, $CDCl_3$) δ: 161.21; 160.4 (d, *J*=249.78); 160.24 (dd, *J*=255.23, 6.36Hz); 159.88 (dd, *J*=276,12, 6.36 Hz); 159.42 (dd, *J*=257.95, 7,27Hz); 150.99 (ddd, *J*=250.61, 10.8, 4.55Hz); 144.70 (m) ;142.18 (t, *J*=9.54 Hz); 141.62 (t, *J*=11.81 Hz); 138.41 (dt, 250.69, 14.54 Hz); 134.19 (t, *J*=11.35 Hz,); 130.45 (d, *J*=4.54 Hz); 120.99 (t, *J*= 117.45 Hz); 117.51 (d, *J*=11.81 Hz); 115.42 (t, *J*=18.62 Hz); 114.15 (dd, *J*=24.98, 2.27 Hz); 114.15 (dd, *J*=24.98, 2.27 Hz); 112.59 (m); 111.33 (m); 110.73 (t, *J*= 18.4 Hz); 109.98 (m,); 107.40 (m, *J*=24.52 Hz); 102.83 (d, *J*=26.34 Hz); 99.17 (t, *J*=2.27 Hz); 72.55; 68.25; 65.73; 33.91; 30.24; 28.57; 22.45; 19.51; 14.15

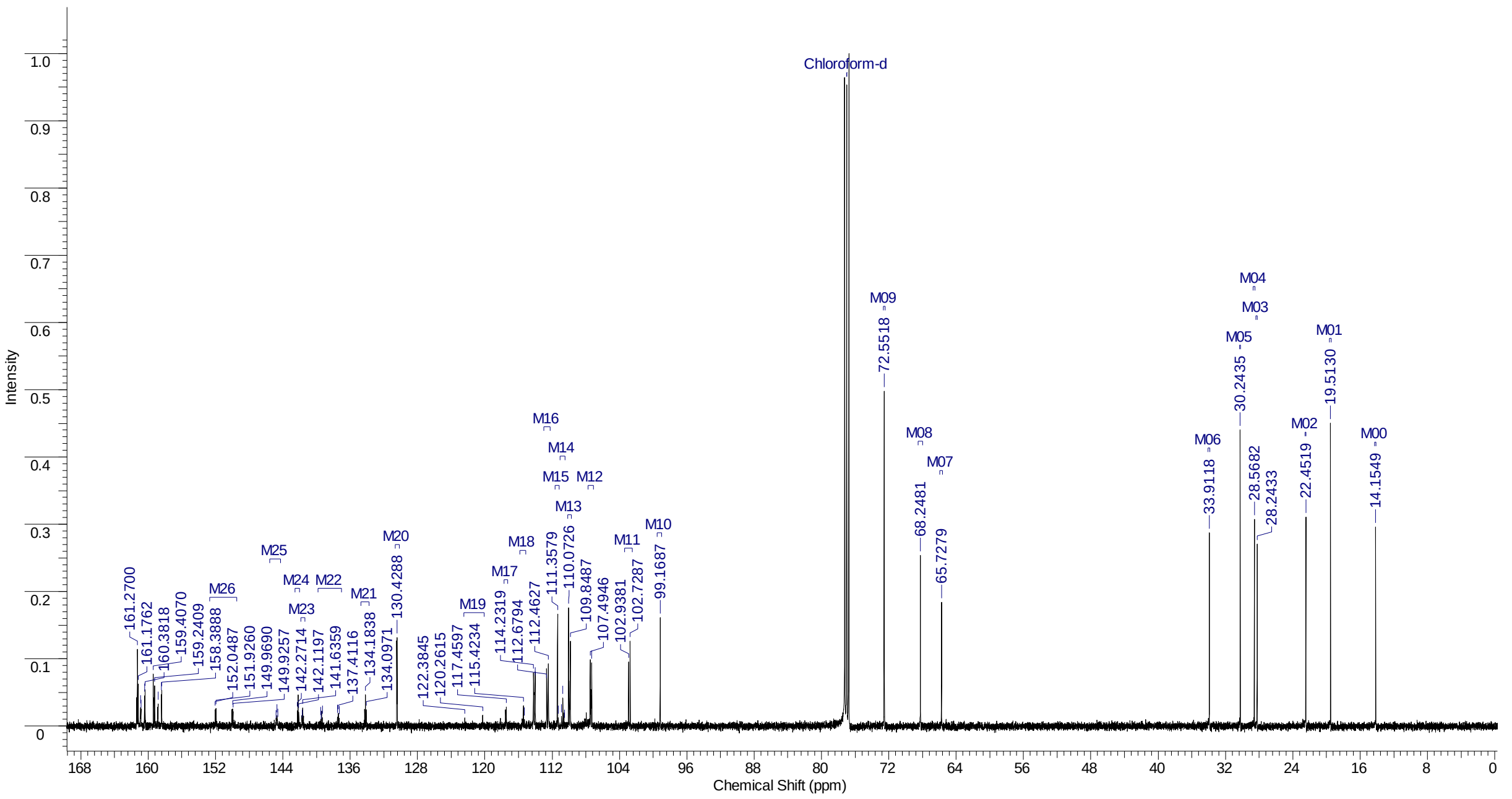

***6-((4'-(difluoro(3,4,5-trifluorophenoxy)methyl)-2,3',5'-trifluoro-[1,1'-biphenyl]-4-yl)oxy)hexyl 2',3,5,6'-tetrafluoro-4'-(5-propyl-1,3-dioxan-2-yl)-[1,1'-biphenyl]-4-carboxylate (I-6)***

The synthesis was carried out in the same manner as in I-7.

Yield 1.98g (52.6%)

Purity 99% (HPLC-MS)

MS(EI) m/z: 900; 873; 199; 781; 653; 625; 381; 281; 273; 253

$^{1}$H NMR (500 MHz, $CDCl_3$) δ: 7.32 (t, *J*=8.85 Hz, 1 H); 7.16 (m, 4 H); 7.06 (d, *J*=8.85 Hz, 2 H); 6.74 (m, 2 H); 5.38 (s, 1 H); 4.40 (t, *J*=6.41 Hz, 2 H); 4.23 (dd, *J*=11.29, 4.58 Hz, 2 H); 3.99 (t, *J*=6.26 Hz, 2 H); 3.53 (t, *J*=11.44 Hz, 2 H); 2.13 (m, 1 H); 1.82 (m, 4 H); 1.54 (m, 4 H, overlapping $H_2O$ singal); 1.33 (m, 2 H); 1.08 (m, 2 H); 0.92 (t, *J*=7.32 Hz, 3 H)

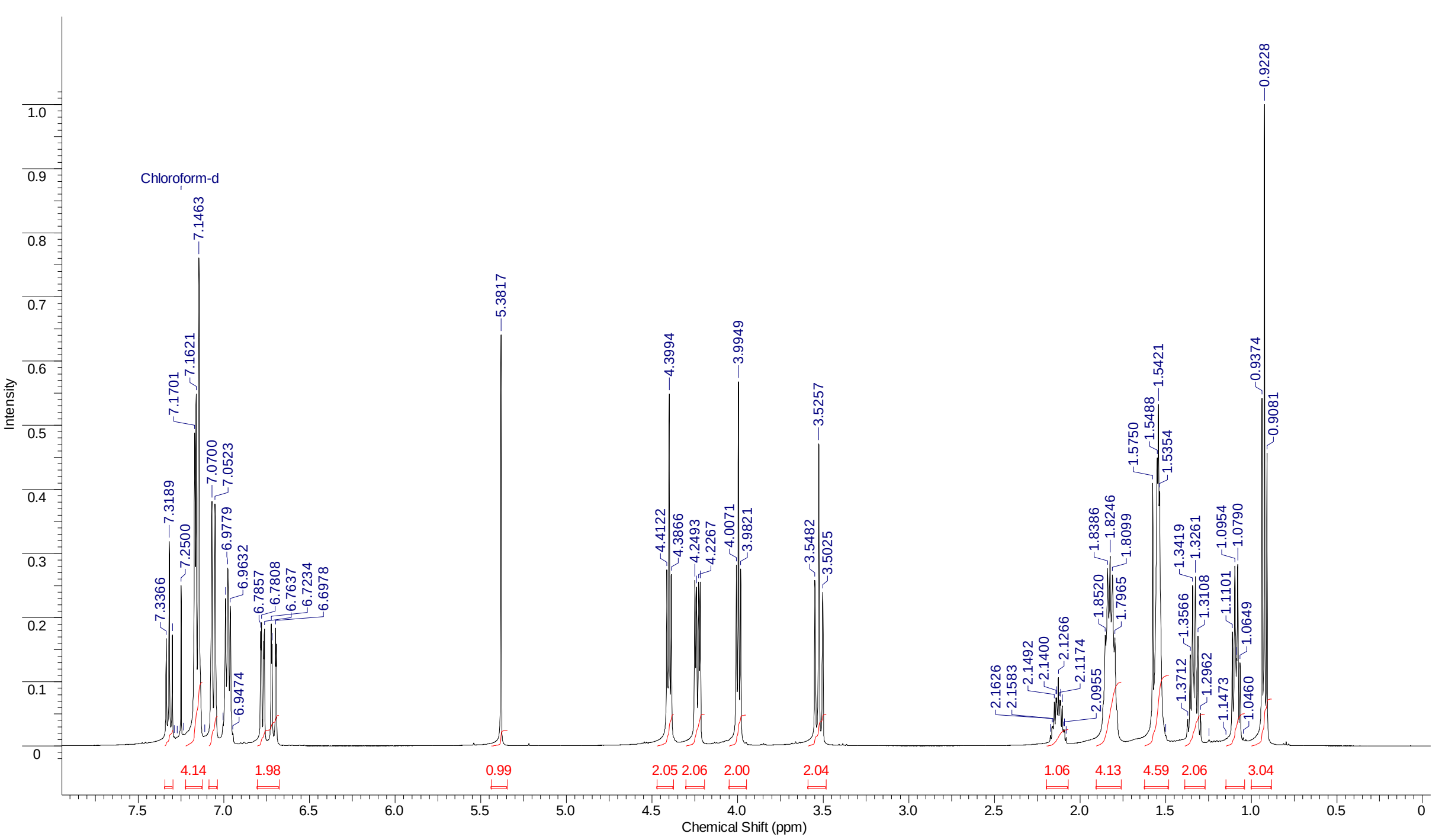

$^{13}$C NMR (126 MHz, $CDCl_3$) δ: 160.43 (d, *J* = 249.78 Hz); 160.30 (dd, *J* = 257.03, 9.54 Hz, overlapping carboxylic carbon); 159.91 (dd, *J* = 256.14, 6.36 Hz); 159.44 (dd, *J* = 252.12, 6.36 Hz); 151.02 (ddd *J*=250.61, 10.8, 4.55 Hz); 144.73 (m); 142.20 (t, *J* = 9.54 Hz); 141.68 (t, *J* = 11.81 Hz); 138.44 (dt, *J*= 249.79, 15.44 H); 134.16 (t, *J* = 10.90 Hz); 130.46 (d, *J* = 4.54 Hz); 120.30 (t, *J*=118 Hz); 117.46 (d, *J* = 11.81 Hz); 115.47 (t, *J* = 18.62 Hz); 114.15 (dd, *J* = 24.98, 2.27 Hz); 112.58 (m); 111.41 (m); 110.84 (t, *J* = 18.17 Hz); 109.99 (m); 107.44 (m); 102.82 (d, *J* = 26.34 Hz); 99.20 (t, *J* = 2.27 Hz); 72.59; 68.38; 65.93; 33.94; 30.28; 28.89; 28.46; 25.58; 19.55; 14.20

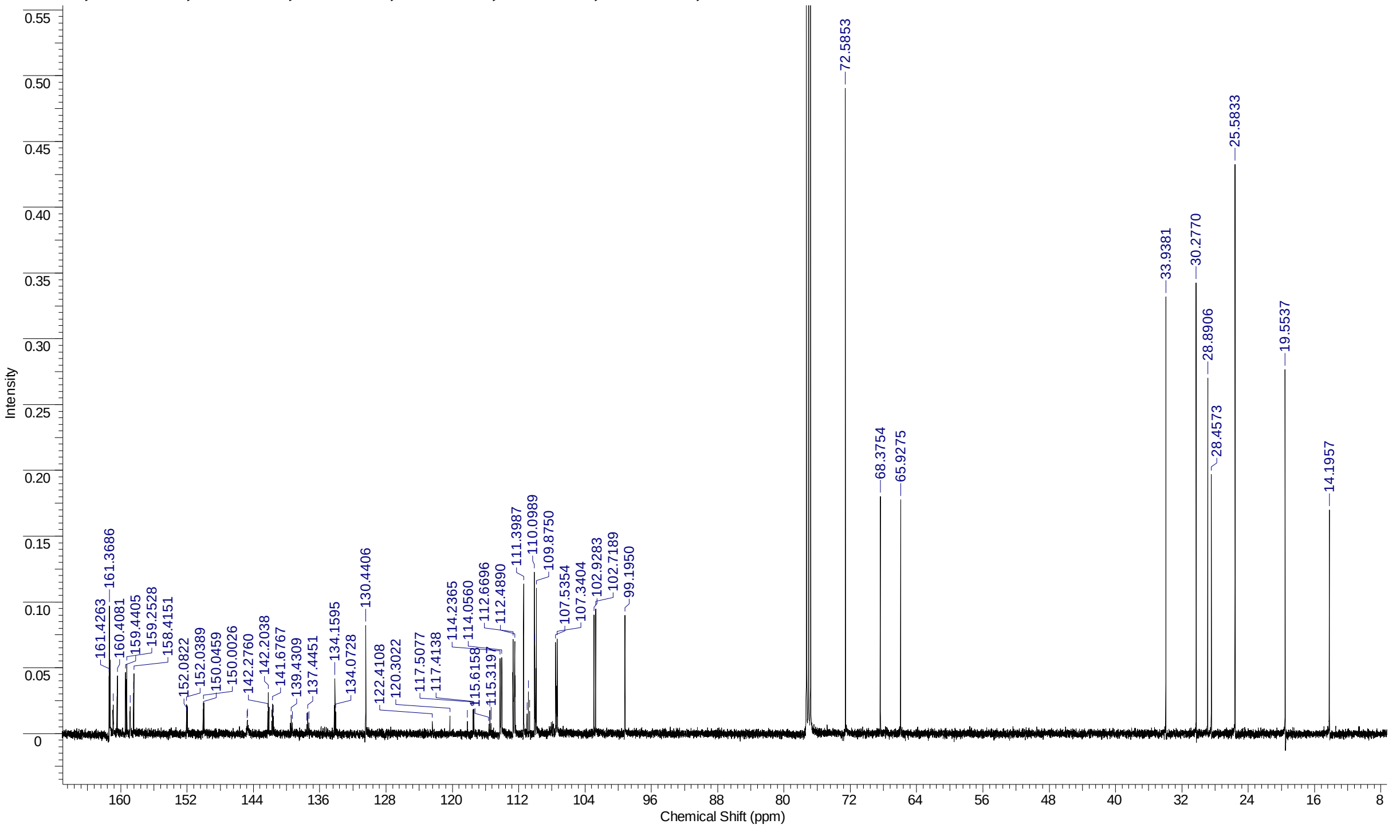

***8-((4'-(difluoro(3,4,5-trifluorophenoxy)methyl)-2,3',5'-trifluoro-[1,1'-biphenyl]-4-yl)oxy)octyl 2',3,5,6'-tetrafluoro-4'-(5-propyl-1,3-dioxan-2-yl)-[1,1'-biphenyl]-4-carboxylate (I-8)***

The synthesis was carried out in the same manner as in I-7.

Yield 0.64g (31.8%)

Purity 98.94% (HPLC-MS)

MS(EI) m/z: 927; 908; 869; 808; 777; 723; 681; 681; 659; 570; 511; 488; 399; 381; 281; 273

$^{1}$H NMR (500 MHz, $CDCl_3$) δ: 7.31 (t, *J*=7.93 Hz, 1 H); 7.18 (m, 4 H); 7.07 (d, *J*=8.55 Hz, 3 H); 6.99 (m, 3 H); 5.38 (s, 1 H); 4.37 (t, *J*=6.41 Hz, 2 H); 4.24 (dd, *J*=11.60, 4.58 Hz, 2 H); 3.53 (t, *J*=11.29 Hz, 2 H); 2.64 (t, *J*=7.63 Hz, 2 H); 2.14 (m, 1 H); 1.76 (m, 2 H); 1.64 (s, 2 H); 1.35 (m, 10 H); 1.09 (m, 2 H); 0.92 (t, *J*=7.32 Hz, 3 H)

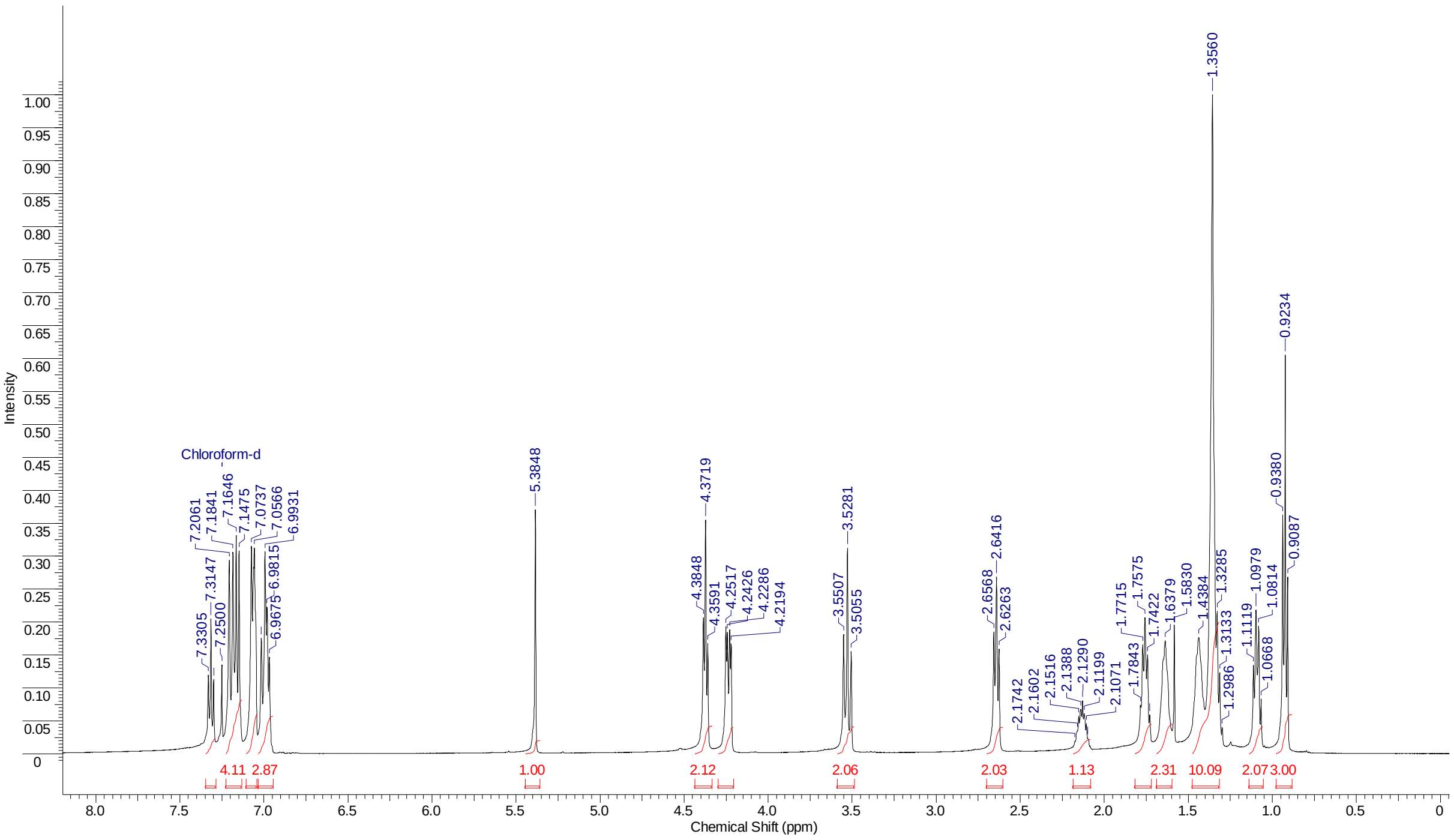

$^{13}$C NMR (126 MHz, $CDCl_3$) δ: 160.93; 159.82 (dd, *J* = 255.68, 6.81 Hz); 159.41 (dd, *J* = 258.86, 6.36 Hz); 159.10 (d, *J* = 249.78 Hz); 158.98 (dd, J = 250.69, 6.36 Hz); 149.24 (ddd, *J* = 250.69, 9.99, 4.54Hz); 144.24 (m); 142.20 (t, *J*=10.9 Hz); 141.50 (t, *J* = 11.81 Hz); 138.21 (dt, *J*= 250.69, 15.44 Hz); 133.62 (t, *J* = 10.90 Hz); 129.27 (d, *J* = 3.63 Hz); 123.08 (t, *J*=117.25 Hz); 117.68 (d, *J* = 11.8 Hz); 115.90 (t, *J* = 17.25); 113.66 (dd, *J*= 24.52, 1.82 Hz); 112.47 (m) ;111.49 (m); 110.46 (t, *J* = 18.17 Hz); 109.54 (dd, *J*=21.9, 5.45Hz); 106.99 (m); 102.82 (d, *J* = 26.34 Hz); 98.74 (t, *J* = 2.27 Hz); 72.12; 65.67; 33.47; 30.49; 29.81; 28.82; 28.61; 28.59; 28.06; 25.33; 19.09 ;13.73;

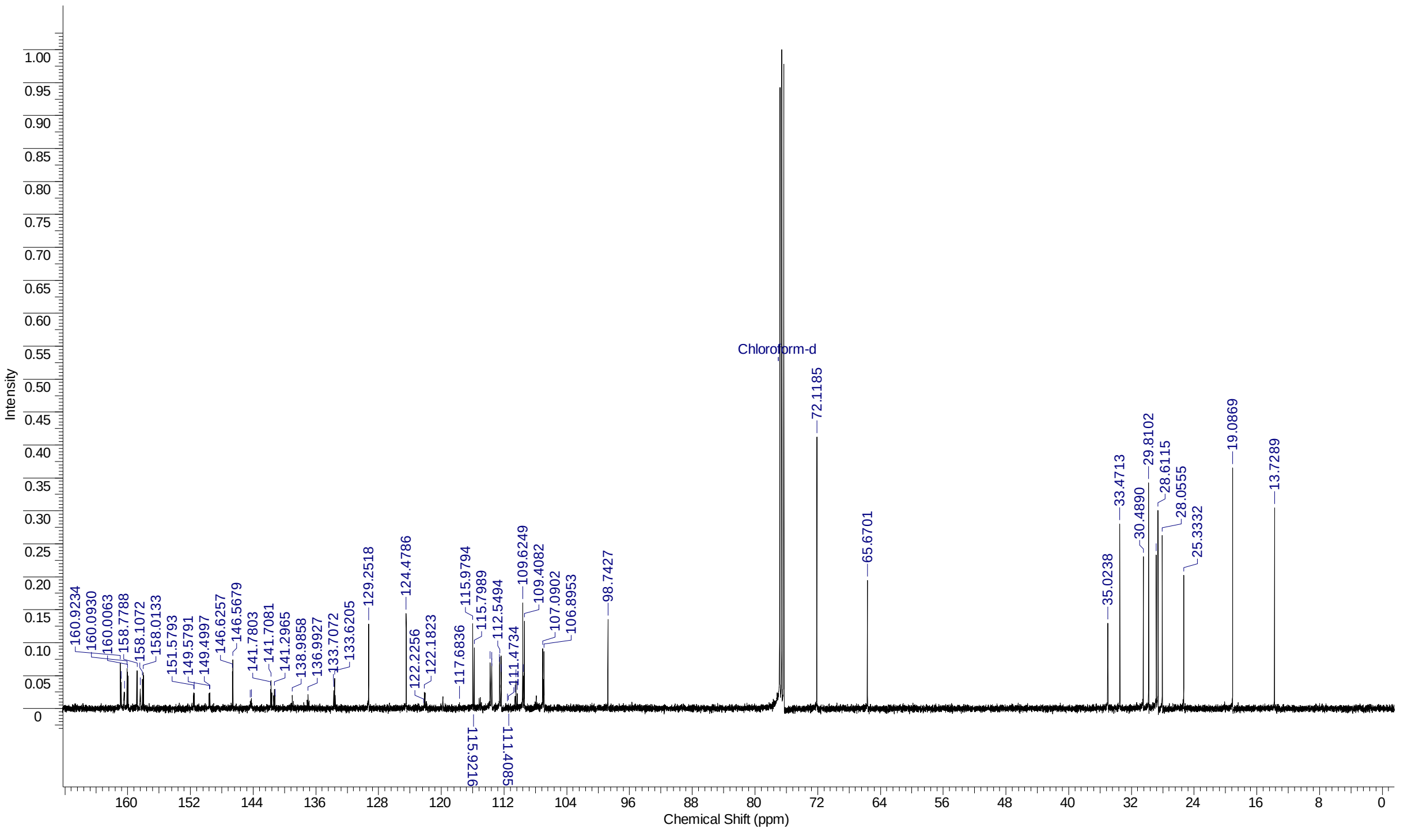

***9-((4'-(difluoro(3,4,5-trifluorophenoxy)methyl)-2,3',5'-trifluoro-[1,1'-biphenyl]-4-yl)oxy)nonyl 2',3,5,6'-tetrafluoro-4'-(5-propyl-1,3-dioxan-2-yl)-[1,1'-biphenyl]-4-carboxylate (I-9)***

The synthesis was carried out in the same manner as in I7.

Yield 1.07 (42.87 %)

Purity 98.66% (HPLC-MS)

MS(EI) m/z: 941; 890; 860; 822; 774; 695; 673; 593; 481; 459; 381; 281; 273; 253

$^{1}H$ NMR (500 MHz, $CDCl_3$) δ: 7.32 (t, *J*=8.85 Hz, 1 H); 7.07 (m, 8 H); 6.74 (m, 2 H); 5.38 (s, 1 H); 4.37 (t, *J*=6.56 Hz, 2 H); 4.23 (dd, *J*=11.75, 4.73 Hz, 2 H); 3.97 (t, *J*=6.56 Hz, 2 H); 3.53 (t, *J*=11.44 Hz, 2 H); 2.13 (m, 1 H); 1.78 (m, 4 H); 1.37 (m, 12 H); 1.09 (m, 2 H); 0.92 (t, *J*=7.32 Hz, 3 H)

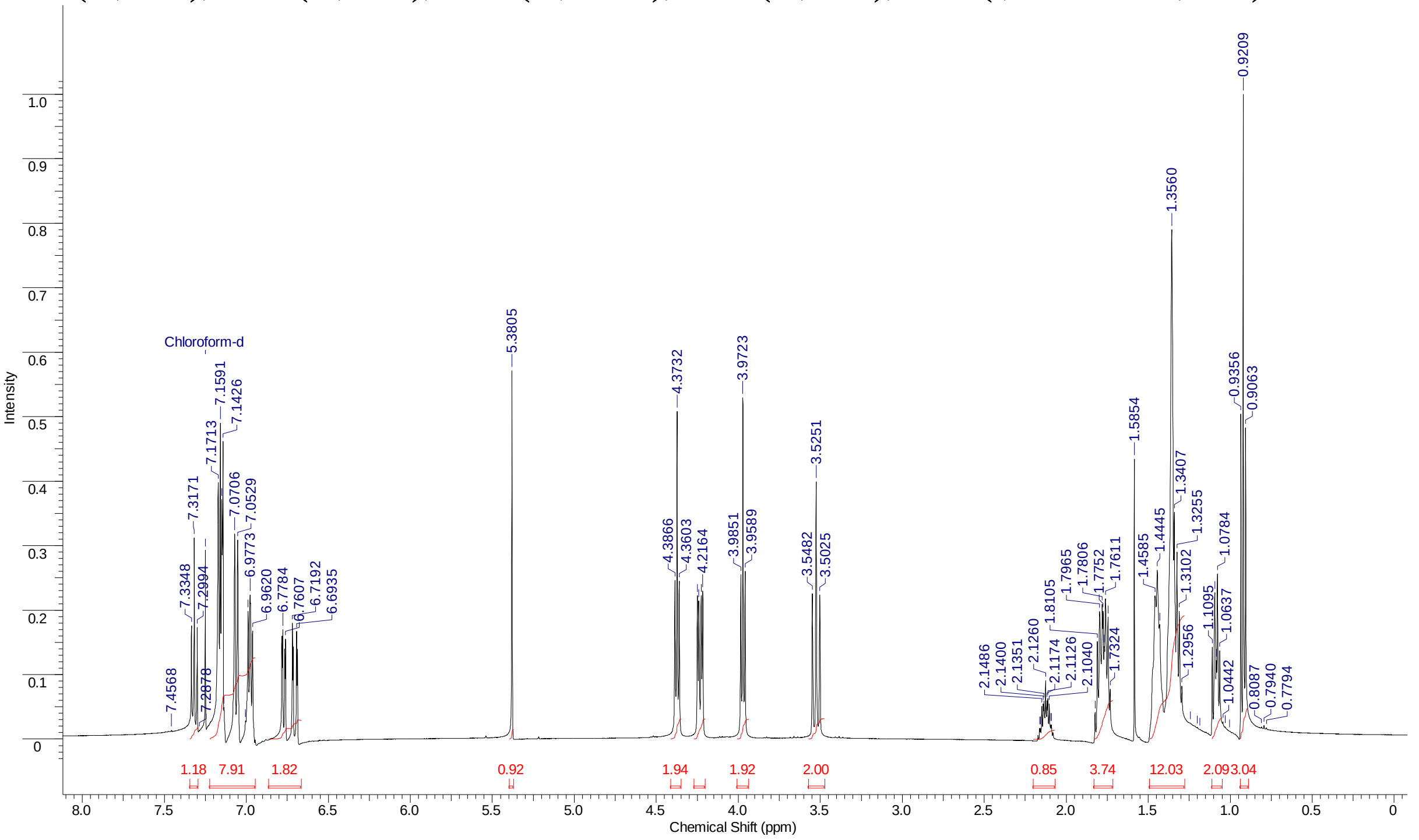

$^{13}C$ NMR (126 MHz, $CDCl_3$) δ: 161.28; 160.44 (d, *J* = 249.78 Hz); 160.29 (dd, *J* = 256.59, 7.72 Hz); 159.91 (dd, *J* = 257.05, 6.36 Hz); 159.41 (dd, *J* = 250.69, 6.36 Hz); 150.98 (ddd, *J*= 250.69, 10.9, 5.45 Hz); 144.71 (m) 142.12 (t, *J* = 9.54 Hz); 141.66 (t, *J* = 10.90 Hz); 138.11 (dt, *J*=249,78, 15.44 Hz); 134.03 (t, *J* = 10.90 Hz); 130.41 (d, *J* = 4.54 Hz); 120.25 (t, *J*=117.72 Hz); 117.35 (d, *J* = 12.72 Hz); 115.41 (t, *J*=9.1Hz); 114.09 (dd, *J* = 24.52, 1.82 Hz); 112.53 (m); 111.38 (m, *J* = 2.72 Hz); 110.90 (t, *J* = 18.17 Hz); 109.94 (dd, *J*=21.8, 5.45 Hz); 107.40 (m); 102.77 (d, *J* = 25.43 Hz); 99.17 (t, *J* = 2.27 Hz); 72.54; 68.58; 66.12; 33.90; 30.24; 29.33; 29.17; 29.03; 28.99; 28.49; 25.89; 25.75; 19.51; 14.16

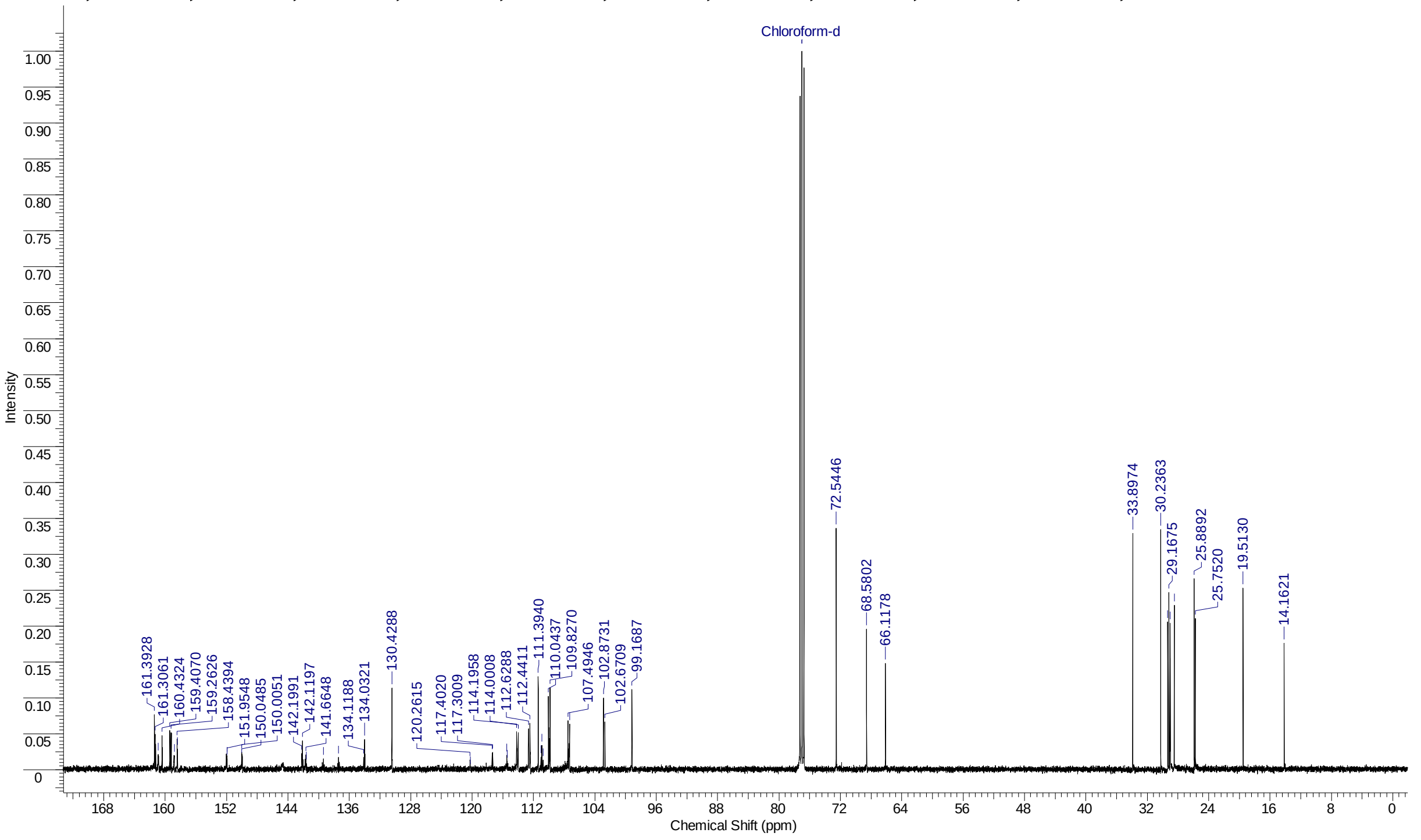

***10-((4'-(difluoro(3,4,5-trifluorophenoxy)methyl)-2,3',5'-trifluoro-[1,1'-biphenyl]-4-yl)oxy)decyl 2',3,5,6'-tetrafluoro-4'-(5-propyl-1,3-dioxan-2-yl)-[1,1'-biphenyl]-4-carboxylate (I-10)***

The synthesis was carried out in the same manner as in I-7.

Yield 0.62g (30.76 %)

Purity 99.71% (HPLC-MS)

MS(EI) m/z: 956; 907; 872; 789; 709; 687; 611; 583; 483; 458; 411; 381; 299; 281; 273;

$^1$H NMR (500 MHz, $CDCl_3$) δ: 7.32 (t, *J*=8.85 Hz, 1 H); 7.16 (m, 4 H); 7.06 (d, *J*=9.16 Hz, 2 H); 6.98 (m, 2 H); 6.73 (m, 2 H); 5.38 (s, 1 H); 4.37 (t, *J*=6.56 Hz, 2 H); 4.23 (dd, *J*=11.44, 4.43 Hz, 2 H); 3.97 (t, *J*=6.41 Hz, 2 H); 3.52 (t, *J*=11.29 Hz, 2 H); 2.12 (m, 1 H); 1.78 (m, 4 H); 1.40 (m, 14 H); 1.09 (m, 2 H); 0.92 (t, *J*=7.32 Hz, 3 H)

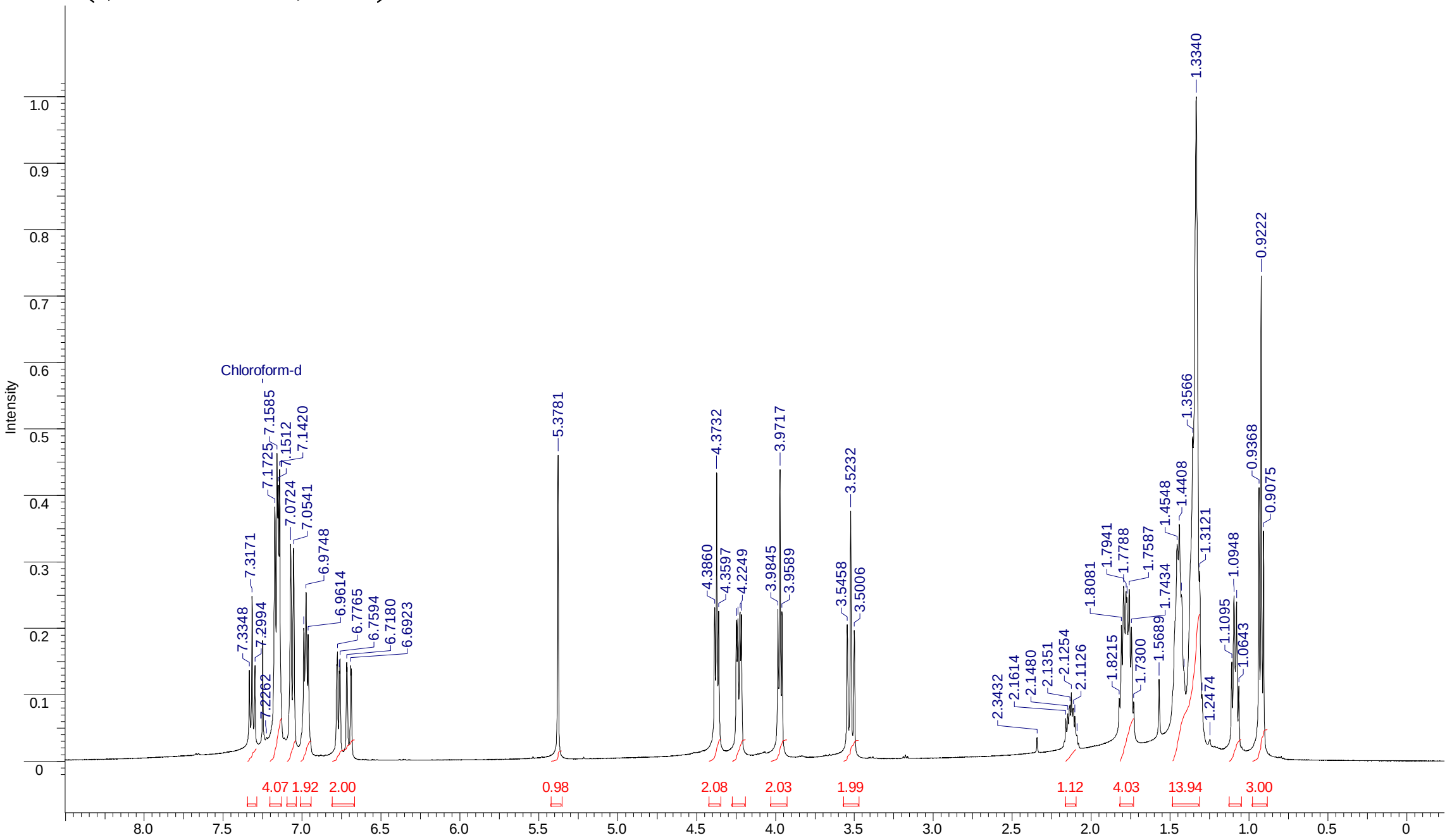

$^{13}$C NMR (126 MHz, $CDCl_3$) δ: 161.41 (dd, *J* = 251.6, 6.36 Hz); 161.24; 160.28 (dd, *J* = 258.86, 7.27 Hz); 159.52 (dd, *J* = 253.13, 6.4 Hz); 159.41 (dd, *J* = 250.69, 6.36 Hz); 150.51 (ddd, *J*=251.6, 10.97, 5.47 Hz ); 144.68 (m);142.09 (t, *J* = 9.99 Hz); 141.66 (t, *J*=11.81 Hz); 137.40 (dt, *J*=249.79, 15.44 Hz); 134.02 (t, *J* = 11.35 Hz); 130.40 (d, *J* = 4.54 Hz); 120.26 (t, *J*=117.5 Hz); 117.14 (d, *J*=11.83Hz); 114.94 (t, *J*=18.76 Hz); 114.08 (dd, *J* = 22.71, 2.41 Hz); 112.57 (m); 111.40 (m); 110.02 (t, *J* = 18.2 Hz); 109.85 (m); 107.40 (m); 102.77 (d, *J* = 26.34 Hz); 99.17 (t, *J*=2.45Hz); 72.54; 68.61; 66.14; 30.24; 29.34; 29.25; 29.07; 29.00; 28.50; 25.92; 25.76; 19.51; 14.15

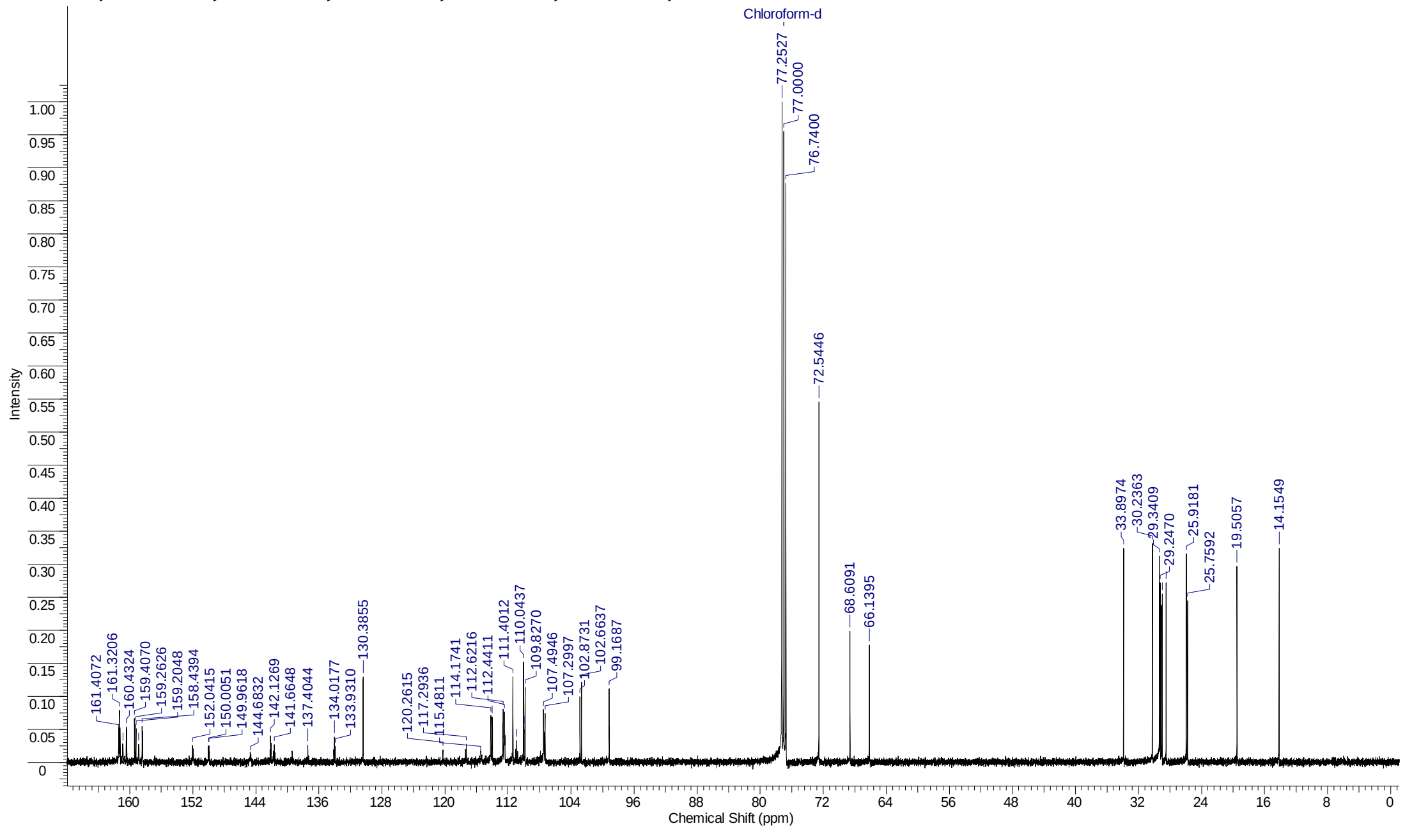

**Series II-m**

Figure S11 presents the stages of the synthesis of symmetric bimesogens – Series **II-m**. The synthesis was initiated by a Fischer esterification between 4-bromo-2-fluorobenzoic acid (**16**), previously converted into the corresponding acid chloride, and the alcohol pentane-1,5-diol. This step determined the length of the spacer linking the symmetrically structured bimesogens.

In the subsequent step, the obtained bromo derivative **(17)** was subjected to a Miyaura borylation reaction, and the product of this transformation **(18)** was then oxidized to afford pentane-1,5-diyl bis(2-fluoro-4-hydroxybenzoate) (**20**). The resulting dihydroxyphenol was subsequently subjected to Steglich esterification, carried out according to a standard protocol, with acids possessing terminal chains of varying length that contain a 1,4-dioxane ring in their structure (**19a-19d**). These acids were obtained via a synthetic route analogous to that described above, leading to the formation of carboxylic acids with an extended core.

**Figure S11** Synthesis of Series **II-m** dimers

***Pentane-1,5-diyl bis(4-bromo-2-fluorobenzoate) (17)***

A solution of 4-bromo-2-fluorobenzoic acid (35.69 g; 0.163 mol), oxalyl chloride (16.65 ml; 0.1939 mol), toluene (600 ml), and 2–3 drops of DMF was stirred for 3 hours. After completion of the reaction, the excess oxalyl chloride was distilled off under reduced pressure. The reaction mixture was then cooled to 50 °C, pentane-1,5-diol (8.29 ml; 0.0793 mol) and pyridine (52.1 ml; 0.64 mol) were added. The mixture was stirred at 50 °C for 4 hours and then allowed to gradually reach room temperature. After the reaction was completed, a 10% aqueous solution of HCl was added. The organic phase was separated, washed three times with water, and dried over anhydrous $MgSO_4$. The solvent was removed under reduced pressure using a rotary evaporator. The crude product was purified by column chromatography using dichloromethane and subjected to further reactions.

Yield 39.21 g (48.63%)

Purity 87.1% (GC-FID)

MS(EI)m/z: 506 (M+); 478; 451; 419; 393; 375; 357; 339; 321; 286; 268; 244; 219; 201; 173; 155

[1]H NMR (500 MHz, $CDCl_3$) δ: 7.78 (t, *J*=8.09 Hz, 2 H); 7.31 (m, 4 H); 4.34 (t, *J*=6.41 Hz, 4 H); 1.82 (m, 4 H); 1.60 (m, 2 H)

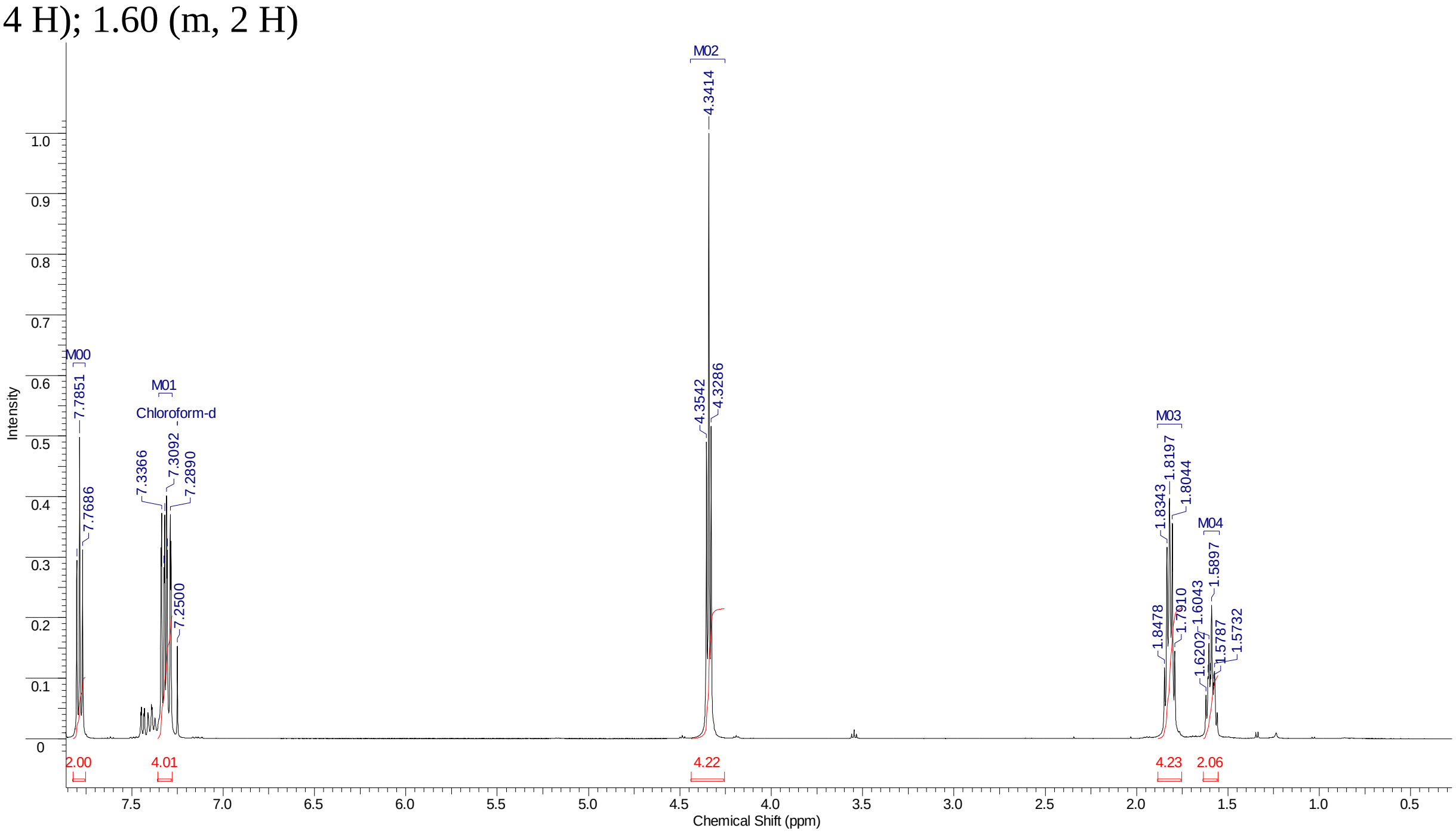

[13]C NMR (126 MHz, $CDCl_3$) δ: 163.75 (d, *J*=3.63 Hz); 161.57 (d, *J*=265.23 Hz); 133.08 (d, *J*=1.82 Hz); 127.78 (d, *J*=9.99 Hz); 127.48 (d, *J*=4.54 Hz); 120.62 (m); 117.89 (d, *J*=9.99 Hz); 65.22; 28.13; 22.46

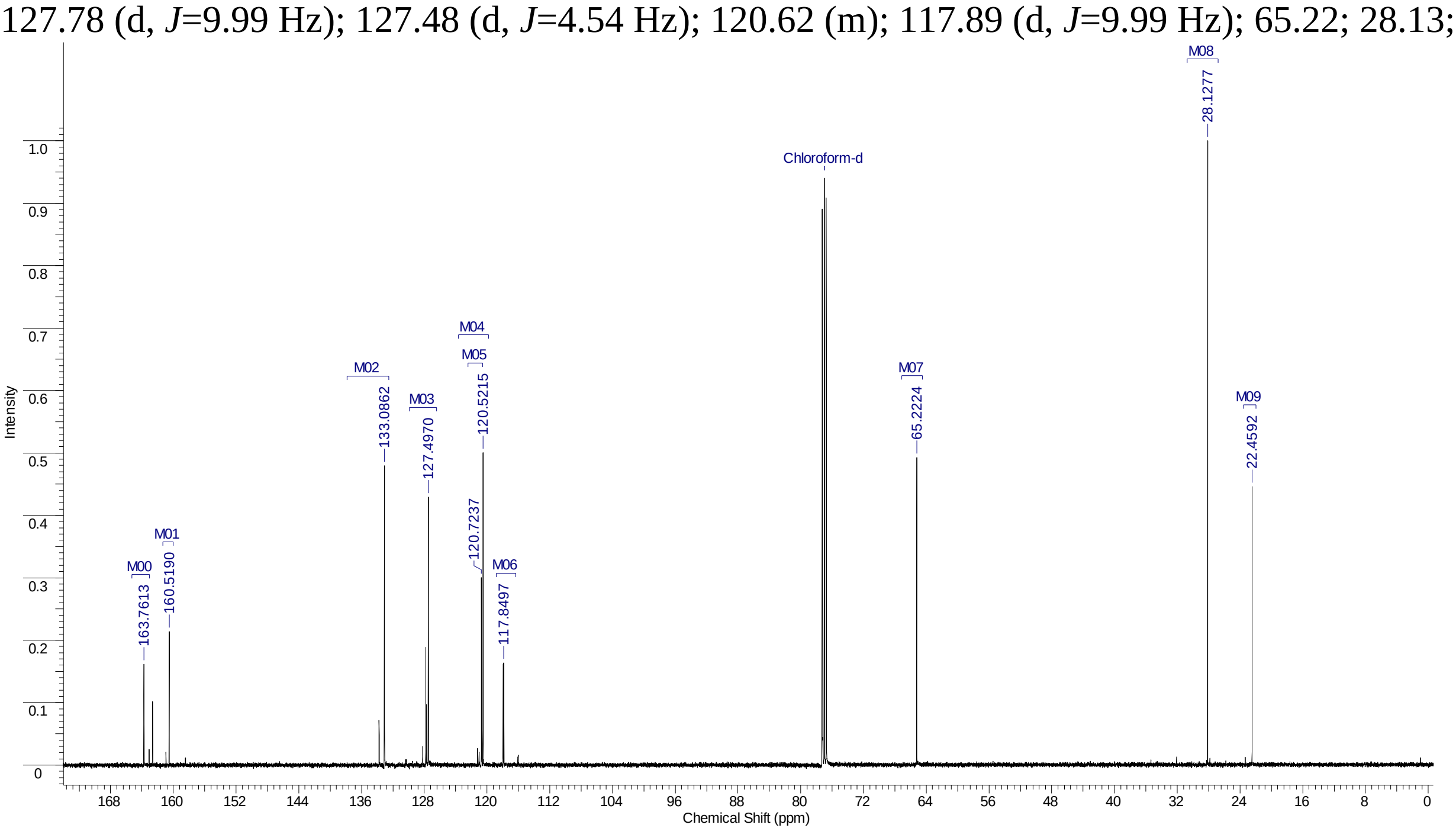

***Pentane-1,5-diyl bis(2-fluoro-4-(4,4,5,5-tetramethyl-1,3,2-dioxaborolan-2-yl)benzoate) (18)***

A solution of pentane-1,5-diyl bis(4-bromo-2-fluorobenzoate) (39.21 g; 0.0793 mol), bis(pinacolato)diboron (44.83 g ; 0.1765 mol), anhydrous potassium acetate (43.11 g ; 0.1439 mol) in toluene (450 ml) and 1,4-dioxane (450 ml) was flushed with nitrogen at 80°C for 1h. Then the catalyst Pd(dppf)$Cl_2$ (3% mol) was added and reaction was stirred at 80°C for 5h. Next reaction was cooled to room temperature and poured into 5% HCl solution. Product was extracted with toluene; organic layer was washed three times with water and dried over $MgSO_4$. Solvent was evaporated under vacuum. Side products were distilled of using short-path vacuum distillation apparatus, and the product was purified by liquid column chromatography (Ethyl acetate : hexane – 2:1 as a mobile phase)

Yield 29.6 g (62.2%)
Purity 76.86% (GC-FID) (the only impurity was $B_2pin_2$ as seen on $^{13}C$ NMR)
MS(EI) m/z: 600 (M+); 579; 543; 519; 485; 457; 428; 399; 371; 334; 306; 273; 249; 220; 180;149; 122

$^1H$ NMR (500 MHz, $CDCl_3$) δ: 1.35 (s, 24 H); 1.64 (m, 2 H); 1.87 (m, 4 H); 4.36 (t, *J*=6.56 Hz, 4 H); 7.57 (m, 4 H); 7.90 (t, *J*=7.32 Hz, 2 H)

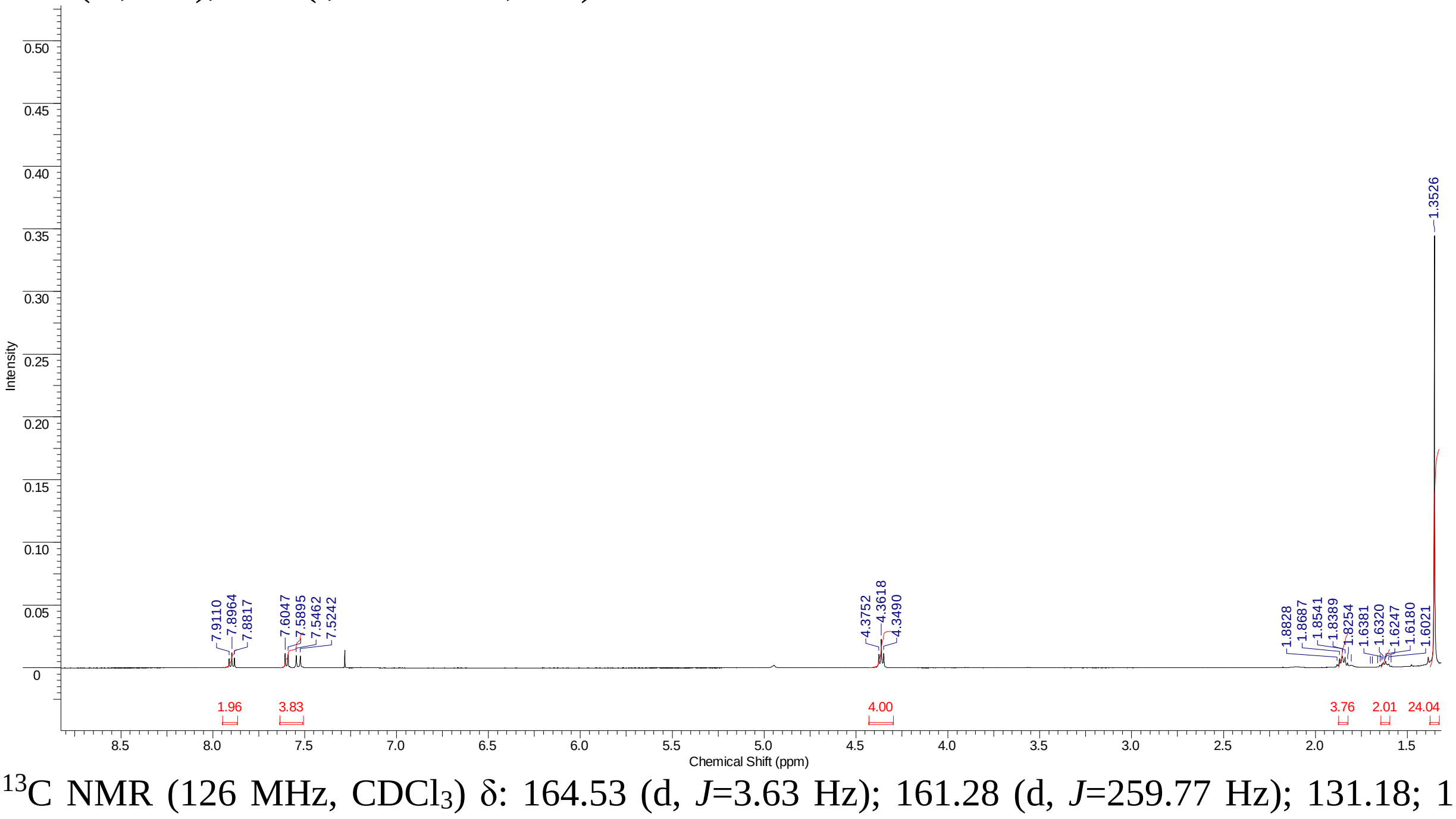

$^{13}C$ NMR (126 MHz, $CDCl_3$) δ: 164.53 (d, *J*=3.63 Hz); 161.28 (d, *J*=259.77 Hz); 131.18; 129.81 (d, *J*=3.63 Hz); 122.61 (d, *J*=20.89 Hz); 120.91 (d, *J*=9.99 Hz); 84.40;m65.12; 28.19; 24.80; 22.50

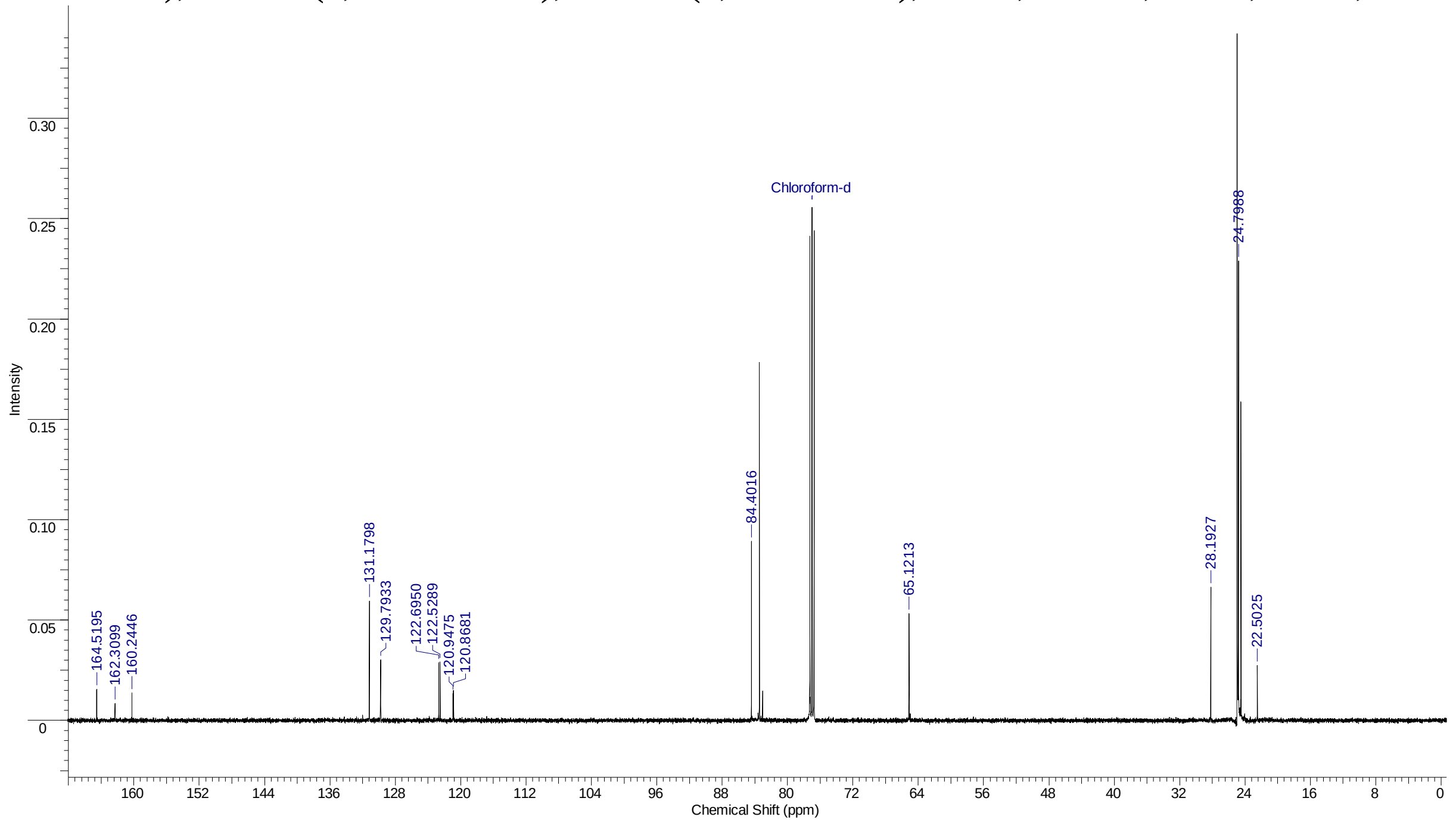

***2,6-difluoro-4-(5-propyl-1,3-dioxan-2-yl)benzoic acid (19a)***

2-(3,5-difluorophenyl)-5-propyl-1,3-dioxane (45g; 0.186mol) was mixed with anhydrous THF (300cm3) under nitrogen and cooled to −78°C in an acetone/dry ice bath. Solution of n- butyllithium dissolved in hexane (0.2mol, 2.5M) was added dropwise, and temperature was kept below −70°C. The reaction mixture was stirred for 2.5h in −78°C. Then the reaction was flushed with gaseous carbon dioxide (not exceeding the -65°C). After the reaction was completed, it was allowed to reach room temperature. 10% solution of $H_2SO_4$ was added to the mixture. Then the crude product was filtered off, washed with water and hexane. Product was recrystallized three times from hexane-toluene mixture (1:1 v/v).

Yield 28 g (52%).

Purity 99.15% (GC-MS)

MS(EI) (MS spectra of corresponding methyl ester) m/z: 299 (M+); 269, 241, 201, 169, 141, 129

$^1$H NMR (500 MHz, $CDCl_3$) δ: 11.61 (s, 1 H); 7.15 (d, *J*=9.16 Hz, 2 H); 5.39 (s, 1 H); 4.25 (dd, *J*=11.90, 4.58 Hz, 2 H); 3.54 (t, *J*=11.44 Hz, 2 H); 2.14 (m, 1 H); 1.34 (m, 2 H); 1.10 (m, 2 H); 0.93 (t *J*=7.48 Hz, 3 H)

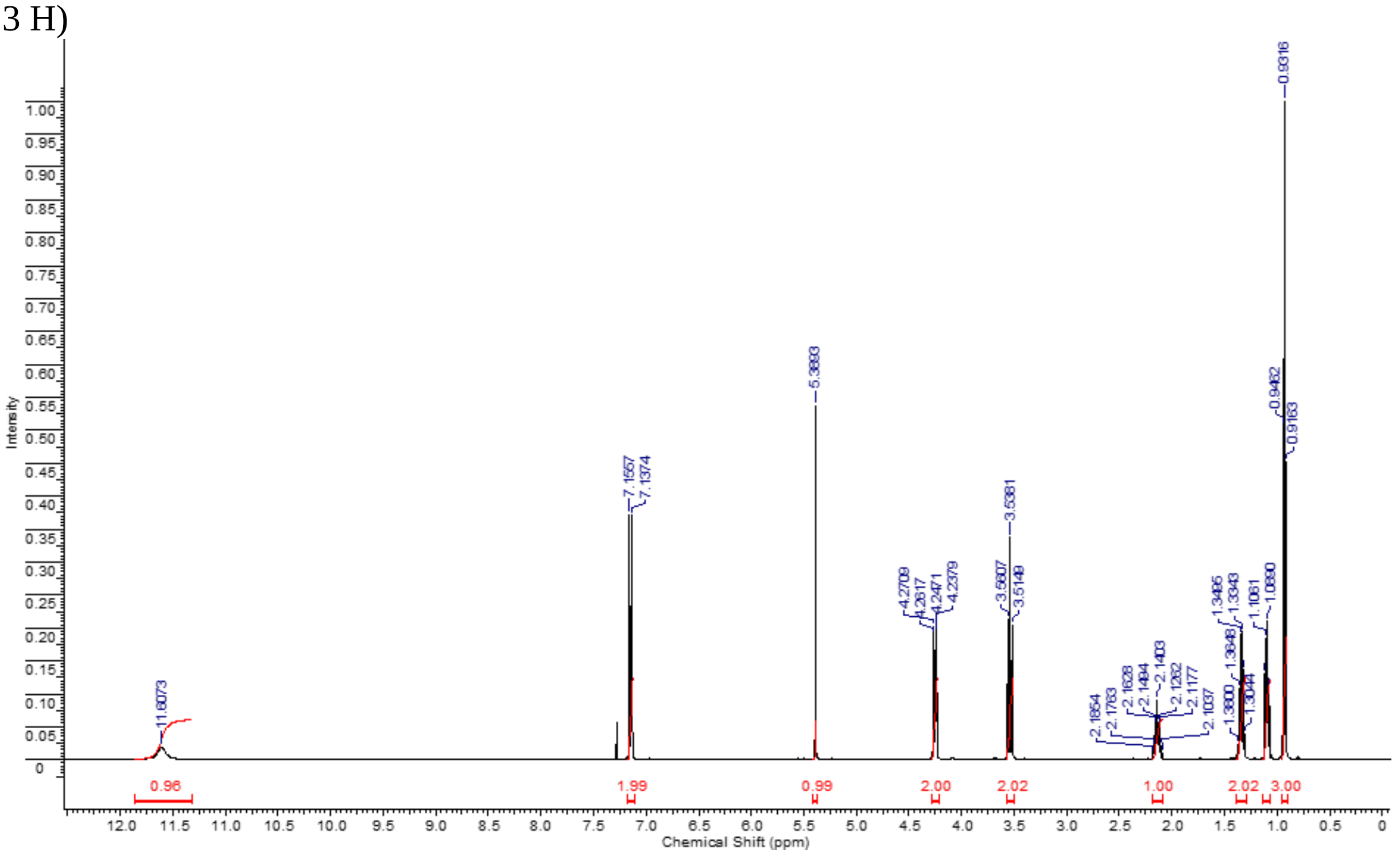

$^{13}C$ NMR (126 MHz, $CDCl_3$) δ: 166.82; 161.20 (dd, *J*=258.86, 5.45 Hz); 145.36 (t, *J*=9.99 Hz); 110.27 (m); 109.43 (t, *J*=16.35 Hz); 98.86; 72.56; 33.88; 30.20; 19.52; 14.18

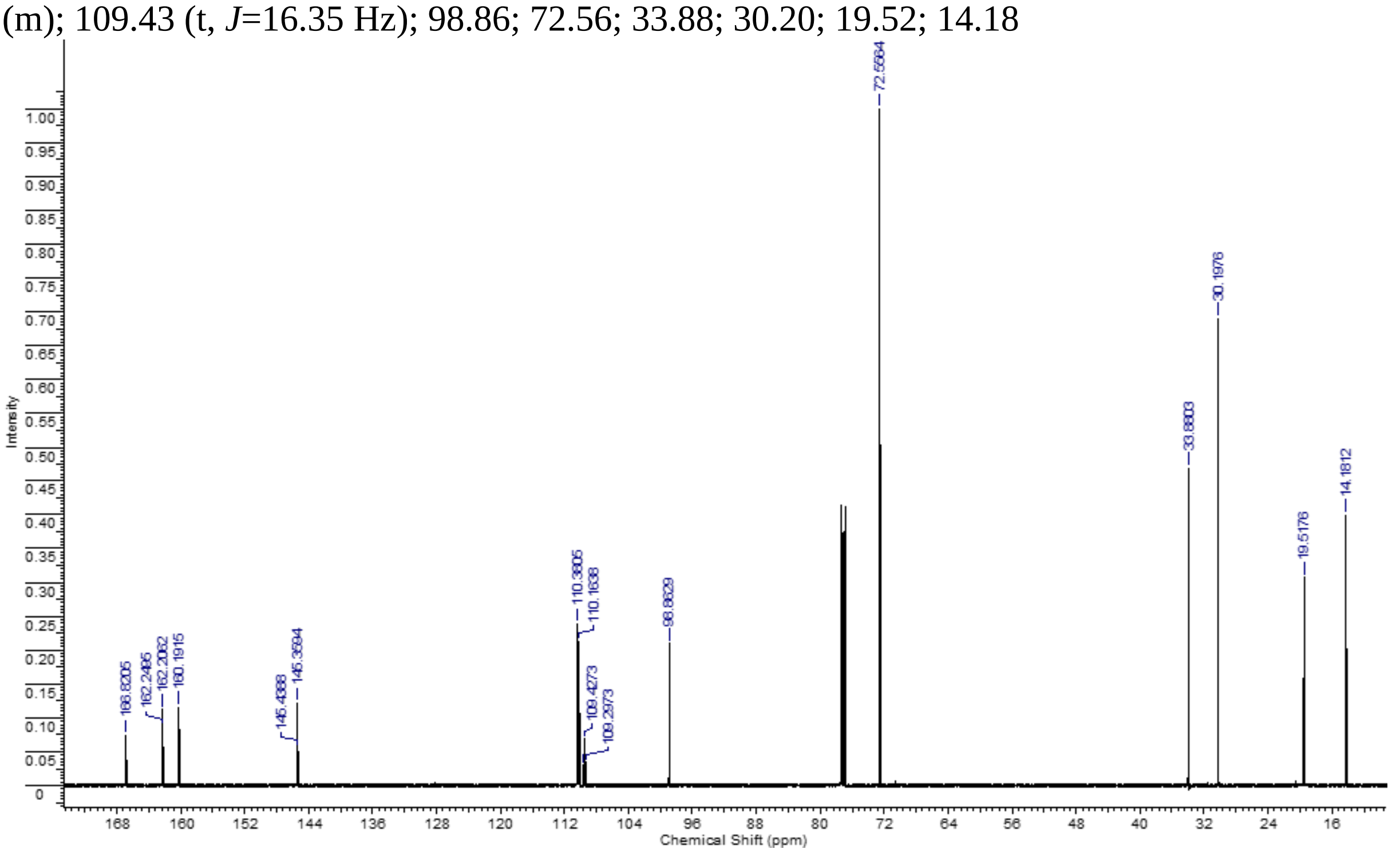

Compounds 19b and 19c were synthesized in the same manner as 19a.

***Pentane-1,5-diyl bis(2-fluoro-4-hydroxybenzoate) (20)***

Flask containing pentane-1,5-diyl bis(2-fluoro-4-(4,4,5,5-tetramethyl-1,3,2-dioxaborolan-2-yl)benzoate) (29.60 g; 0.049 mol), 30% solution of $H_2O_2$ (100 ml), tetrahydrofuran (300 ml) and water (300 ml) was mixed at room temperature for 4h. After the reaction was completed THF was evaporated and 5% HCl solution was added. Product was extracted with dichloromethane, organic layer was washed three times with water, dried over $MgSO_4$ and concentrated. The crude product purified by flash column chromatography (ethyl acetate:hexane as a mobile phase.)

Yield 7 g (37 %)

Purity 80.6% (GC-FID)

MS(EI) (MS spectra of corresponding trimethylsilyl ether) m/z: 524 (M+); 524; 494; 473; 452; 423; 393; 369; 351; 331; 306; 285; 266; 249; 228; 211; 193; 168

[1]H NMR (500 MHz, DMSO-*d6*) δ: 10.77 (s, 2 H); 7.73 (t, *J*=8.85 Hz, 2 H); 6.63 (m, 4 H); 4.21 (t, *J*=6.41 Hz, 4 H); (3.93 and 1.97 - THF, 3,38 - $H_2O$ signal); 1.70 (dd, *J*=14.50, 6.87 Hz, 4 H); 1.49 (m, 2 H)

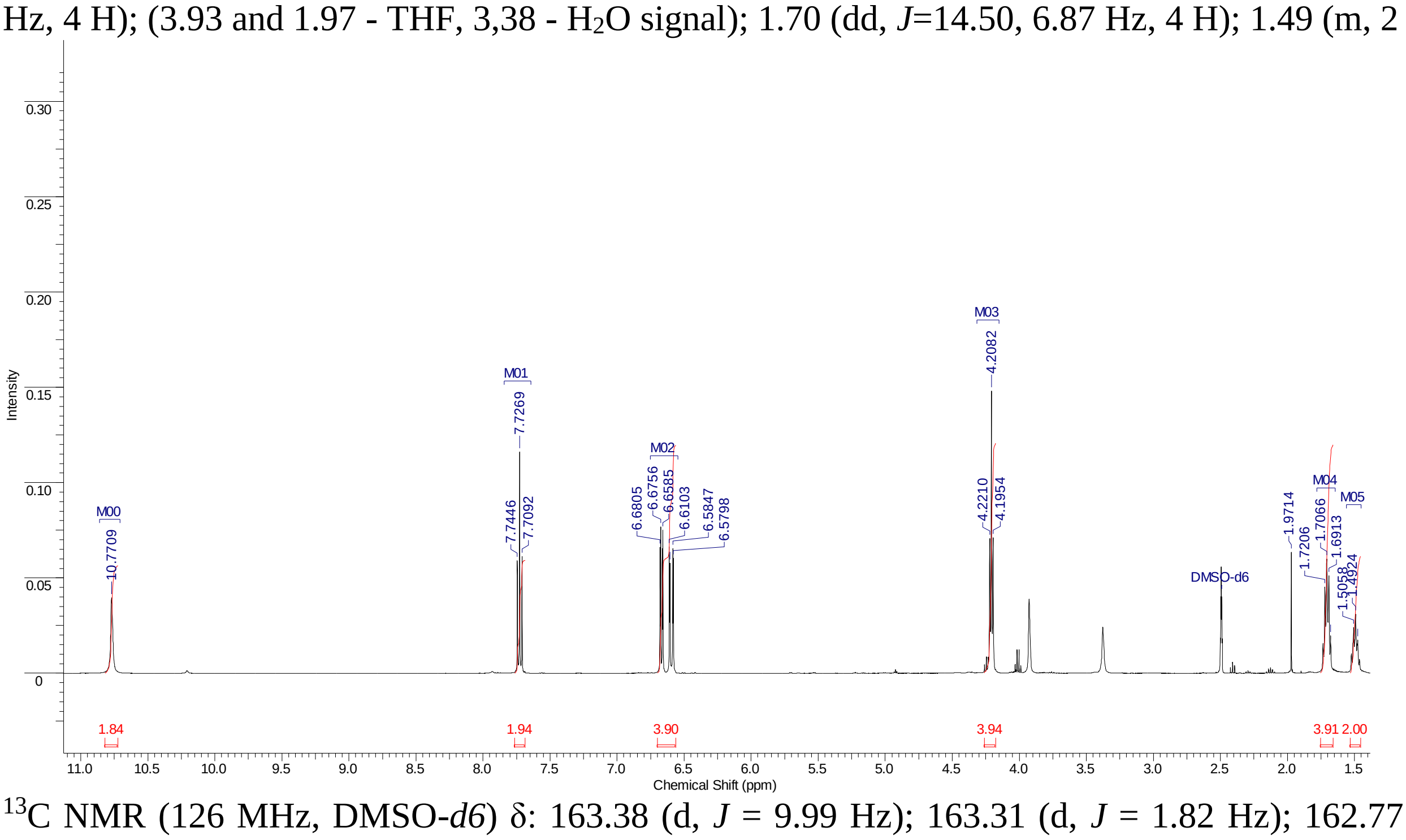

[13]C NMR (126 MHz, DMSO-*d6*) δ: 163.38 (d, *J* = 9.99 Hz); 163.31 (d, *J* = 1.82 Hz); 162.77 (d, *J* = 257.04 Hz); 133.22 (d, *J* = 3.63 Hz); 111.87 (d, *J* = 2.72 Hz); 108.74 (d, *J* = 9.99 Hz); 103.46 (d, *J* = 24.53 Hz); 64.05; 27.74; 24.92

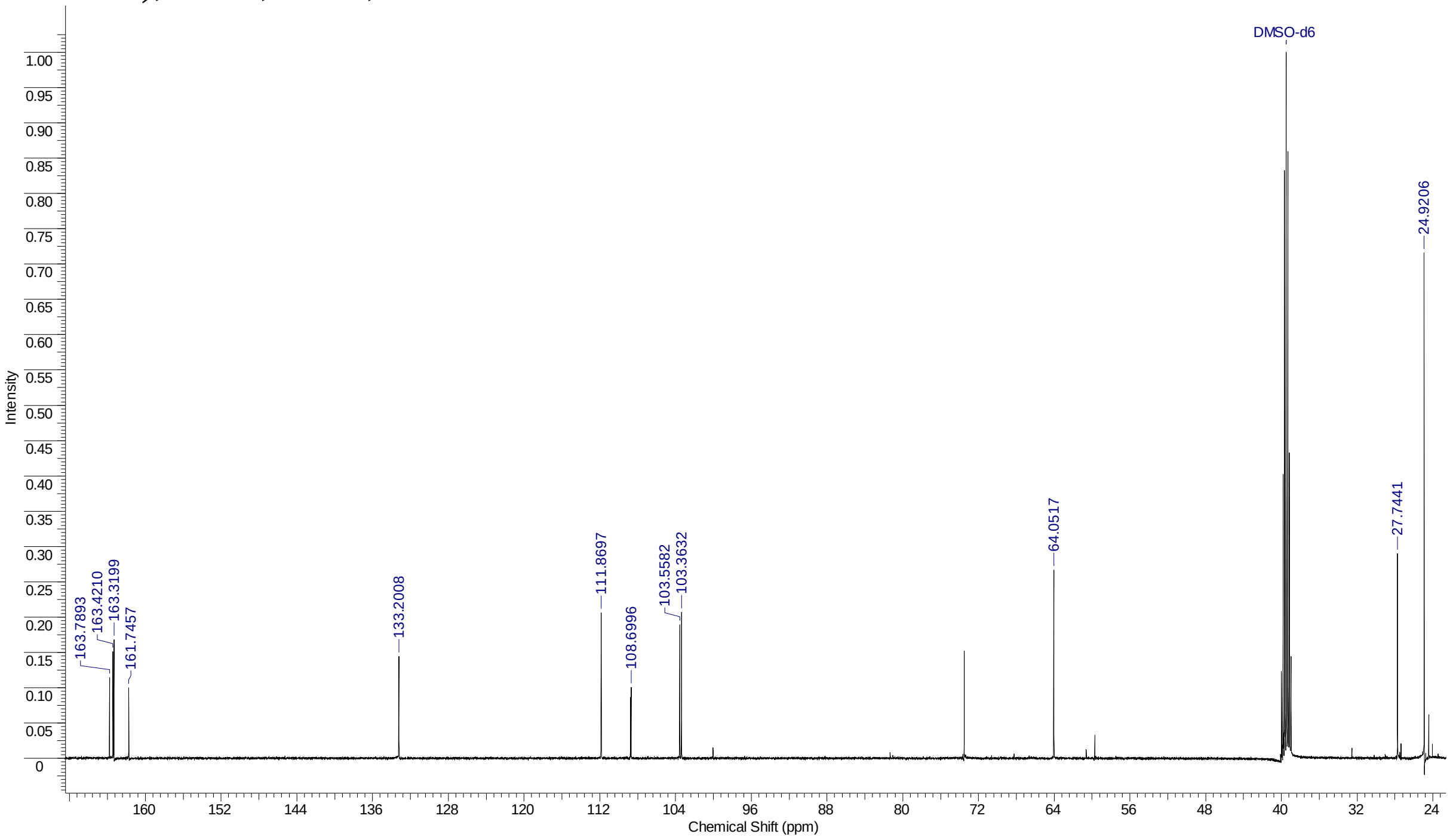

***((Pentane-1,5-diylbis(oxy))bis(carbonyl))bis(3-fluoro-4,1-phenylene) bis(2,6-difluoro-4-(5-propyl-1,3-dioxan-2-yl)benzoate) (II-3)***

To a stirred solution of 2,6-difluoro-4-(5-propyl-1,3-dioxan-2-yl)benzoic acid (1.48 g; 0.0052 mol), Pentane-1,5-diyl bis(2-fluoro-4-hydroxybenzoate) (2.2 g; 0.0057 mol) and DCC (1.29 g; 0.0057 mol) in DCM (50 ml), DMAP (0.1 g) was added and the solution was stirred at room temperature overnight. The reaction mixture was filtered through silica pad and the filtrate was concentrated under vacuum. Crude product was purified by column chromatography (DCM, hexane as a mobile phase). The residue was then recrystallized from ethanol to give white solid.

Yield 0.6 g (11.4%)
Purity 96.85% (HPLC)
MS(ESI) m/z: 939 (M+Na) 918; 799; 693; 602; 543
$^{1}$H NMR (500 MHz, $CDCl_3$) δ: 8.00 (t, *J*=8.39 Hz, 2 H); 7.15 (m, 8 H); 5.37 (s, 2 H); 4.37 (t, *J*=6.41 Hz, 4 H); 4.23 (dd, *J*=11.60, 4.58 Hz, 4 H); 3.52 (t, *J*=11.44 Hz, 4 H); 2.12 (m, 2 H); 1.87 (m, 4 H); 1.64 (m, 2 H); 1.34 (m, 4 H); 1.07 (m, 4 H); 0.92 (t, *J*=7.32 Hz, 6 H)

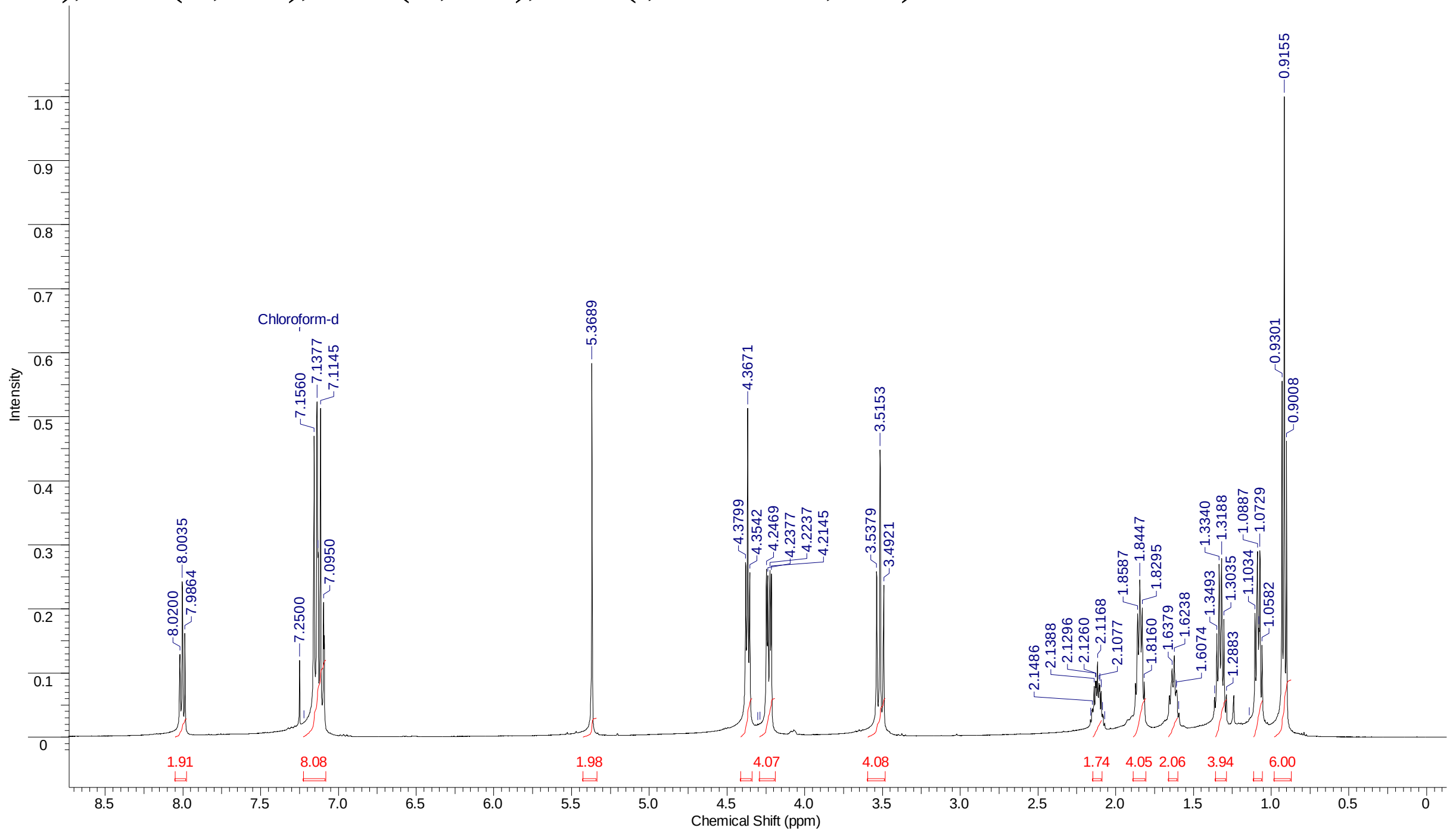

$^{13}$C NMR (126 MHz, $CDCl_3$) δ: 163.68 (d, *J*=4.54 Hz); 162.26 (d, *J*=262.49 Hz); 160.90 (dd, *J*=258.41, 5.90 Hz); 158.67; 154.28 (d, *J*=10.90 Hz); 145.70 (t, *J*=9.99 Hz); 133.04 (d, *J*=1.82 Hz); 117.38 (d, *J*=4.54 Hz); 116.88 (d, *J*=9.99 Hz); 110.95 (d, *J*=26.34 Hz); 110.27 (m, 1 C); 109.17 (t, *J*=16.81 Hz); 98.73; 72.52; 65.11; 33.85; 30.19; 28.19; 22.50; 19.48; 14.13

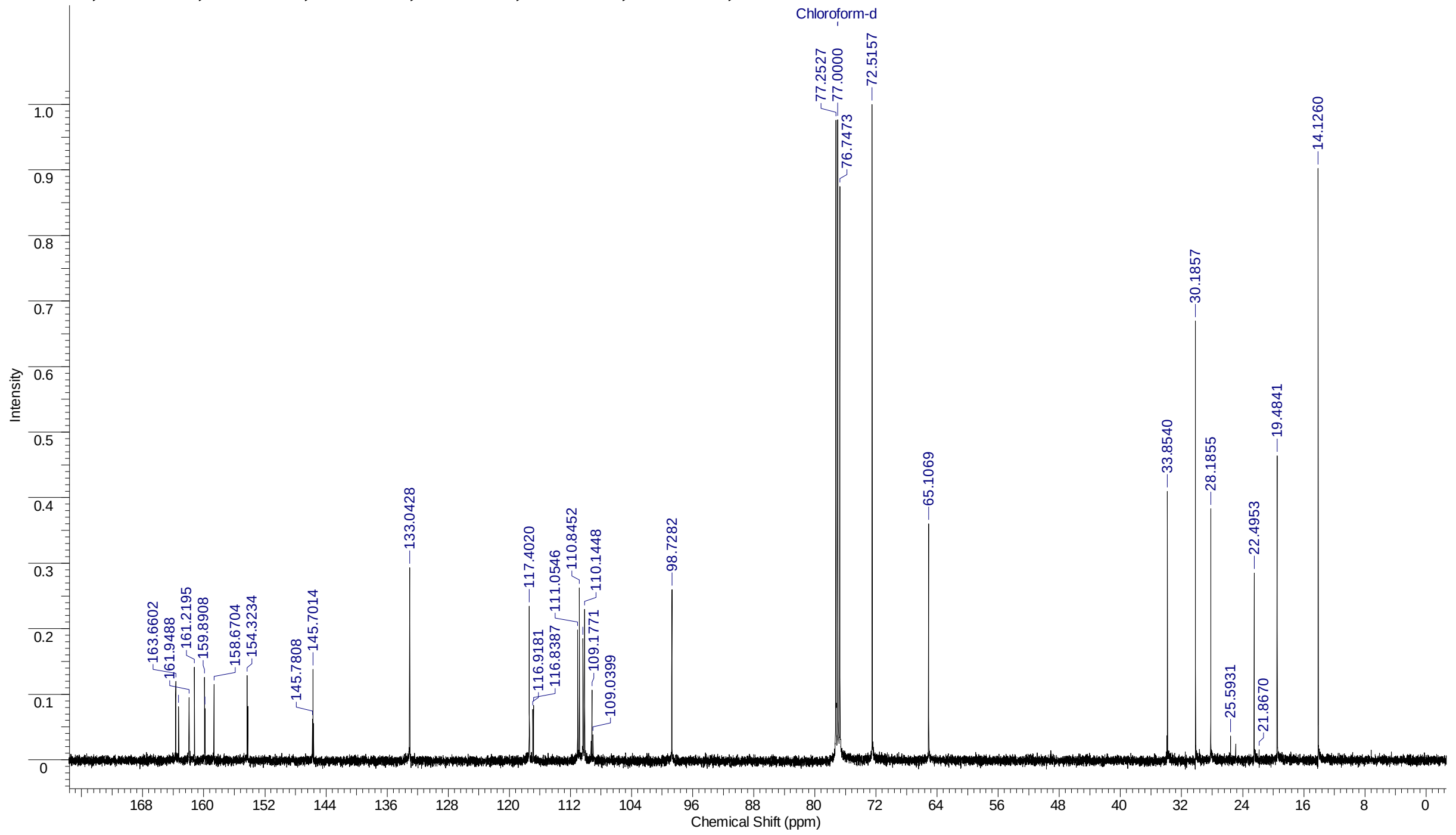

***((pentane-1,5-diylbis(oxy))bis(carbonyl))bis(3-fluoro-4,1-phenylene) bis(4-(5-butyl-1,3-dioxan-2-yl)-2,6-difluorobenzoate) (II-4)***

The synthesis was carried out in the same manner as in II-3.

Yield 1.1 g (11.4%)

Purity 99.45% (HPLC-MS)

MS(ESI) m/z: 967 (M+Na) 882; 795; 707; 640; 569; 472; 328;

$^{1}$H NMR (500 MHz, $CDCl_3$) δ: 8.01 (t, *J*=8.39 Hz, 2 H); 7.12 (m, 8 H); 5.37 (s, 2 H) 4.37; (t, *J*=6.56 Hz, 4 H); 4.24 (dd, *J*=11.75, 4.73 Hz, 4 H); 3.52 (t, *J*=11.44 Hz, 4 H); 2.10 (m, 2 H); 1.84 (m, 4 H); 1.63 (m, 2 H); 1.30 (m, 8 H); 1.10 (m, 4 H); 0.89 (t, *J*=7.02 Hz, 6 H);

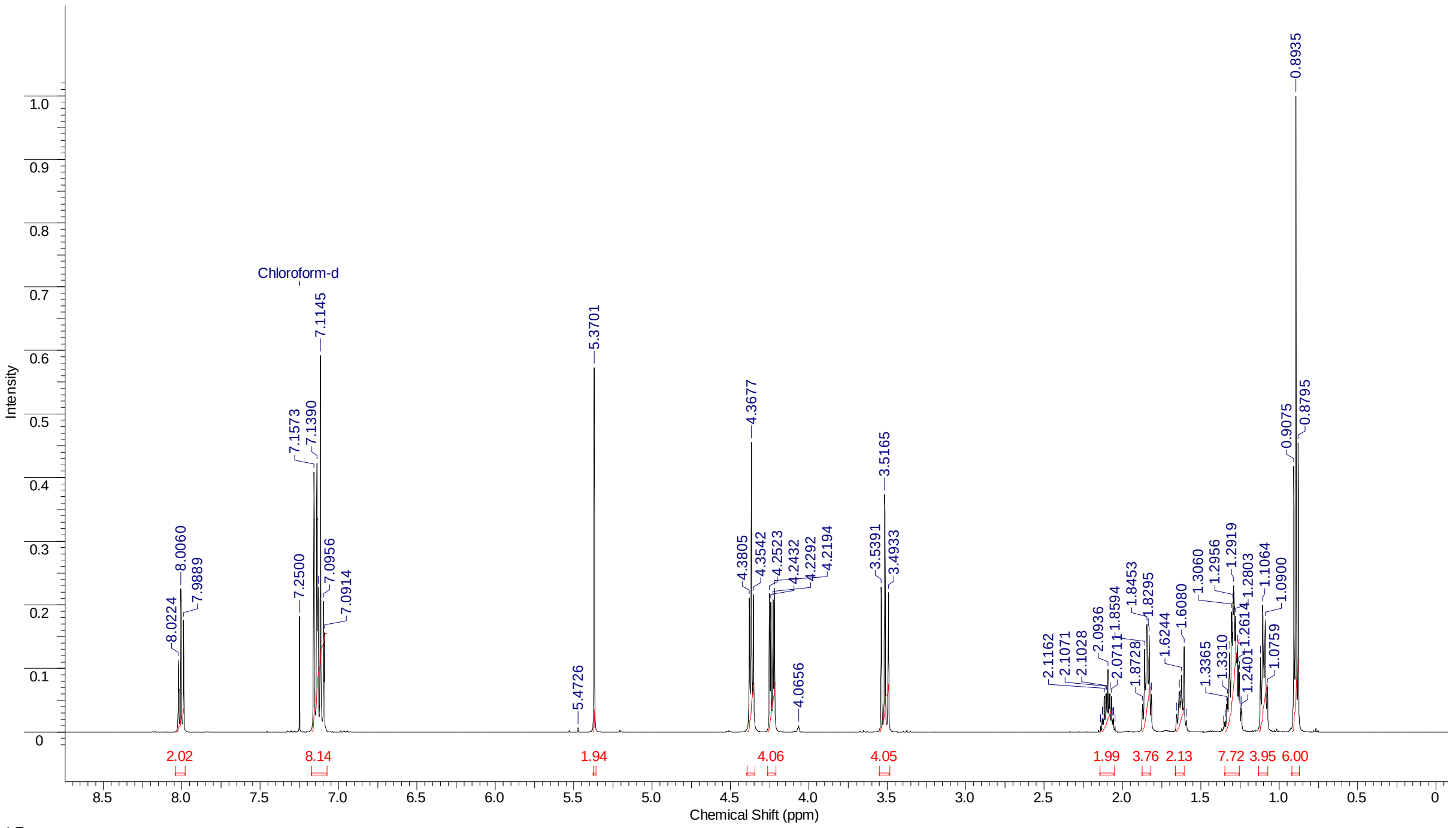

$^{13}$C NMR (126 MHz, $CDCl_3$) δ: 163.7 (d, *J*=4.5 Hz); 162.1(d, *J*=262.49 Hz); 161.92 (dd, *J*=158.4, 5.45 Hz); 158.67; 154.27 (d, *J*=10.90 Hz); 145.73 (t, *J*=9.99 Hz); 133.03 (d, J= 1.82Hz); 117.38 (d, *J*=4.54 Hz); 116.87 (d, *J*=9.99 Hz); 111.05 (d, *J*=26.34 Hz); 110.35 (m); 109.16 (t, *J*=16.80 Hz); 98.72; 72.54; 65.10; 34.08; 28.30; 28.18; 27.73; 22.76; 22.49 ;13.82

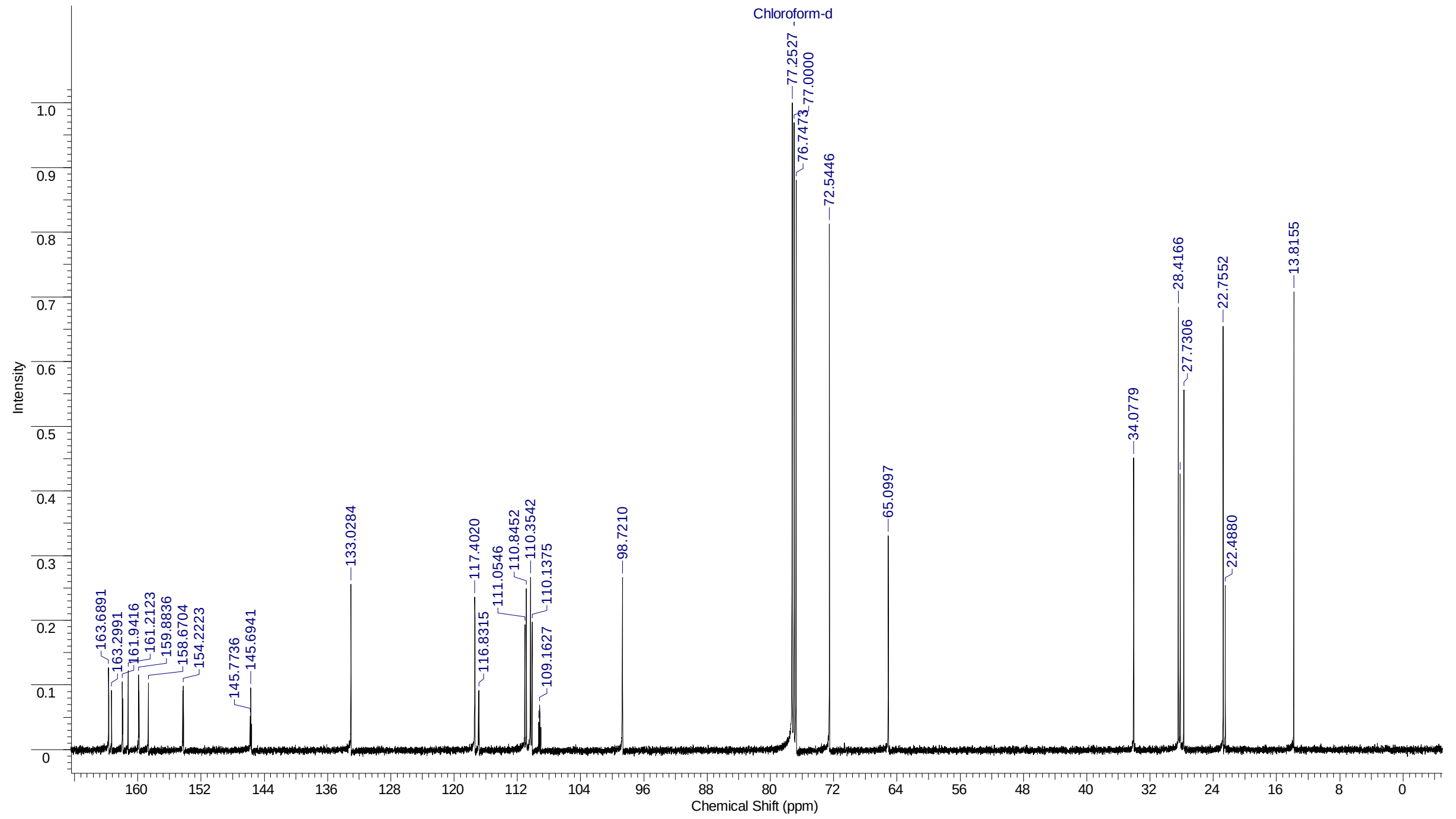

***((pentane-1,5-diylbis(oxy))bis(carbonyl))bis(3-fluoro-4,1-phenylene) bis(2,6-difluoro-4-(5-pentyl-1,3-dioxan-2-yl)benzoate) (II-5)***

The synthesis was carried out in the same manner as in II-3.

Yield 1.4 g (54%)

Purity 98.2% (HPLC-MS)

MS(ESI) m/z: 995 (M+Na) 875; 783; 720; 667; 554;405; 328; 251;

$^1$H NMR (500 MHz, $CDCl_3$) δ: 8.01 (t, *J*=8.39 Hz, 2 H); 7.12 (m, 8 H); 5.37 (s, 2 H); 4.36 (t, *J*=6.41 Hz, 4 H); 4.23 (dd, *J*=11.60, 4.58 Hz, 4 H); 3.51 (t, *J*=11.44 Hz, 4 H); 2.09 (m, 2 H); 1.84 (m, 4 H); 1.63 (m, 2 H); 1.28 (m, 12 H); 1.10 (m, 4 H); 0.88 (m, 6 H)

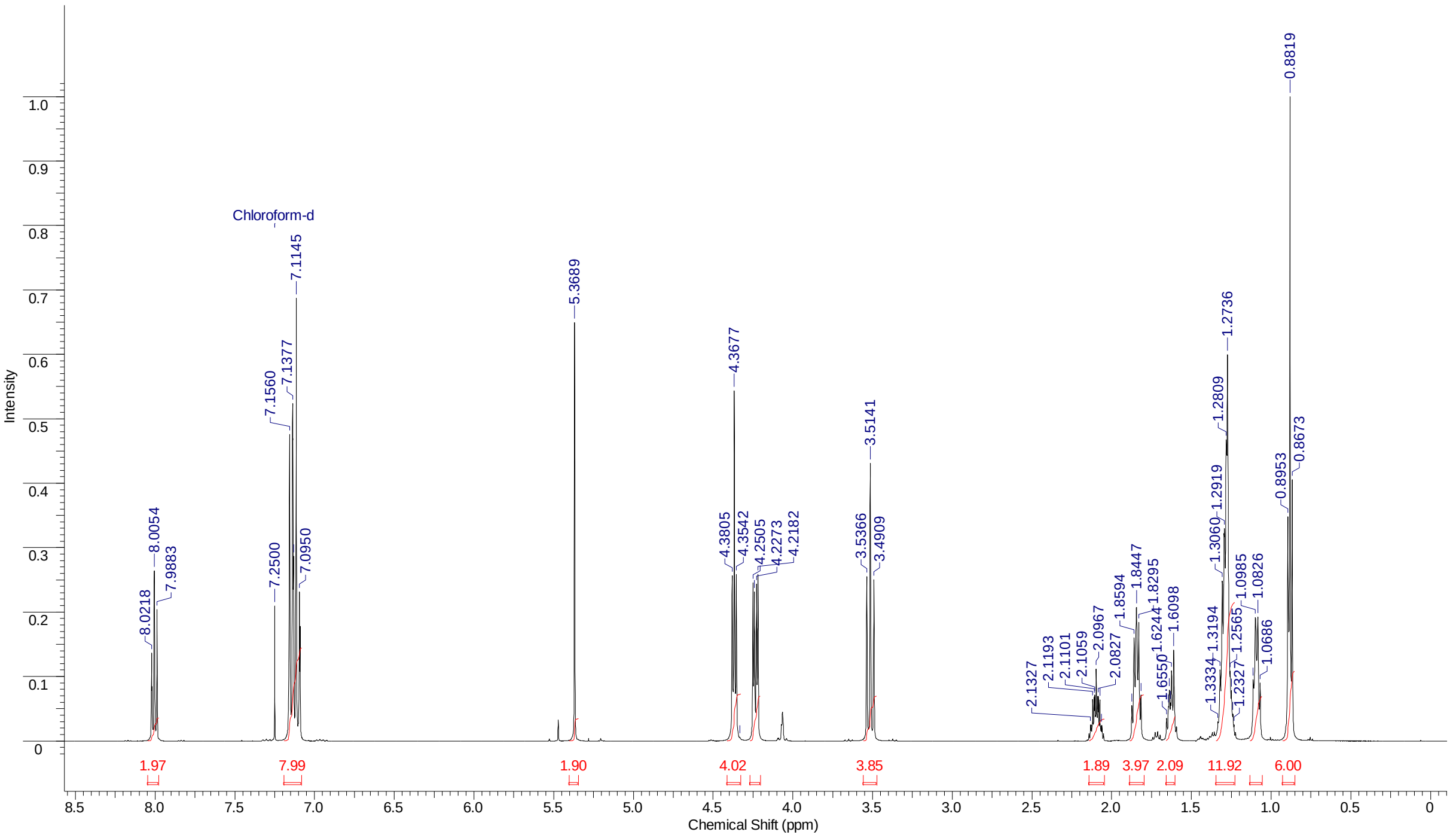

$^{13}$C NMR (126 MHz, $CDCl_3$) δ: 163.7 (d, *J*=3.6 Hz); 162.1 (d, J= 262.49 Hz); 161.0 (dd, *J*=258.87, 5.45 Hz);158.7 (s); 154.3 (d, *J*=10.9 Hz); 145.65 (t, *J*=9.99 Hz); 133.04 (d, *J*=1.9 Hz); 117.39 (d, *J*=3.63 Hz); 116.88 (d, *J*=9.99 Hz); 110.35 (d, *J*=26.5 Hz); 110.15 (m); 110.15 (t, *J*=16.81 Hz); 98.74; 72.56; 65.11; 34.15; 31.84; 28.19; 28.00; 25.93; 22.60; 22.46; 13.98

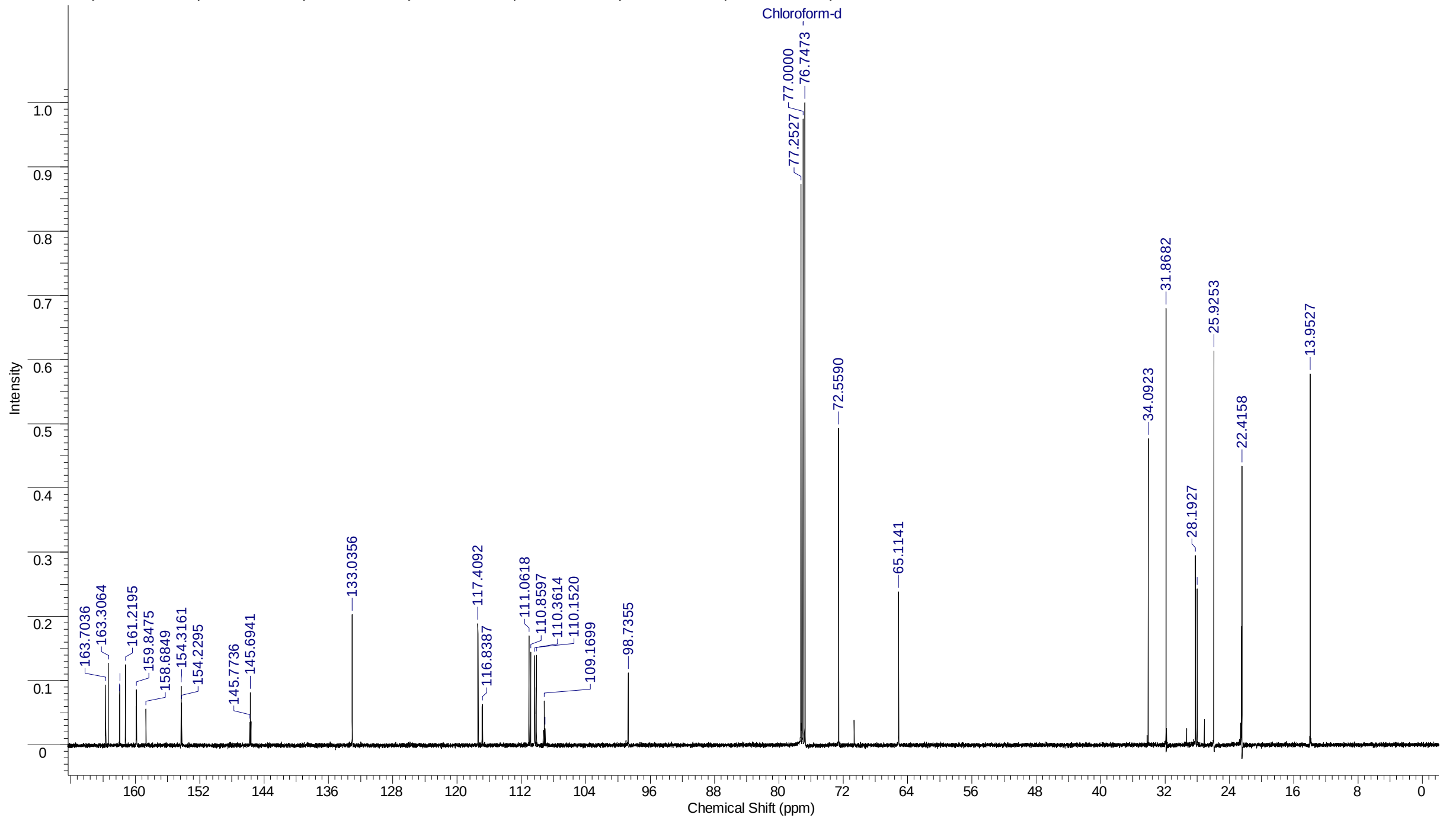